\documentclass[a4paper,11pt]{article}
\pdfoutput=1 

\usepackage{jheppub_mod} 

\usepackage[T1]{fontenc} 
\usepackage{afterpage}

\usepackage{amsmath,amsthm,verbatim,amssymb,amsfonts,amscd,graphicx,collectbox,mathtools,float}

\renewcommand{\vec}[1]{\mathbf{#1}}

\newcommand{\norm}[1]{\lVert#1\rVert}

\usepackage{subcaption}

\usepackage{tikz}
\usetikzlibrary{math}
\usetikzlibrary{arrows,shapes,positioning}
\usetikzlibrary{decorations.markings}
\tikzstyle arrowstyle=[scale=1]
\tikzstyle directed=[postaction={decorate,decoration={markings,
    mark=at position .65 with {\arrow[arrowstyle]{stealth}}}}]
\tikzstyle reverse directed=[postaction={decorate,decoration={markings,
    mark=at position .65 with {\arrowreversed[arrowstyle]{stealth};}}}]

\newlength{\mywidth}
\newcommand{\vv}{D}

\newtheorem{Lemma}{Lemma}

\usepackage[colorlinks=true
,urlcolor=blue
,anchorcolor=blue
,citecolor=blue
,filecolor=blue
,linkcolor=blue
,menucolor=blue
,pagecolor=blue
,linktocpage=true
,pdfproducer=medialab
,pdfa=true
]{hyperref}

\definecolor{ao(english)}{rgb}{0.0, 0.5, 0.0}
\definecolor{bittersweet}{rgb}{1.0, 0.44, 0.37}
\definecolor{blue-violet}{rgb}{0.54, 0.17, 0.89}
\definecolor{blue(munsell)}{rgb}{0.0, 0.5, 0.69}
\definecolor{battleshipgrey}{rgb}{0.52, 0.52, 0.51}
\definecolor{bole}{rgb}{0.47, 0.27, 0.23}
\definecolor{cottoncandy}{rgb}{1.0, 0.74, 0.85}
\definecolor{mediumblue}{rgb}{0.0, 0.0, 0.8}
\definecolor{amber}{rgb}{1.0, 0.75, 0.0}
\definecolor{bostonuniversityred}{rgb}{0.8, 0.0, 0.0}
\definecolor{cadetgrey}{rgb}{0.57, 0.64, 0.69}
\definecolor{capri}{rgb}{0.0, 0.75, 1.0}

\title{\boldmath From tree- to loop-simplicity in affine Toda theories II: higher-order poles and cut decompositions}

\author[a]{Patrick Dorey,} 
\author[a,b,c]{Davide Polvara}

\affiliation[a]{Department of Mathematical Sciences, Durham University, South Road, Durham DH1 3LE, United Kingdom }
\affiliation[b]{Dipartimento di Fisica e Astronomia,
Universita degli Studi di Padova, via Marzolo 8, 35131 Padova, Italy.}
\affiliation[c]{INFN,
Sezione di Padova, via Marzolo 8, 35131 Padova, Italy.}

\emailAdd{p.e.dorey@durham.ac.uk}
\emailAdd{davide.polvara@unipd.it}

\abstract{ 
Recently we showed how, in two-dimensional scalar theories, one-loop threshold diagrams can be cut into the product of one or more tree-level diagrams~\cite{First_loop_paper_sagex}.
Using this method on the ADE series of Toda models, we computed the double- and single-pole coefficients of the Laurent expansion of the S-matrix around a pole of arbitrary even order, finding agreement with the bootstrapped results. 
Here we generalise the cut method explained in~\cite{First_loop_paper_sagex} to multiple loops and use it to simplify large networks of singular diagrams. We observe that only a small number of cut diagrams survive and contribute to the expected bootstrapped result, while most of them cancel each other out through a mechanism inherited from the tree-level integrability of these models.  The simplification mechanism between
cut diagrams inside networks is reminiscent of Gauss’s theorem in the space of Feynman
diagrams.
}

\begin{document} 
\maketitle
\flushbottom


\section{Introduction}\label{sect_Intro}

Affine Toda models are a well-known class of (1+1)-dimensional quantum field theories, describing $r$ bosonic scalar fields $\phi=(\phi_1, \ldots, \phi_r)$ interacting through a Lagrangian 
\begin{equation}
\label{Toda_theory_lagrangian_defined_in_terms_of_roots}
\mathcal{L}=\frac{1}{2} \partial_\nu \phi_a  \partial^\nu \phi_a - \frac{\mu^2}{\beta^2} \sum_{i=0}^r n_i e^{\beta \alpha_i^a  \phi_a},
\end{equation}
which are known to be classically integrable \cite{Mikhailov:1980my,Olive:1984mb}. 
The quantities $\mu$ and $\beta$ are real constants setting the mass and the interaction scales of the model, while the vectors $\alpha_i$ ($i=1,\dots,r$) are the simple roots of a rank $r$ Lie algebra, and $\alpha_0$ is the lowest root defined through the linear combination
$$
\alpha_0= - \sum_{i=1}^r n_i \alpha_i .
$$
The integers $n_i$ depend on the Lie algebra considered, and we set $n_0=1$. In this paper, we focus on simply-laced  affine Toda theories, i.e.\ models based on Lie algebras belonging to the ADE series. Moreover, we assume the length common to all the roots (including $\alpha_0$) to be $\sqrt{2}$.
The additional root $\alpha_0$ is necessary to obtain a massive model with stable vacuum $\phi=0$, around which, Taylor-expanding the potential, we can compute scattering processes using standard perturbation theory.  The first non-trivial term that we find expanding the Lagrangian \eqref{Toda_theory_lagrangian_defined_in_terms_of_roots} is the mass matrix
\begin{equation}
\label{mass_spectrum_in_terms_of_roots_not_diagonalised}
M_{ab}^2=\mu^2 \sum_{i=0}^r n_i \alpha_i^a \alpha_i^b .
\end{equation}
All the higher-order couplings can be obtained from an expansion of the potential around $\phi=0$ after having properly diagonalised \eqref{mass_spectrum_in_terms_of_roots_not_diagonalised}.

There is an interesting correspondence between particles and Coxeter orbits in the root system, obtained as follows \cite{Dorey:1990xa,Freeman:1991xw}. We can suitably choose $r$ roots $\gamma_a$ ($a=1,\dots,r$) and to each of them associate a closed orbit $\Gamma_a$ composed of $h$ elements, where $h$ is  the Coxeter number of the Lie algebra considered. 
Each orbit is obtained by acting up to $h{-}1$ times on the root $\gamma_a$ with a Coxeter element $w$ of the Weyl group, which is a composition of Weyl reflections with respect to the planes orthogonal to the simple roots \cite{Kostant1}.
The duals of Feynman diagrams at particular imaginary rapidities of the external momenta for which both the external particles and the internal propagators are on shell, correspond to projections of root polytopes onto the eigenplane of $w$ for the eigenvalues $\exp \bigl( \pm \frac{2 \pi i}{h} \bigr)$. Therefore, the space on which the on-shell interacting particles live is a projection of the root space, generally known as spin-$1$ eigenplane of $w$, and the energy-momentum conservation law is obtained by projecting the root system onto this plane. 
The perturbative integrability of these theories has been recently proved at the tree level \cite{Davide_Patrick_tree_level_paper} and emerges from the underlying geometry, through which all the Feynman diagrams contributing to inelastic processes sum to zero. 
An important property used in the proof in~\cite{Davide_Patrick_tree_level_paper} is the fusing rule for the three-point couplings $C_{abc}$ \cite{Dorey:1990xa}, which says that 
$$
C_{abc} \ne 0 \ \  \text{iff} \ \  \exists \ \alpha \in \Gamma_a, \beta \in \Gamma_b \ \text{and} \ \gamma \in \Gamma_c \ \text{with} \ \alpha+\beta+\gamma=0, 
$$
where $\Gamma_a$, $\Gamma_b$ and $\Gamma_c$ are Coxeter orbits obtained by acting on the representative roots $\gamma_a$, $\gamma_b$ and $\gamma_c$ with $w$. Moreover, whenever a $3$-point coupling is different from zero in a simply-laced theory, it is equal to \cite{Braden:1989bu,Fring:1991me}
\begin{equation}
\label{Connection_among_three_point_couplings_and_areas}
C_{abc}=\frac{4 \beta}{\sqrt{h}} \sigma_{abc} \Delta_{abc} \hspace{4mm} \text{with} \hspace{4mm}  \sigma_{abc} = \pm 1 \ ,
\end{equation}
where $\Delta_{abc}$ is the area of the triangle with sides equal to the masses of the particles $a$, $b$ and $c$, which is the projection of the triangle composed of the roots $\alpha$, $\beta$ and $\gamma$ onto the spin-$1$ eigenplane of the Coxeter element. The set of masses and $3$-point couplings are sufficient  to  identify the Toda theory considered; indeed, all the higher-point couplings can be uniquely determined in terms of these parameters and can be derived from universal properties of root systems~\cite{Fring:1992tt}. Higher-order couplings can also be determined by imposing recursively the absence of particle production at the tree level \cite{Dorey:1996gd, Gabai:2018tmm,Bercini:2018ysh}. A remarkable fact highlighted in~\cite{Davide_Patrick_tree_level_paper} is that, for the class of affine Toda field theories, these two separate constructions lead to the same set of recursion relations for higher-order couplings. This means that, in principle, Lie algebra information can be read out from tree-level scattering requirements.

Despite there being some freedom in the choice of the basis adopted to diagonalise \eqref{mass_spectrum_in_terms_of_roots_not_diagonalised}, not all the signs $\sigma_{abc}$ can be freely chosen. These signs are related to the structure constants of the Lie algebra and the geometry of the root system in such a way to ensure the cancellation of all poles appearing in $4$-point off-diagonal processes at the tree level. 
In the ADE series of Toda models, such cancellations happen between doublets of Feynman diagram with on-shell propagating particles in pairs of channels: $s/t$, $s/u$ or $t/u$. 
The procedure under which these copies of poles cancel one another is known as flipping~\cite{Braden:1990wx}, since it corresponds to flipping one polygon diagonal in an on-shell dual Feynman diagram to get another diagonal and therefore a new diagram. In the next sections, we follow the convention used in \cite{Davide_Patrick_tree_level_paper}, where three different types of flips are distinguished according to the on-shell geometry of the tree-level flipped diagrams\footnote{Note that in~\cite{Davide_Patrick_tree_level_paper} the notation $C_{abc}=f_{abc} \Delta_{abc}$ was used for the area rule. Clearly, the constants $f_{abc}$ are connected to the phases $\sigma_{abc}$ in~\eqref{Connection_among_three_point_couplings_and_areas} through $f_{abc} = \frac{4 \beta}{\sqrt{h}} \sigma_{abc}$.}. It will be shown that the flipping rule, used in~\cite{Davide_Patrick_tree_level_paper} to prove the tree-level integrability of these theories, turns out to be a fundamental property to understand also the simplification underlying loop-level Feynman diagrams.

While all the off-diagonal processes $a+b \to c+d$ are zero (for $\{a,b\} \ne \{c,d\}$) this is not the case if the final particles are equal to the initial ones. In such a case the scattering is allowed and the two-body S-matrix can be exactly conjectured by imposing, additionally to unitarity, crossing-symmetry and analyticity, certain bootstrap properties \cite{Zamolodchikov:1978xm,Zamolodchikov:1989fp}. Following this axiomatic procedure, exact S-matrices for a variety of Toda models have been found in the past \cite{Dorey:1990xa,Braden:1989bu,a1,a2,a3,a4,a5,a6,a7,a8,a9,a10,a11,Corrigan:1993xh,Fring:1991gh}. Among them, the ADE series represent particularly simple theories since at one loop they preserve all the mass ratios \cite{Braden:1989bu} suggesting that the geometrical structure obtained from the bootstrap is maintained at the quantum level in perturbation theory. The S-matrix elements for such a class of quantum field theories can be written in terms of roots and weights of their underlying Lie algebras~\cite{Dorey:1990xa} and take one of the two following forms~\cite{Dorey:1990xa,Braden:1989bu}: 
\begin{subequations}
    \label{S_matrix_bootstrap_result_structure}
\begin{align}
    \label{S_matrix_bootstrap_result_structure_1}
 &S_{ab}(\theta)=\prod_{\substack{x=1 \\ \text{step $2$}}}^{h-1} \{ x \}_{\theta}^{N_{ab}(x)},\\
    \label{S_matrix_bootstrap_result_structure_form_2}
    &S_{ab}(\theta)=\prod_{\substack{x=2 \\ \text{step $2$}}}^{h-2} \{ x \}_{\theta}^{N_{ab}(x)}.
\end{align}
\end{subequations}
The parameter $\theta$ in the expressions above is the difference between the rapidities of the particles $a$ and $b$ involved in the scattering.
The non-negative integer coefficients ${N_{ab}(x)}$ are functions of $x$, depending in addition on the theory and types of particles $\{a, b\}$. They correspond to the multiplicities of the building blocks $\{ x \}$. These blocks, also named bricks, are $2\pi i$-periodic functions of $\theta$ that automatically satisfy the unitary condition of the S-matrix. They are defined in terms of hyperbolic functions \cite{Braden:1989bu}
\begin{equation}
\label{brink_definition_curly_bracket}
\{ x \}_{\theta} \equiv \frac{(x+1)_{\theta} (x-1)_{\theta}}{(x-1+B)_{\theta} (x+1-B)_{\theta}} 
\end{equation}
with
$$
(x)_{\theta}\equiv \frac{\sinh \bigl( \frac{\theta}{2} + \frac{i \pi x}{2h} \bigr)}{\sinh \bigl( \frac{\theta}{2} - \frac{i \pi x}{2h} \bigr)} \hspace{6mm} \text{and} \hspace{6mm} B \equiv \frac{1}{2 \pi} \frac{\beta^2}{1+\frac{\beta^2}{4 \pi}}.
$$
The blocks \eqref{brink_definition_curly_bracket} 
can be Taylor-expanded in $\beta^2$, and at leading order have 
two simple poles in $\theta$:
\begin{equation}
\label{brick_expansion}
\{ x \}_{\theta}= 1+ \frac{i \beta^2}{2 h} \Bigl( - \frac{1}{\theta-\frac{i \pi}{h} (x-1)}+ \frac{1}{\theta-\frac{i \pi}{h} (x+1)} + \ldots \Bigr) + O (\beta^4) \ .
\end{equation}
The ellipses in~\eqref{brick_expansion} contain terms that are finite at the pole positions $\theta=\frac{i \pi}{h}(x\pm1)$. At higher orders in the coupling expansion both finite and singular terms are contained. In this paper, continuing the study started in \cite{Braden:1990wx}, we are interested in the analysis of leading order singularities arising from the brick structure of the S-matrix. By leading order singularities we mean that we perform the Laurent expansion of the S-matrix around a pole $\theta=i\theta_0$ of arbitrary order $P$ and at each order in the expansion in $(\theta-i\theta_0)$ we consider only the leading order in $\beta^2$. In other words, we can write the S-matrix around the pole as 
\begin{equation}
\label{General_Laurent_expansion_on_the_pole}
\begin{split}
S_{ab}(\theta) &= \sum_{p=0}^P \frac{1}{(\theta- i \theta_0)^p} \Bigl(\frac{\beta^{2}}{2h}\Bigr)^p \bigl( a_p+ b_p \beta^2+\ldots \bigr)+ O(\theta- i \theta_0).
\end{split}
\end{equation}
A factor $\Bigl(\frac{\beta^{2}}{2h}\Bigr)^p$, one for each $p$, has been introduced for convenience.
For each coefficient of the Laurent series we are interested only in the leading order term in coupling expansion, therefore $a_0, \ a_1 , \ldots , a_P$, neglecting what is the structure of $b_p$. Universal results for the coefficients $b_1$, in the case of arbitrary even $P$, were recently obtained in~\cite{First_loop_paper_sagex} for the entire class of ADE affine Toda models \footnote{Previous results for $b_1$ were known from some years~\cite{Braden:1992gh}, but were obtained for the only $A_r^{(1)}$ affine Toda theories and $P=2$.}. 

\begin{figure}
\medskip

\begin{center}
\begin{tikzpicture}
\tikzmath{\y=0.8;}

\draw[] (0*\y-6.5*\y,0*\y) -- (0*\y-6.5*\y,0.5*\y);
\draw[] (0*\y-6.5*\y,0.5*\y) -- (2*\y-6.5*\y,0.5*\y);
\draw[] (2*\y-6.5*\y,0.5*\y) -- (2*\y-6.5*\y,0*\y);

\draw[] (2.1*\y-6.5*\y,0.5*\y) -- (4.1*\y-6.5*\y,0.5*\y);
\draw[] (2.1*\y-6.5*\y,0.5*\y) -- (2.1*\y-6.5*\y,0*\y);
\draw[] (4.1*\y-6.5*\y,0.5*\y) -- (4.1*\y-6.5*\y,0*\y);

\filldraw[black] (0.1*\y-6.5*\y,-0.4*\y)  node[anchor=west] {\tiny{$\{ x-1 \}^{N}$}};
\filldraw[black] (2.4*\y-6.5*\y,-0.4*\y)  node[anchor=west] {\tiny{$\{ x+1 \}^{N}$}};
\filldraw[black] (1.5*\y-6.5*\y,-1.3*\y)  node[anchor=west] {\tiny{$\nu = 0 $}};

\draw[] (0*\y,0*\y) -- (0*\y,1*\y);
\draw[] (0*\y,0.5*\y) -- (2*\y,0.5*\y);
\draw[] (2*\y,1*\y) -- (2*\y,0*\y);
\draw[] (0*\y,1*\y) -- (2*\y,1*\y);

\draw[] (2.1*\y,0.5*\y) -- (4.1*\y,0.5*\y);
\draw[] (2.1*\y,0.5*\y) -- (2.1*\y,0*\y);
\draw[] (4.1*\y,0.5*\y) -- (4.1*\y,0*\y);

\filldraw[black] (0.1*\y,-0.4*\y)  node[anchor=west] {\tiny{$\{ x-1 \}^{N+1}$}};
\filldraw[black] (2.4*\y,-0.4*\y)  node[anchor=west] {\tiny{$\{ x+1 \}^{N}$}};
\filldraw[black] (1.5*\y,-1.3*\y)  node[anchor=west] {\tiny{$\nu = i $}};

\draw[] (0*\y+6.5*\y,0*\y) -- (0*\y+6.5*\y,0.5*\y);
\draw[] (0*\y+6.5*\y,0.5*\y) -- (2*\y+6.5*\y,0.5*\y);
\draw[] (2*\y+6.5*\y,0.5*\y) -- (2*\y+6.5*\y,0*\y);

\draw[] (2.1*\y+6.5*\y,1*\y) -- (4.1*\y+6.5*\y,1*\y);
\draw[] (2.1*\y+6.5*\y,1*\y) -- (2.1*\y+6.5*\y,0*\y);
\draw[] (4.1*\y+6.5*\y,1*\y) -- (4.1*\y+6.5*\y,0*\y);
\draw[] (2.1*\y+6.5*\y,0.5*\y) -- (4.1*\y+6.5*\y,0.5*\y);

\filldraw[black] (0.1*\y+6.5*\y,-0.4*\y)  node[anchor=west] {\tiny{$\{ x-1 \}^{N}$}};
\filldraw[black] (2.4*\y+6.5*\y,-0.4*\y)  node[anchor=west] {\tiny{$\{ x+1 \}^{N+1}$}};
\filldraw[black] (1.5*\y+6.5*\y,-1.3*\y)  node[anchor=west] {\tiny{$\nu = -i $}};

\end{tikzpicture}
\end{center}
\caption{Higher order singularities at $\theta=i \frac{\pi x}{h}$. Working from left to right the pictures show a pole of order $2N$, and forward and crossed channel poles of order $2N+1$. The assignment of forward and crossed channels is in line with the `uphill/downhill' rule mentioned in \cite{a8}. All the situations are encoded in equation \eqref{S_matrix_contributions_on_the_poles_expanded_and_containing_all_the_leading_singularities} substituting the corresponding values of $\nu$.}
\label{brick_towers_close_each_other}
\end{figure}
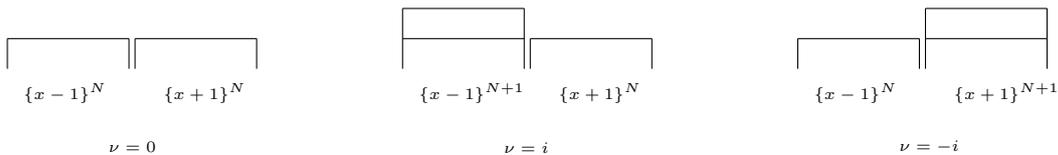

It is straightforward to obtain the coefficients $a_p$ from the bootstrapped results by substituting the building block expansion \eqref{brick_expansion} into \eqref{S_matrix_bootstrap_result_structure}. A pole of order $P$ at $\theta=i\theta_0= \frac{i \pi}{h} x$, with $P=M+N$, arises by terms in the S-matrix of the form
\begin{equation}
\label{close_brick_general_expansion_with_M_and_N_explicit}
\{ x-1 \}^M \{ x+1 \}^N \sim \biggl(1+ \frac{i \beta^2}{2 h} \frac{1}{\theta-\frac{i \pi}{h} x}  \biggr)^M \biggl(1- \frac{i \beta^2}{2 h} \frac{1}{\theta-\frac{i \pi}{h} x}  \biggr)^N \ .
\end{equation}
Each term obtained by expanding~\eqref{close_brick_general_expansion_with_M_and_N_explicit} corresponds to a leading order coefficient of~\eqref{General_Laurent_expansion_on_the_pole}.  
Since the multiplicity of adjacent bricks can only differ by $0$ or $\pm1$ \cite{a8} the leading order contributions to the Laurent expansion around the pole $i \theta_0$ can be written as
\begin{equation}
\label{S_matrix_contributions_on_the_poles_expanded_and_containing_all_the_leading_singularities}
S^{(l)}_{ab}(\theta) = \sum_{n=0}^N  \binom{N}{n} \Bigl(\frac{\beta^2}{2 h}\Bigr)^{2n} \frac{1}{(\theta- i \theta_0)^{2n}} + \nu \sum_{n=0}^N  \binom{N}{n} \Bigl(\frac{\beta^2}{2 h}\Bigr)^{2n+1} \frac{1}{(\theta-i \theta_0)^{2n+1}} \ ,
\end{equation}
where the superscript  $l$ attached to the S-matrix element stands for `leading'. The expansion has been truncated so as to contain only singular contributions at the pole location.
The coefficient $\nu$ is equal to zero if two bricks of the same multiplicity $N$ touch each other, generating a pole of even order $P=2N$. Otherwise, $\nu=+i/-i$ when there is a jump from a brick of multiplicity $N+1 / N$ on the left to a brick of multiplicity $N / N+1$ on the right. The poles are odd-order singularities since $P=2N+1$. The different situations are depicted in figure \ref{brick_towers_close_each_other}.

The main purpose of this paper is to understand the origin of the universal formula \eqref{S_matrix_contributions_on_the_poles_expanded_and_containing_all_the_leading_singularities} using perturbation theory.
The $n=0$ contributions in \eqref{S_matrix_contributions_on_the_poles_expanded_and_containing_all_the_leading_singularities} are naturally interpreted as the identity (i.e. free propagators), in the first sum, and a bound state particle propagating at the tree level in the forward/crossed channel ($\nu=+i/-i$), in the second sum. From $n=1$ onwards, loop contributions begin. 
These higher-order singularities   
can be explained in perturbation theory in terms of the Coleman-Thun mechanism~\cite{Coleman:1978kk}: they are anomalous thresholds coming from Feynman diagrams in which all the internal propagators, within the integration loops, can go simultaneously on-shell for particular choices of the external kinematics. 
Following the technique adopted in \cite{Braden:1990wx} we draw closed networks of singular on-shell Feynman diagrams connected by flipping internal propagators. Then, based on observations highlighted in~\cite{First_loop_paper_sagex}, we explain how each multiple-loop diagram, evaluated at the pole, can be decomposed into a sum of products of tree-level graphs obtained by cutting some of its propagators.  From this perspective, we call the loop diagrams entering the network `molecules' while we call their smaller constituents, the products of tree diagrams obtained by cutting their propagators, `atoms'.  
Not all the Feynman diagrams composing the network contribute to the result: many of the atoms cancel between one another in a way that can be traced back to tree-level properties of the model. The only relevant terms come from some particular cuts of the diagrams, that in the cases considered are located at the boundary of the network. 

In the next sections this analysis will be carried out for second- and third-order poles (which is $P=2$ and $P=3$ in \eqref{General_Laurent_expansion_on_the_pole}) in diagonal scattering processes. In all the cases the coefficient $a_p$ of each term of the Laurent expansion~\eqref{General_Laurent_expansion_on_the_pole} are computed, displaying perfect agreement with the bootstrapped formula \eqref{S_matrix_contributions_on_the_poles_expanded_and_containing_all_the_leading_singularities}.
In~\cite{Braden:1990wx} poles of order $P=2$ and $P=3$ were studied, finding in both cases the coefficient $a_P$ of the maximal singularity in~\eqref{General_Laurent_expansion_on_the_pole}. While the results found matched the bootstrap expectations,
the underlying mechanism that led to the answer was not completely clear, particularly for $P=3$. 
In that case, the Feynman diagrams contributing to the pole generated a collection of complicated numbers which, only after the sum, collapsed to the correct residue expected from the bootstrap. We will highlight better the reason why this happens using the cut method of~\cite{First_loop_paper_sagex}.
Other aspects of higher-order singularities are also covered in this paper, and even though we have not yet achieved the computation of all the leading coefficients $a_p$ of the Laurent expansion at arbitrary $P$, due to the huge number of singular Feynman diagrams arising at higher loops, we do see some signs of how a general proof might be achieved.

The paper is structured as follows. In section~\ref{Review_of_Landau_sinngularities_and_geometrical_cuts} we review the mechanism responsible for the generation of higher-order poles in perturbation theory. We show that certain loops, singular for some values of the external momenta, can be cut into sums of particular products of tree-level Feynman diagrams; each product of tree-level graphs entering these sums will be called an atom. We explain how the different ways in which the loops can be cut are determined by the on-shell geometry of the loops and follow from certain `positivity' conditions. In sections~\ref{Chapter_on_second_order_poles} and \ref{Chapter_on_third_order_poles} we use such a cutting technique to simplify large networks of Feynman diagrams contributing to $2^{\text{nd}}$- and $3^{\text{rd}}$-order poles respectively. We show that many atoms cancel against each other and the residues at the poles are obtained by summing a few surviving atoms located at the boundary of the network. In this way, we reproduce the Laurent coefficients of the S-matrix at the poles. In section~\ref{Universal_properties_of_higher_order_singularities} we discuss some aspects of higher-order singularities. In particular
by extending the argument presented in~\cite{First_loop_paper_sagex} to odd-order singularities, we compute the coefficient $a_2$ in equation~\eqref{General_Laurent_expansion_on_the_pole} for arbitrary $P$, showing that it is indeed equal to the binomial factor $\binom{N}{1}$ present in the bootstrapped formula~\eqref{S_matrix_contributions_on_the_poles_expanded_and_containing_all_the_leading_singularities}. 
The one-particle reducible part of the S-matrix on a $5^{\text{th}}$-order pole is computed as well, returning a universal result not depending on the model considered.
 
A number of previous papers~\cite{Bianchi:2013nra,Engelund:2013fja,Bianchi:2014rfa,Polvara:2023vnx} have linked higher-loop S-matrices of integrable models in 1+1 dimensions to tree-level S-matrices, but using unitarity-like methods.
Even though this type of approach will not be investigated here, it would be interesting to check whether some of the results discussed in the next sections could be obtained from this different perspective; 
in particular, a fascinating direction would be to generate expressions for higher-order Landau poles in terms of simple poles of tree-level S-matrices 
by writing sums of Feynman diagrams in terms of products and integrals of tree-level S-matrices as discussed in~\cite{Polvara:2023vnx}.

\section{Landau singularities and cut diagrams}
\label{Review_of_Landau_sinngularities_and_geometrical_cuts}

In this section, we review the mechanism responsible for the generation of higher-order poles in perturbation theory following the approach used in \cite{First_loop_paper_sagex}. This method, compared with the technique used in \cite{Braden:1990wx}, makes it evident how the loops, on the pole position, factorise into particular products of singular tree-level diagrams. 
We are interested in finding the leading order coefficients $a_p$ in~\eqref{General_Laurent_expansion_on_the_pole}. This corresponds to finding on-shell Feynman diagrams with the maximum number of internal on-shell momenta and carrying the minimum power of $\beta$. Such a situation occurs if the only vertices entering the diagrams are $3$-point couplings.

Consider a diagram $D$ contributing to a diagonal process involving the scattering of two particles of types $a$ and $b$ at $L$ loops, 
close to a higher-order pole at $\theta=i\theta_0$.
Since there are two degrees of freedom for each loop, we can always find values of the loop integration variables for which two momenta inside each loop are on-shell, and satisfy
\begin{equation}
\label{on_shell_condition_on_the_momenta_qj_around_which_the_loop_is_expanded}
q_j^2-m_j^2=0
\end{equation}
where $j=1, \ldots , 2 L$. 
This is always possible if we are close to the pole $\theta=i\theta_0$, since actually \textit{at} the pole, there is a choice of loop momenta for which \textit{all} internal lines of the diagram are on-shell. 
Then we expand the corresponding momenta entering into the loop integrals around these values
\begin{equation}
Q_j=q_j+\sum_{i=1}^L \lambda^{(Q)}_{j i} l_i \equiv q_j + k^{(Q)}_j.
\end{equation}
The vectors $l_1 , \ldots , l_L$ are the $L$ loop integration variables and $\lambda^{(Q)}_{j i}$ are certain real coefficients whose values are determined by the way that momentum conservation is implemented across the diagram. This permits the relevant propagators to be written in the following form
\begin{equation}
\frac{i}{Q^2_j-m^2_j + i \epsilon}=\frac{i}{2 q_j \cdot k^{(Q)}_j + (k^{(Q)}_j)^2 +  i \epsilon} .
\end{equation}
All the loop integration variables are contained in the vectors $k^{(Q)}_j$.
If we define $I$ to be the total number of propagators then the remaining $I-2L$ momenta (those that are not expanded around their on-shell values when $\theta\neq \theta_0$) are given by
\begin{equation}
\label{remaining_set_of_momenta_on_shell_only_for_a_particular_choice_of_s_expansion_of_the_loop}
P_j=p_j+\sum_{i=1}^L \lambda^{(P)}_{j i} l_i \equiv p_j + k^{(P)}_j .
\end{equation}
The momenta $p_j$ around which the $P_j$  are expanded are fixed, since we have already used all the loop freedom to set the momenta $q_j$ on-shell in \eqref{on_shell_condition_on_the_momenta_qj_around_which_the_loop_is_expanded}. 
Unlike the $q_j$, the values $p_j$ are generally off-shell and go on-shell only at the pole position $\theta=i\theta_0$. Labelling by $s_0$ the value of the Mandelstam variable $s=(p_a + p_b)^2$ at the pole, 
we can expand $p^2_j$ in $(s-s_0)$ as follows
\begin{equation}
\label{derivative_of_propagators_on_the_pole_positions_general_formula_for_loop_computation}
p_j^2-m_j^2= \frac{dp_j^2}{ds}\Bigl|_{s=s_0}(s-s_0)+ \ldots
\end{equation}
Therefore up to extra multiplicative factors, irrelevant to the purpose of the discussion, the value of the Feynman diagram in the neighbourhood of $s_0$ can be written as 
\begin{equation}
\int \prod_{i=1}^L \frac{d^2 l_i}{(2 \pi)^2} \prod_{j=1}^{2L} \frac{1}{ 2 q_j \cdot k^{(Q)}_j + (k^{(Q)}_j)^2 + i \epsilon}    \prod_{j=1}^{I-2L} \frac{1}{\frac{dp_j^2}{ds}\Bigl|_{s=s_0} (s-s_0) + 2 p_j \cdot k^{(P)}_j + (k^{(P)}_j)^2 + i \epsilon} 
\label{carrot}
\end{equation}
where the $k^{(Q)}_j$ and $k^{(P)}_j$ are linear functions of the loop momenta $l_i$.
Scaling the integration variables in the following manner
\begin{equation}
l_i=(s-s_0)\tilde{l}_i
\end{equation}
and omitting subleading terms in $s-s_0$ we can isolate the singular part of (\ref{carrot}) at the pole position, which we denote by $I_D$:
\begin{equation}
\label{general_formula_for_the_pole_integrating_over_tilde_variables}
I_D= \frac{1}{(s-s_0)^{I-2L}} \int \prod_{i=1}^L \frac{d^2 \tilde{l}_i}{(2\pi)^2} \prod_{j=1}^{2L} \frac{1}{2 q_j \cdot \tilde{k}^{(Q)}_j + i \epsilon}  \prod_{j=1}^{I-2L} \frac{1}{\frac{d p_j^2}{d s}\Bigl|_{s=s_0}+2 p_j \cdot \tilde{k}^{(P)}_j + i \epsilon}.
\end{equation}
Clearly the order of the pole is equal to $p=I-2L$ where, as already specified, $I$ is the total number of propagators in the diagram and $L$ is the total number of loops.
Since all the denominators in the propagators of \eqref{general_formula_for_the_pole_integrating_over_tilde_variables} are linear functions of the loop variables, the integrations can be simply performed by closing a contour in the complex plane and using Cauchy's theorem. Depending on the choice of variables adopted when we perform the integral the number of poles in the integration contour can change and the same result can be written as sums over different residues. Each one of these sums is a sum over different products of tree-level diagrams where each propagator with respect to which we take the residue is cut. We will show this explicitly in section~\ref{Box_computation_example} with an example.

To obtain the 
singular part of the Feynman diagram at the pole, which we will denote by $D_D$, we need to multiply the result of the integral by the following factors, which were omitted in \eqref{general_formula_for_the_pole_integrating_over_tilde_variables}:
\begin{equation}
\label{extra_multiplicative_factors_to_be_added_in_the_end}
\begin{split}
&\text{for each vertex:} \hspace{40mm} - i C_{abc} \ ;\\
&\text{for each propagator:} \hspace{37mm} i.
\end{split}\end{equation}
After summing over all Feynman diagrams contributing to the Landau singularity,
we express $s-s_0$ in terms of the rapidity at the pole
\begin{equation}
\label{s_as_function_of_the_rapidity}
s-s_0=2m_a m_b \bigl(\cosh \theta - \cosh i \theta_0 \bigr)=4 i \Delta_{ab} (\theta -i \theta_0) + O((\theta-i\theta_0)^2),
\end{equation}
where $\Delta_{ab}$ is the area of the triangle formed by the vectors $p_a$, $p_b$ at the pole position. 
In the last equality we have expanded $\cosh \theta$ around $\theta=i \theta_0$ and used $\Delta_{ab}=\frac{1}{2}m_a m_b \sin \theta_0$. If we do so we find the singular part of the amplitude, expressed as a function of the rapidity, at the leading order in the coupling. To obtain the coefficient in the expansion~\eqref{General_Laurent_expansion_on_the_pole} of the associated S-matrix element  we need to multiply the result by
\begin{equation}
\label{overall_multiplicative_factor_coming_from_momentum_conservation}
    \frac{1}{4 m_a m_b \sinh i \theta_0}= \frac{1}{8 i \Delta_{ab}}.
\end{equation}
This factor comes by expressing the overall energy-momentum conservation delta function $(2 \pi)^2 \delta^{(2)} (p_a^{in}+p_b^{in}-p_a^{out}-p_b^{out})$ in terms of the rapidities at the pole position and inserting a normalization factor $\frac{1}{\sqrt{4 \pi}}$ for each external particle.

Once we understand which residues to take, the difficult part in the evaluation of the integral \eqref{general_formula_for_the_pole_integrating_over_tilde_variables} is computing the derivatives of propagators at the pole position, which are the terms on the RHS of \eqref{derivative_of_propagators_on_the_pole_positions_general_formula_for_loop_computation}.
The rapidity values at which poles can appear are purely imaginary, and so Feynman diagrams at $l_1=l_2=\ldots=l_L=(0,0)$ (the point at which all loop propagators are singular simultaneously) can be represented as geometrical figures in Euclidean space. Then evaluating the propagator derivatives amounts to a geometrical problem, best formulated in terms of the dual diagrams to be discussed in the next section. The following lemma, used in \cite{Davide_Patrick_tree_level_paper} to evaluate the residues of tree-level diagrams, will be useful:
\begin{Lemma}\label{Formula_for_propgator_derivatives_in_loop_diagrams}
Let $\vec{x}^0$,
$\vec{x}^1$, $\vec{x}^2$, $\vec{x}^3$ be $4$ points in $\mathbb{R}^2$, and set $X_{ij}=\norm{\vec{x}^i-\vec{x}^j}^2$ for $0\le i,j\le 3$. 
Then the following identity holds:
\begin{equation}\label{relation_amonf_squaredistances_of_points_in_a_plane}
\sum_{i\ne j \ne k \ne n} dX_{i j} \ \varepsilon_{ij}(k,n) \ \Delta_{kni} \Delta_{knj}=0 .
\end{equation}
$\varepsilon_{ij}(k,n) =+1/-1$ according to whether $\vec{x}^i$ and $\vec{x}^j$  lie on the same / opposite sides of the line identified by the diagonal connecting the points $\vec{x}^k$ and $\vec{x}^n$.
\end{Lemma}
In lemma~\ref{Formula_for_propgator_derivatives_in_loop_diagrams} we used the conventions of appendix A of \cite{Davide_Patrick_tree_level_paper} and labelled by $\Delta_{ijk}$ the area of the triangle having vertices defined by the vectors $\vec{x}^i$, $\vec{x}^j$, $\vec{x}^k$. 
In this paper, we will instead label triangles using their sides (corresponding in dual diagrams to propagating particles) rather than vertices.
Lemma~\ref{Formula_for_propgator_derivatives_in_loop_diagrams} is straightforwardly adapted to such situations with little change of notations.
We can also define the oriented area of the triangle identified by a pair of vectors $A$, $B$ in $\mathbb{R}^2$ as the determinant
\begin{equation}
\label{oriented_triangle_area_definition}
\bigl \langle A B \bigr \rangle \equiv \frac{1}{2} 
\begin{vmatrix}
~A_x & B_x~     \\[3pt]
~A_y & B_y~ 
\end{vmatrix}~.
\end{equation}
It will sometimes be necessary to write vectors in a non-orthogonal basis. To do this the following identity can be used
\begin{equation}\label{writing_D_in_terms_of_A_and_B_general_relation}
\bigr \langle A B \bigr \rangle \ D  + \bigr \langle B D \bigr \rangle \ A + \bigr \langle D A \bigr \rangle \ B  = 0
\end{equation}
on any triplet of vectors $A$, $B$ and $D$ in $\mathbb{R}^2$.
Performing the vector product with a fourth element $C$ we obtain the Pl\"ucker relation connecting areas of different triangles
\begin{equation}
\bigr \langle A B \bigr \rangle  \bigr \langle C D \bigr \rangle + \bigr \langle B D \bigr \rangle  \bigr \langle C A \bigr \rangle + \bigr \langle D A \bigr \rangle \bigr \langle C B \bigr \rangle = 0 .
\end{equation}

\subsection{An example: the box integral}
\label{Box_computation_example}
\begin{figure}
\begin{center}
\includegraphics[width=0.8\linewidth]{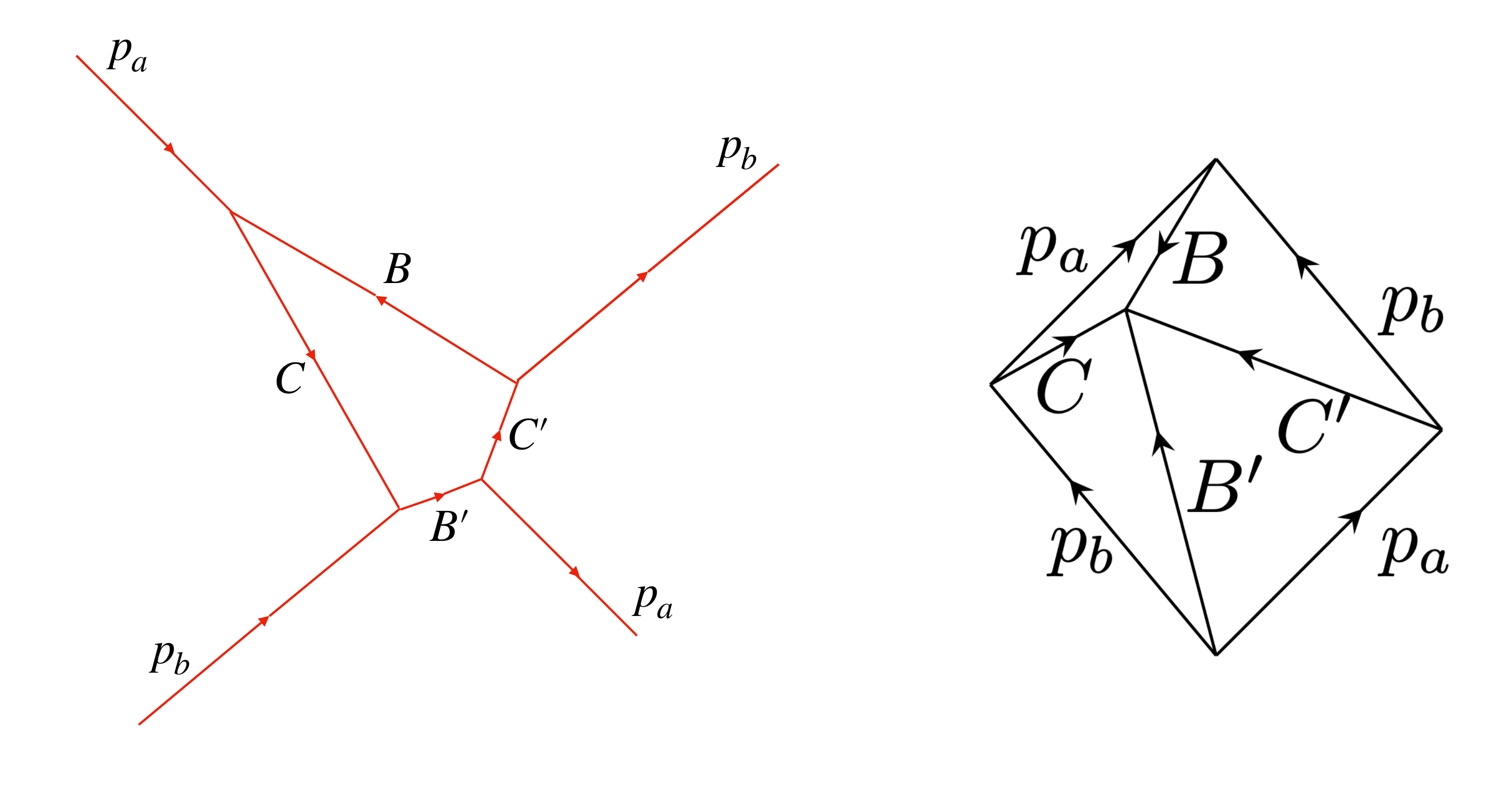}
\end{center}
\caption{On-shell (box) diagram on the left and its dual on the right.}
\label{Example_of_on_shell_one_loop_box_diagram}
\end{figure}

We show an example of a Feynman diagram in which the Coleman-Thun mechanism \cite{Coleman:1978kk} for the generation of Landau singularities is manifest. The example is simple and already well known but it contains some subtleties that are also characteristic of more difficult cases and it is worth explaining them here. Consider the box diagram in figure \ref{Example_of_on_shell_one_loop_box_diagram}. 
The LHS shows the box diagram at the singularity:
the directions of the momenta are the imaginary values of the rapidities at the pole and for this reason, we refer to this diagram as an on-shell diagram. The RHS shows its dual.
The dual diagram is a tiled parallelogram, composed of vectors in the complex plane having lengths and arguments equal to the masses and imaginary rapidities of the associated particles respectively.
From the dual diagram the associated on-shell diagram is easily obtained: for each triangle $\Delta_{ABC}$ in the dual diagram,
draw a vertex with the vectors $A$, $B$ and $C$ attached, each rotated by $90^\circ$ in a clockwise direction. Even though the directions of the momenta are different in the on-shell diagram and its dual, they are related by a Lorentz rotation with an imaginary angle and the values of the two pictures are the same; this should be clear by the fact that all the fusing angles remain unchanged by an overall rotation. 
We stress that in figure~\ref{Example_of_on_shell_one_loop_box_diagram} the dual diagram on the RHS contains more information than the on-shell diagram on the LHS. Indeed, in the dual diagram, the sides have lengths equal to the values of the associated masses and as vectors they match the on-shell momenta at the pole $l=(0,0)$. Instead, the vectors in the on-shell diagram are not equal to the (rotated) on-shell momenta since their lengths do not correspond to the masses of the propagating particles. For this reason, in the following discussion, the on-shell momenta should always be read from the dual diagram and not from the on-shell Feynman diagram.
 
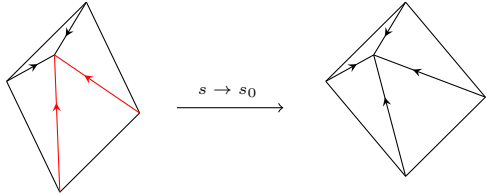
\begin{figure}
\medskip
\begin{center}
\begin{tikzpicture}
\tikzmath{\y = 0.7;}

\draw[] (5.8*\y,1*\y) -- (7.3*\y,2.5*\y);
\draw[directed] (5.8*\y,1*\y) -- (6.7*\y,1.5*\y);
\draw[directed] (7.3*\y,2.5*\y) -- (6.7*\y,1.5*\y);
\draw[] (6.8*\y,-1.1*\y) -- (5.8*\y,1*\y);
\draw[] (6.8*\y,-1.1*\y) -- (8.3*\y,0.4*\y);
\draw[] (7.3*\y,2.5*\y) -- (8.3*\y,0.4*\y);

\draw[directed][red] (6.8*\y,-1.1*\y) -- (6.7*\y,1.5*\y);
\draw[directed][red] (8.3*\y,0.4*\y) -- (6.7*\y,1.5*\y);

\draw [->] (9*\y,0.5*\y) -- (11*\y,0.5*\y);
\filldraw[black] (9.2*\y,0.8*\y)  node[anchor=west] {\tiny{$s \to s_0$}};

\draw[] (5.8*\y+6*\y,1*\y) -- (7.3*\y+6*\y,2.5*\y);
\draw[directed] (5.8*\y+6*\y,1*\y) -- (6.7*\y+6*\y,1.5*\y);
\draw[directed] (7.3*\y+6*\y,2.5*\y) -- (6.7*\y+6*\y,1.5*\y);
\draw[] (7.3*\y+6*\y,-0.8*\y) -- (5.8*\y+6*\y,1*\y);
\draw[] (7.3*\y+6*\y,-0.8*\y) -- (8.8*\y+6*\y,0.7*\y);
\draw[] (8.8*\y+6*\y,0.7*\y) -- (7.3*\y+6*\y,2.5*\y);
\draw[directed][] (8.8*\y+6*\y, 0.7*\y) -- (6.7*\y+6*\y,1.5*\y);
\draw[directed][] (7.3*\y+6*\y,-0.8*\y) -- (6.7*\y+6*\y,1.5*\y);

\end{tikzpicture}
\end{center}

\caption{Loop expansion around the on-shell values $B^2=m^2_B$ and $C^2=m^2_C$. The remaining momenta $B'$ and $C'$ (drawn in red in the figure) are not on-shell for general external kinematics (on the LHS) and go on-shell only on the pole position $s=s_0$ (on the RHS).}
\label{Example_of_on_shell_one_loop_box_diagram_how_to_perform_the_limit}
\end{figure}

\begin{figure}
\begin{center}
\includegraphics[width=0.8\linewidth]{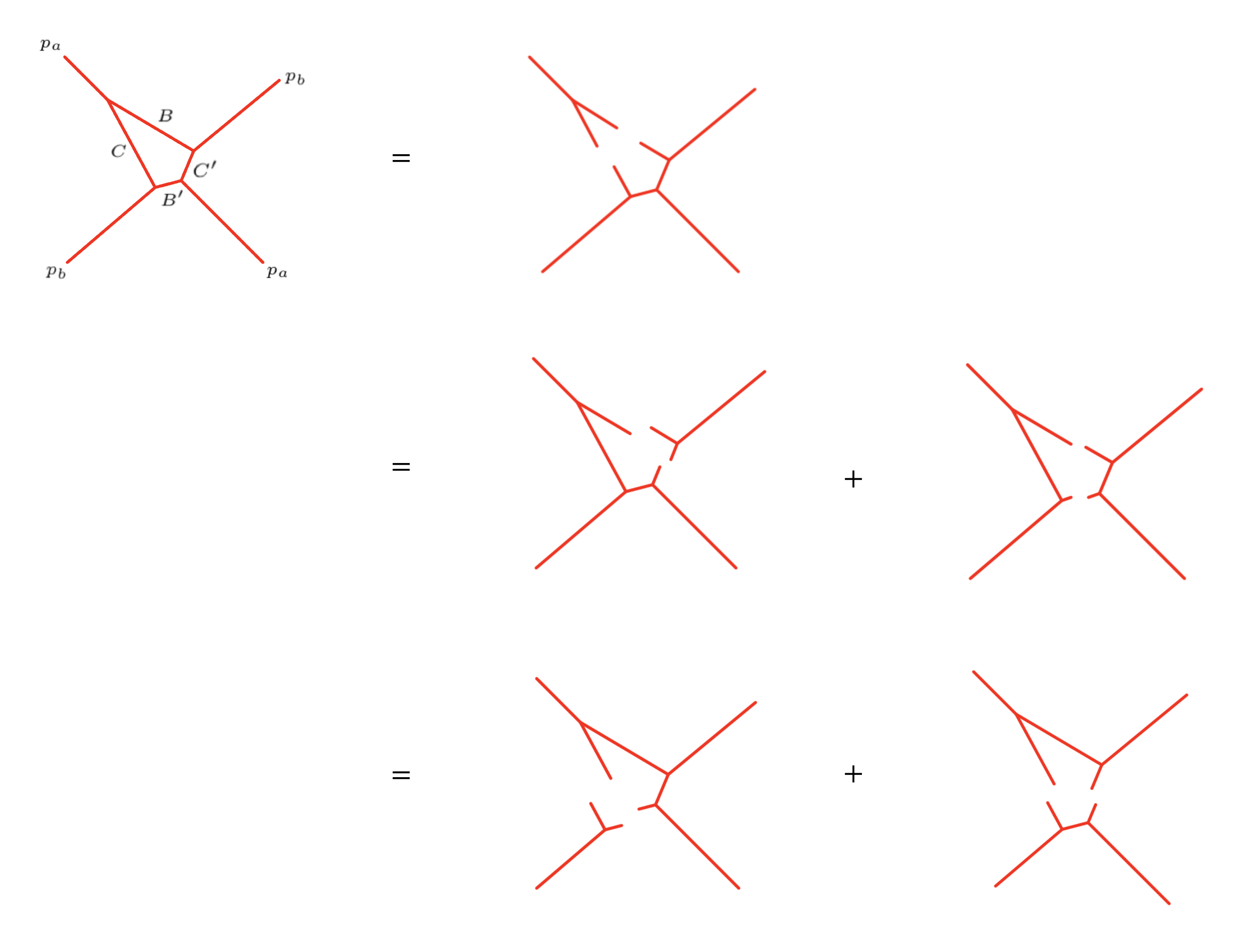}
\end{center}
\caption{Box integral at the pole written as sums of different atoms depending on the way we close the integration path. 
The single contribution on the RHS of the first equality corresponds to closing both the $u$ and $v$ contours on the lower half complex plane. The sum over the two terms in the second (third) row corresponds to closing $u$ in the lower (upper) half plane and $v$ in the upper (lower) half plane.}
\label{Example_of_different_possible_cuts_of_the_box_diagrams_nondual}
\end{figure}

\begin{figure}
\medskip
\begin{center}
\begin{tikzpicture}
\tikzmath{\x = 1;\y = 0.7;}

\draw[] (0.8*\y,1*\y-15*\y) -- (2.3*\y,2.5*\y-15*\y);
\draw[] (0.8*\y,1*\y-15*\y) -- (1.7*\y,1.5*\y-15*\y);
\draw[] (2.3*\y,2.5*\y-15*\y) -- (1.7*\y,1.5*\y-15*\y);
\draw[] (2.3*\y,-0.8*\y-15*\y) -- (0.8*\y,1*\y-15*\y);
\draw[] (2.3*\y,-0.8*\y-15*\y) -- (1.7*\y,1.5*\y-15*\y);
\draw[] (2.3*\y,-0.8*\y-15*\y) -- (3.8*\y,0.7*\y-15*\y);
\draw[] (3.8*\y,0.7*\y-15*\y) -- (2.3*\y,2.5*\y-15*\y);
\draw[] (3.8*\y, 0.7*\y-15*\y) -- (1.7*\y,1.5*\y-15*\y);

\filldraw[black] (0.8*\y,2*\y-15*\y)  node[anchor=west] {\tiny{$p_a$}};
\filldraw[black] (1*\y,0*\y-15*\y)  node[anchor=west] {\tiny{$p_b$}};
\filldraw[black] (3.2*\y,0*\y-15*\y)  node[anchor=west] {\tiny{$p_a$}};
\filldraw[black] (3.2*\y,1.6*\y-15*\y)  node[anchor=west] {\tiny{$p_b$}};

\filldraw[black] (0.9*\y,1*\y-15*\y)  node[anchor=west] {\tiny{$C$}};
\filldraw[black] (1.9*\y,0.4*\y-15*\y)  node[anchor=west] {\tiny{$B'$}};
\filldraw[black] (2.5*\y,0.8*\y-15*\y)  node[anchor=west] {\tiny{$C'$}};
\filldraw[black] (1.8*\y,1.9*\y-15*\y)  node[anchor=west] {\tiny{$B$}};

\filldraw[black] (5.2*\y,0.6*\y-15*\y)  node[anchor=west] {\small{$=$}};
\draw[] (0.8*\y+6.2*\y,1*\y-15*\y) -- (2.3*\y+6.2*\y,2.5*\y-15*\y);
\draw[] (0.8*\y+6.2*\y,1*\y-15*\y) -- (1.7*\y+6.2*\y,1.5*\y-15*\y);
\draw[] (2.3*\y+6.2*\y,2.5*\y-15*\y) -- (1.7*\y+6.2*\y,1.5*\y-15*\y);

\draw[] (0.8*\y+7*\y,1*\y-15*\y) -- (1.7*\y+7*\y,1.5*\y-15*\y);
\draw[] (2.3*\y+7*\y,2.5*\y-15*\y) -- (1.7*\y+7*\y,1.5*\y-15*\y);
\draw[] (2.3*\y+7*\y,-0.8*\y-15*\y) -- (0.8*\y+7*\y,1*\y-15*\y);
\draw[] (2.3*\y+7*\y,-0.8*\y-15*\y) -- (1.7*\y+7*\y,1.5*\y-15*\y);
\draw[] (2.3*\y+7*\y,-0.8*\y-15*\y) -- (3.8*\y+7*\y,0.7*\y-15*\y);
\draw[] (3.8*\y+7*\y,0.7*\y-15*\y) -- (2.3*\y+7*\y,2.5*\y-15*\y);
\draw[] (3.8*\y+7*\y, 0.7*\y-15*\y) -- (1.7*\y+7*\y,1.5*\y-15*\y);

\filldraw[black] (5.2*\y,0.6*\y-20*\y)  node[anchor=west] {\small{$=$}};

\draw[] (0.3*\y+7*\y,1*\y-20*\y) -- (1.8*\y+7*\y,2.5*\y-20*\y);
\draw[] (0.3*\y+7*\y,1*\y-20*\y) -- (1.2*\y+7*\y,1.5*\y-20*\y);
\draw[] (1.8*\y+7*\y,2.5*\y-20*\y) -- (1.2*\y+7*\y,1.5*\y-20*\y);
\draw[] (1.8*\y+7*\y,-0.8*\y-20*\y) -- (0.3*\y+7*\y,1*\y-20*\y);
\draw[] (1.8*\y+7*\y,-0.8*\y-20*\y) -- (1.2*\y+7*\y,1.5*\y-20*\y);
\draw[] (1.8*\y+7*\y,-0.8*\y-20*\y) -- (3.3*\y+7*\y,0.7*\y-20*\y);
\draw[] (3.3*\y+7*\y, 0.7*\y-20*\y) -- (1.2*\y+7*\y,1.5*\y-20*\y);

\draw[] (3.8*\y+7*\y, 0.7*\y-20*\y) -- (1.7*\y+7*\y,1.5*\y-20*\y);
\draw[] (3.8*\y+7*\y,0.7*\y-20*\y) -- (2.3*\y+7*\y,2.5*\y-20*\y);
\draw[] (2.3*\y+7*\y,2.5*\y-20*\y) -- (1.7*\y+7*\y,1.5*\y-20*\y);

\filldraw[black] (11.8*\y,-19*\y)  node[anchor=west] {\tiny{$+$}};

\draw[] (0.3*\y+13*\y,1*\y-20*\y) -- (1.8*\y+13*\y,2.5*\y-20*\y);
\draw[] (0.3*\y+13*\y,1*\y-20*\y) -- (1.2*\y+13*\y,1.5*\y-20*\y);
\draw[] (1.8*\y+13*\y,2.5*\y-20*\y) -- (1.2*\y+13*\y,1.5*\y-20*\y);
\draw[] (1.8*\y+13*\y,-0.8*\y-20*\y) -- (0.3*\y+13*\y,1*\y-20*\y);
\draw[] (1.8*\y+13*\y,-0.8*\y-20*\y) -- (1.2*\y+13*\y,1.5*\y-20*\y);

\draw[] (2.3*\y+13*\y,2.5*\y-20*\y) -- (1.7*\y+13*\y,1.5*\y-20*\y);
\draw[] (2.3*\y+13*\y,-0.8*\y-20*\y) -- (1.7*\y+13*\y,1.5*\y-20*\y);
\draw[] (2.3*\y+13*\y,-0.8*\y-20*\y) -- (3.8*\y+13*\y,0.7*\y-20*\y);
\draw[] (3.8*\y+13*\y,0.7*\y-20*\y) -- (2.3*\y+13*\y,2.5*\y-20*\y);
\draw[] (3.8*\y+13*\y, 0.7*\y-20*\y) -- (1.7*\y+13*\y,1.5*\y-20*\y);


\filldraw[black] (5.2*\y,0.6*\y-25*\y)  node[anchor=west] {\small{$=$}};

\draw[] (0.8*\y+7*\y,1.5*\y-25*\y) -- (2.3*\y+7*\y,3*\y-25*\y);
\draw[] (0.8*\y+7*\y,1.5*\y-25*\y) -- (1.7*\y+7*\y,2*\y-25*\y);
\draw[] (2.3*\y+7*\y,3*\y-25*\y) -- (1.7*\y+7*\y,2*\y-25*\y);
\draw[] (3.8*\y+7*\y,1.2*\y-25*\y) -- (2.3*\y+7*\y,3*\y-25*\y);
\draw[] (3.8*\y+7*\y, 1.2*\y-25*\y) -- (1.7*\y+7*\y,2*\y-25*\y);
\draw[] (2.3*\y+7*\y,-0.3*\y-25*\y) -- (3.8*\y+7*\y,1.2*\y-25*\y);
\draw[] (2.3*\y+7*\y,-0.3*\y-25*\y) -- (1.7*\y+7*\y,2*\y-25*\y);

\draw[] (2.3*\y+6.5*\y,-0.8*\y-25*\y) -- (1.7*\y+6.5*\y,1.5*\y-25*\y);
\draw[] (2.3*\y+6.5*\y,-0.8*\y-25*\y) -- (0.8*\y+6.5*\y,1*\y-25*\y);
\draw[] (0.8*\y+6.5*\y,1*\y-25*\y) -- (1.7*\y+6.5*\y,1.5*\y-25*\y);

\filldraw[black] (11.8*\y,-24*\y)  node[anchor=west] {\tiny{$+$}};

\draw[] (0.8*\y+13*\y,1.5*\y-25*\y) -- (2.3*\y+13*\y,3*\y-25*\y);
\draw[] (0.8*\y+13*\y,1.5*\y-25*\y) -- (1.7*\y+13*\y,2*\y-25*\y);
\draw[] (2.3*\y+13*\y,3*\y-25*\y) -- (1.7*\y+13*\y,2*\y-25*\y);
\draw[] (3.8*\y+13*\y,1.2*\y-25*\y) -- (2.3*\y+13*\y,3*\y-25*\y);
\draw[] (3.8*\y+13*\y, 1.2*\y-25*\y) -- (1.7*\y+13*\y,2*\y-25*\y);

\draw[] (0.8*\y+13*\y,1*\y-25*\y) -- (1.7*\y+13*\y,1.5*\y-25*\y);
\draw[] (2.3*\y+13*\y,-0.8*\y-25*\y) -- (0.8*\y+13*\y,1*\y-25*\y);
\draw[] (2.3*\y+13*\y,-0.8*\y-25*\y) -- (1.7*\y+13*\y,1.5*\y-25*\y);
\draw[] (2.3*\y+13*\y,-0.8*\y-25*\y) -- (3.8*\y+13*\y,0.7*\y-25*\y);
\draw[] (3.8*\y+13*\y, 0.7*\y-25*\y) -- (1.7*\y+13*\y,1.5*\y-25*\y);

\end{tikzpicture}
\end{center}

\caption{Dual depictions of the atoms shown in figure~\ref{Example_of_different_possible_cuts_of_the_box_diagrams_nondual}. }
\label{Example_of_different_possible_cuts_of_the_box_diagrams_dual}
\end{figure}
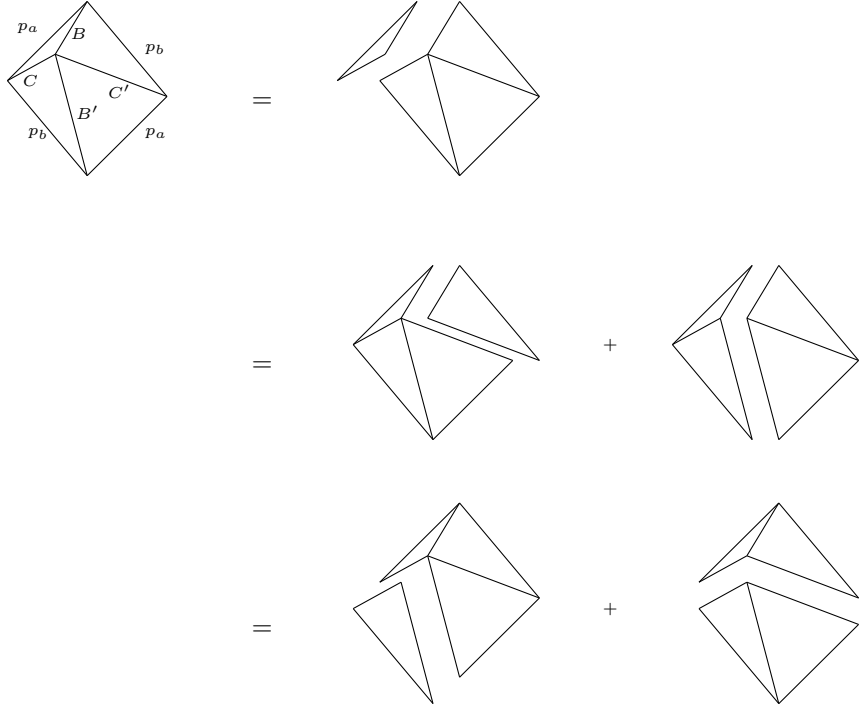

Note that the masses involved are such that $B'$ and $C'$ are negative linear combinations of $B$ and $C$ at this point.
With a small abuse of notation, we use the letters $B$, $C$, $B'$, and $C'$ for both the on-shell vectors and the particle labels. 
As explained previously, since the loop carries two degrees of freedom, even away from the singularity we can shift $l$ so that two momenta, say $B$ and $C$, are on-shell at $l=(0,0)$ 
which means that at this point we have
\begin{equation}
\label{B_C_box_integral_on_shell_example_to_explain_the_method}
B^2-m^2_B=0 \hspace{4mm}, \hspace{4mm} C^2-m^2_C=0.
\end{equation}
Even though we can choose any pair of momenta to be on-shell at $l=(0,0)$ this choice will be particularly convenient for the computation of the residue since, as just mentioned, both $B'$ and $C'$ are negative linear combinations of the chosen momenta $B$ and $C$ on the pole. We will return to this point later.
On the other hand for a general choice of the external kinematics, $B'$ and $C'$ are off-shell at $l=(0,0)$ and their propagators diverge there only at the special point $s=s_0$ where the box diagram is singular.
If we expand the lengths of vectors $B'$ and $C'$ around $s_0$ we obtain
\begin{equation}
\label{B'_C'_box_integral_expansion_around_s0_example_to_explain_the_method}
\begin{split}
&B'^2-m^2_{B'}=\frac{d B'^2}{ds}\Bigr|_{B,C} (s-s_0) + \ldots\\
&C'^2-m^2_{C'}=\frac{d C'^2}{ds}\Bigr|_{B,C} (s-s_0) + \ldots
\end{split}
\end{equation}
where the ellipses contain higher powers of $(s-s_0)$. The subscripts $B$ and $C$ in the derivatives indicate that we are differentiating $B'^2$ and $C'^2$ with respect to $s$ keeping the lengths of $B$ and $C$ fixed at their on-shell values. The situation is shown in figure \ref{Example_of_on_shell_one_loop_box_diagram_how_to_perform_the_limit} where on the LHS we see that near the pole position only $B$ and $C$ are on-shell (the black internal propagators) while the red momenta $B'$ and $C'$ are off-shell; indeed their lengths do not correspond to the values of their masses. We perform the integral in this position and then we take the limit $s \to s_0$ (RHS part of figure \ref{Example_of_on_shell_one_loop_box_diagram_how_to_perform_the_limit}) at which the Feynman diagram becomes singular.
Expanding the loop around the values in \eqref{B_C_box_integral_on_shell_example_to_explain_the_method}, \eqref{B'_C'_box_integral_expansion_around_s0_example_to_explain_the_method} and scaling the loop variable $l=(s-s_0) \tilde{l}$ we obtain
\begin{equation}
\label{box_integral_rescaled_B_C_onshell_B'C'_off_shell}
I_\Box= \frac{1}{(s-s_0)^{2}} \int  \frac{d^2 \tilde{l}}{(2\pi)^2}  \frac{1}{2 B \cdot \tilde{l} + i \epsilon} \ \frac{1}{2 C \cdot  \tilde{l} + i \epsilon} \   \frac{1}{\frac{d B'^2}{d s}\Bigl|_{B,C}+2 B' \cdot  \tilde{l} + i \epsilon} \  \frac{1}{\frac{d C'^2}{d s}\Bigl|_{B,C}+2 C' \cdot  \tilde{l} + i \epsilon}.
\end{equation}
Equality~\eqref{box_integral_rescaled_B_C_onshell_B'C'_off_shell} is a particular case of the general formula \eqref{general_formula_for_the_pole_integrating_over_tilde_variables}. The residue at the pole is isolated in the integral expression above and can be easily computed using Cauchy's theorem. With the purpose of using the scalar products $2 B \cdot \tilde{l}$ and $2 C \cdot \tilde{l}$ as integration variables we express all the momenta in \eqref{box_integral_rescaled_B_C_onshell_B'C'_off_shell} as linear combinations of $B$ and $C$. This is easily done by using \eqref{writing_D_in_terms_of_A_and_B_general_relation}
\begin{equation}
\label{linear_combinations_B'_C'_in_terms_of_B_and_C}
B'=-\frac{\Delta_{B'C}}{\Delta_{BC}} B - \frac{\Delta_{B'B}}{\Delta_{BC}} C \hspace{4mm},\hspace{4mm} C'=-\frac{\Delta_{C'C}}{\Delta_{BC}} B - \frac{\Delta_{C'B}}{\Delta_{BC}} C .
\end{equation}
In these expressions $\Delta_{XY}$ denotes the area of the triangle having the two vectors $X$ and $Y$ as sides, which means, referring to definition \eqref{oriented_triangle_area_definition}, $\Delta_{XY}= \sqrt{\langle X Y \rangle^2}$. We note that all the coefficients of the linear combinations in \eqref{linear_combinations_B'_C'_in_terms_of_B_and_C} are negative numbers. This can be understood pictorially if we look at the right-hand part of figure \ref{Example_of_on_shell_one_loop_box_diagram}. There we see that both $B'$ and $C'$ belong to the section of the plane spanned by positive linear combinations of the vectors $-B$ and $-C$, since the oriented lines identifying their directions are contained in the angle formed between the arrows of $B$ and $C$. If we write the integral \eqref{box_integral_rescaled_B_C_onshell_B'C'_off_shell} in terms of the integration variables 
$$
u=2 B \cdot \tilde{l}  \hspace{4mm} ,\hspace{4mm}  v=2 C \cdot \tilde{l}
$$
we obtain
\begin{equation}
\label{box_integral_rescaled_B_C_onshell_B'C'_off_shell_expressed_in_terms_of_u_v}
\begin{split}
I_\Box= \frac{1}{(s-s_0)^{2}} \frac{1}{(2\pi)^2 8 i \Delta_{BC}} &\int du \ dv  \underbrace{ \frac{1}{u + i \epsilon}}_{B}  \ \underbrace{ \frac{1}{v + i \epsilon}}_{C}  \\
&\underbrace{\frac{1}{\frac{d B'^2}{d s}\Bigl|_{B,C}-\frac{\Delta_{B'C}}{\Delta_{BC}} u - \frac{\Delta_{B'B}}{\Delta_{BC}} v + i \epsilon}}_{B'} \  \underbrace{\frac{1}{\frac{d C'^2}{d s}\Bigl|_{B,C}-\frac{\Delta_{C'C}}{\Delta_{BC}} u - \frac{\Delta_{C'B}}{\Delta_{BC}} v + i \epsilon}}_{C'} ,
\end{split}
\end{equation}
where the factor in front of the integral comes from the Jacobian associated with the change of variables
$$
d^2 \tilde{l}= \frac{1}{8 i \Delta_{BC}} du \ dv .
$$
Each term in the integral \eqref{box_integral_rescaled_B_C_onshell_B'C'_off_shell_expressed_in_terms_of_u_v} corresponds to a particular propagator whose particle label is indicated under it. The different ways in which we can close the $u$/$v$-contours in the complex plane to write the result as sums over different residues correspond to different possible decompositions of the loop as sums over products of tree-level graphs. Let us consider the simplest
situation in which we close both the $u$- and $v$- contours in the lower half complex plane with semicircles of infinite radius.  Using Cauchy's theorem the result is simply given by
\begin{equation}
\label{Box_integral_result_still_derivatives_to_be_substituted_still_additional_factors_to_be_inserted}
I_\Box= \frac{i}{(s-s_0)^{2}} \frac{1}{ 8 \Delta_{BC}} \underbrace{\frac{1}{\frac{d B'^2}{d s}\Bigl|_{B,C}}}_{B'}  \underbrace{\frac{1}{\frac{d C'^2}{d s}\Bigl|_{B,C}}}_{C'}.
\end{equation}
The propagators corresponding to the $B$ and $C$ particles are removed when we take the residues since they have simple poles in the lower half plane. On the other hand, propagators corresponding to $B'$ and $C'$ survive since their poles are not contained in the integration contour. These surviving propagators are evaluated at fixed on-shell lengths of the external momenta and fixed on-shell values of $B$ and $C$. This implies that the loop integral is reduced to a tree-level diagram in which $B$ and $C$ play the role of external on-shell particles. 
After multiplying by the remaining factors in~\eqref{extra_multiplicative_factors_to_be_added_in_the_end}, the box diagram can be written as a single atom which we denote as $D(B,C)$ (as already specified in the introduction, we refer to the various cut diagrams into which a loop can be decomposed as atoms):
\begin{equation}
\label{explicit_decomposition_of_box_into_product_of_tree_level_quantities}
    D_\Box = D(B,C)= \frac{1}{8 i \Delta_{BC}} X_1 \times X_2 \,,
\end{equation}
where
\begin{equation}
\label{X1_X2_explicit_tree_expressions}
    \begin{split}
        X_1&\equiv (-iC_{aB\bar{C}}),\\
        X_2&\equiv  (-iC_{Cb\bar{B'}}) \frac{i}{\frac{d B'^2}{d s} (s-s_0)} (-iC_{B' \bar{a}\bar{C'}}) \frac{i}{\frac{d C'^2}{d s} (s-s_0)} (-iC_{C' \bar{b}\bar{B}}).
    \end{split}
\end{equation}
The quantities $X_1$ and $X_2$ are two tree-level Feynman diagrams with three and five external legs respectively, evaluated in the neighbourhoods of the points where all their internal propagators are singular. However, it is important to remark that we are following a particular direction to reach these singular points, something which is left implicit in the notation. 
Note that a diagram with five external legs has two degrees of freedom in general: two rapidities of the external particles are fixed by the overall energy-momentum conservation and one rapidity can be removed by Lorentz invariance. Therefore, among the five external rapidities of the particles involved in a five-point amplitude,  two are actual degrees of freedom of the amplitude. 
A natural question would then be why in the second line of~\eqref{X1_X2_explicit_tree_expressions} the expansion of the denominators is performed with respect to a single degree of freedom $s$. The reason is that we are moving to the pole position $s=s_0$, where all the propagators in $X_2$ are on-shell, keeping the difference between the rapidities of the particles on the lines that have been cut, that is $B$ and $C$, fixed. 
In this way, there is only one degree of freedom left ($s$) and 
the denominators in the second equality in~\eqref{X1_X2_explicit_tree_expressions} correspond to the leading order expansions of $B'^2 -m_{B'}^2$ and $C'^2 -m_{C'}^2$ with respect to this degree of freedom. $X_2$ can then be written as
\begin{equation}
X_2=(-iC_{Cb\bar{B'}}) \frac{i}{B'^2 -m_{B'}^2} (-iC_{B' \bar{a}\bar{C'}}) \frac{i}{C'^2 -m_{C'}^2} (-iC_{C' \bar{b}\bar{B}})+ O((s-s_0)^{-1}) \, ,
\end{equation}
where it is understood that $B'^2$ and $C'^2$ are both functions of a single parameter $s$ for the reason just explained.
A pictorial representation of (\ref{explicit_decomposition_of_box_into_product_of_tree_level_quantities}) is given by the first equality in figures~\ref{Example_of_different_possible_cuts_of_the_box_diagrams_nondual} and~\ref{Example_of_different_possible_cuts_of_the_box_diagrams_dual}, where the cut of the diagram and the associated dual are reported.

At this point, the problem reduces to finding the derivatives of $B'^2$ and $C'^2$ with respect to the Mandelstam variable $s$. We focus on the triangle $\Delta_{bC}$ on the RHS of figure \ref{Example_of_on_shell_one_loop_box_diagram} with sides $p_b$, $C$ and $B'$, and on the triangle $\Delta_{ab}$ with sides $p_a$, $p_b$ and $p_a+p_b$ corresponding to half of the parallelogram on the RHS of figure~\ref{Example_of_on_shell_one_loop_box_diagram}. The values of $B'^2$ and $s$ are given by
\begin{equation}
\label{expression_of_B'2_and_s_before_differentiating_box_integral}
B'^2=m_b^2+m_C^2-2m_b m_C \cos \bar{U}_{bC} \hspace{4mm},\hspace{4mm} s=m_a^2+m_b^2-2m_a m_b \cos \bar{U}_{ab} 
\end{equation}
where $\bar{U}_{XY}$ is the angle between sides $X$ and $Y$. Since the momenta $p_a$, $B$ and $C$ defining the thin triangle $\Delta_{aBC}$ are on-shell, the angle $\bar{U}_{aC}$ is fixed along the direction we follow to take the limit $s \to s_0$ and the following variations correspond
$$
d \bar{U}_{bC} = d \bar{U}_{ab}.
$$
Thanks to this fact the derivative of $B'^2$ respect to $s$ can be written as
$$
\frac{d B'^2}{ds}=  \frac{\frac{d B'^2}{d \bar{U}_{bC} }}{\frac{d s}{d \bar{U}_{ab} }} = \frac{\Delta_{bC}}{\Delta_{ab}}.
$$
The last equality in the relation above has been obtained using the expressions in \eqref{expression_of_B'2_and_s_before_differentiating_box_integral}. Similarly, it is possible to show that 
$$
\frac{d C'^2}{ds}=-\frac{\Delta_{bB}}{\Delta_{ab}}.
$$
Plugging the values of these derivatives into \eqref{X1_X2_explicit_tree_expressions} and using the area rule  \eqref{Connection_among_three_point_couplings_and_areas} the equality in~\eqref{explicit_decomposition_of_box_into_product_of_tree_level_quantities} becomes
\begin{equation}
\label{final_result_for_the_box_integral_number_1_expressed_with_f_non_explicit_single_cut_BC}
D_\Box= D(B,C)=\frac{\Delta^2_{ab}}{8i (s-s_0)^2} \Delta_{B' C'} \Bigl(\frac{4 \beta}{\sqrt{h}} \Bigl)^4 \sigma_{Ba\bar{C}} \sigma_{Cb\bar{B}'} \sigma_{B' \bar{a}\bar{C}'} \sigma_{C' \bar{b}\bar{B}}.
\end{equation}
By substituting the values of the signs $\sigma$ defined in \eqref{Connection_among_three_point_couplings_and_areas} into \eqref{final_result_for_the_box_integral_number_1_expressed_with_f_non_explicit_single_cut_BC} we obtain  the residue of the diagram at the pole. 

Now we discuss what happens if we choose a different integration contour when we use Cauchy's theorem. We perform the integral \eqref{box_integral_rescaled_B_C_onshell_B'C'_off_shell_expressed_in_terms_of_u_v} in two steps. First
we integrate the $u$-variable closing as before
its contour with a semicircle in the lower half complex plane. We obtain
\begin{equation}
\label{box_integral_rescaled_B_C_onshell_B'C'_off_shell_expressed_in_terms_of_u_v_u_integrated}
\begin{split}
I_\Box= \frac{1}{(s-s_0)^{2}} \frac{2 \pi i}{(2\pi)^2 8 i \Delta_{BC}} \int  dv  \ \underbrace{ \frac{1}{v + i \epsilon}}_{C} \underbrace{\frac{1}{\frac{d B'^2}{d s}\Bigl|_{B,C} - \frac{\Delta_{B'B}}{\Delta_{BC}} v + i \epsilon}}_{B'} \  \underbrace{\frac{1}{\frac{d C'^2}{d s}\Bigl|_{B,C} - \frac{\Delta_{C'B}}{\Delta_{BC}} v + i \epsilon}}_{C'}.
\end{split}
\end{equation}
The $B$-propagator has disappeared from the integral, i.e.\ it has been cut. 
At this point, instead of also closing the $v$-path in the region below the real axis (generating only one residue), we close the path in the upper half complex plane where the terms $B'$ and $C'$ in \eqref{box_integral_rescaled_B_C_onshell_B'C'_off_shell_expressed_in_terms_of_u_v_u_integrated} both have simple poles. The final result is a sum over two residues, one for each propagator having a pole in the region enclosed in the integration path:
\begin{equation}
\label{I_box_integral_written_as_sum_of_two_different_cuts_still_to_be_added_vertex_factors}
\begin{split}
I_\Box= \frac{1}{(s-s_0)^{2}} \frac{i}{8 } \biggl[  &\frac{1}{\Delta_{BB'}}  \underbrace{\frac{1}{\frac{\Delta_{BC}}{\Delta_{B'B}}\frac{d B'^2}{d s}\Bigl|_{B,C}}}_{C} \underbrace{\frac{1}{\frac{d C'^2}{d s}\Bigl|_{B,C}-\frac{\Delta_{C'B}}{\Delta_{B'B}}\frac{d B'^2}{d s}\Bigl|_{B,C}}}_{C'}+\\
&\frac{1}{\Delta_{C'B}} \underbrace{\frac{1}{\frac{\Delta_{BC}}{\Delta_{C'B}}\frac{d C'^2}{d s}\Bigl|_{B,C}}}_{C} \underbrace{\frac{1}{\frac{d B'^2}{d s}\Bigl|_{B,C}-\frac{\Delta_{B'B}}{\Delta_{C'B}}\frac{d C'^2}{d s}\Bigl|_{B,C}}}_{B'} \biggr] .
\end{split}
\end{equation}
The factor $\frac{1}{\Delta_{XY}}$ in front of each term in the sum above is a Jacobian depending on which variables have been used in the integration; it can be used as a pointer indicating which propagators $X$, $Y$ have been cut. In the first term for example we read $\frac{1}{\Delta_{BB'}}$ indicating that the propagators $B$ and $B'$ have been cut, indeed we take the residues with respect to their poles, and such propagators disappear in the final result. The remaining terms $C$ and $C'$ are related to propagators evaluated close to their on-shell values making the diagram singular.

Making use of lemma~\ref{Formula_for_propgator_derivatives_in_loop_diagrams} it is not difficult to check that each combination of derivatives in the denominators of~\eqref{I_box_integral_written_as_sum_of_two_different_cuts_still_to_be_added_vertex_factors} can be written in a simpler form and the result for the singular diagram is given by
\begin{equation*}
I_\Box= \frac{1}{(s-s_0)^{2}} \frac{i}{8 } \biggl[  \frac{1}{\Delta_{BB'}}  \frac{1}{\frac{d C^2}{d s}\Bigl|_{B,B'}}  \frac{1}{\frac{d C'^2}{d s}\Bigl|_{B,B'}}+ \frac{1}{\Delta_{C'B}} \frac{1}{\frac{d C^2}{d s}\Bigl|_{C',B}} \frac{1}{\frac{d B'^2}{d s}\Bigl|_{C',B}} \biggr].
 \end{equation*}
In the first term the particles $B$ and $B'$ are on-shell and the derivatives are taken at fixed values $p_a^2=m_a^2$,  $p_b^2=m_b^2$ , $B^2=m^2_B$, $B'^2=m^2_{B'}$, in the second term the vectors having fixed lengths are instead $p_a$, $p_b$, $C'$ and $B$.

Multiplying by the usual vertex and propagator factors in~\eqref{extra_multiplicative_factors_to_be_added_in_the_end}, we can write the result for the diagram as a sum of two atoms
\begin{equation}
\label{explicit_double_atom_decomposition_of_box_integral}
    D_{\Box}=D(B,B')+D(B,C')=\frac{1}{8 i \Delta_{B B'}} X_3 \times X_4 + \frac{1}{8 i \Delta_{B C'}} X_5 \times X_6.
\end{equation}
The quantities $X_3$, $X_4$, $X_5$ and $X_6$ are defined as follows
\begin{subequations}
\label{expressions_for_X3_X4_X5_X6}
    \begin{align}
    \label{X3_value_atom_decomposition}
    X_3 &\equiv (-i C_{aB\bar{C}})\frac{i}{\frac{d C^2}{d s}(s-s_0)} (-i C_{C b\bar{B'}})  \sim (-i C_{aB\bar{C}})\frac{i}{C^2 - m^2_C} (-i C_{C b\bar{B'}}),\\
    \label{X4_value_atom_decomposition}
    X_4&\equiv (-i C_{B' \bar{a} \bar{C}'})\frac{i}{\frac{d C'^2}{d s}(s-s_0)} (-i C_{C' \bar{b}\bar{B}})  \sim (-i C_{B' \bar{a} \bar{C}'})\frac{i}{C'^2 - m^2_{C'}} (-i C_{C' \bar{b}\bar{B}}),\\
    X_5&\equiv -i C_{C' \bar{b} \bar{B}},\\
    \begin{split}
     X_6&= (-i C_{B a \bar{C}})\frac{i}{\frac{d C^2}{d s}(s-s_0)} (-i C_{C b\bar{B'}})\frac{i}{\frac{d B'^2}{d s}(s-s_0)} (-i C_{B' \bar{a} \bar{C'}})\\
     &\sim (-i C_{B a \bar{C}})\frac{i}{C^2 - m^2_C} (-i C_{C b\bar{B'}})\frac{i}{B'^2 - m^2_{B'}} (-i C_{B' \bar{a} \bar{C'}}).
    \end{split}
\end{align}
\end{subequations}
In some of the relations above we use the symbol $\sim$ instead of $=$ to represent the fact that the equality is valid only at the leading order in $s-s_0$ (or equivalently in $\theta-i\theta_0$). We will keep using this symbol also in the next sections.
Using the area rule~\eqref{Connection_among_three_point_couplings_and_areas} and computing  the derivatives of the squared of the momenta appearing in the propagators we obtain
\begin{multline}
\label{fineal_result_for_the_box_integral_number_1_expressed_with_f_non_explicit_double_cut_BBprime_BCprime}
D_\Box= D(B,B')+D(B,C')\\
=\frac{\Delta^2_{ab}}{8i (s-s_0)^2} (\Delta_{B B'}+\Delta_{B' C}) \Bigl( \frac{4 \beta}{\sqrt{h}}\Bigl)^4 \sigma_{Ba\bar{C}} \sigma_{Cb\bar{B}'} \sigma_{B' \bar{a}\bar{C}'} \sigma_{C' \bar{b}\bar{B}}.
\end{multline}
To verify that the two results in \eqref{fineal_result_for_the_box_integral_number_1_expressed_with_f_non_explicit_double_cut_BBprime_BCprime} and \eqref{final_result_for_the_box_integral_number_1_expressed_with_f_non_explicit_single_cut_BC} are equal we write the triangle areas in terms of vector products and note that 
$$
{\bigr \langle B' B \bigr \rangle} + {\bigr \langle C B' \bigr \rangle}  = {\bigr \langle (C-B) B' \bigr \rangle} =  {\bigr \langle p_a B' \bigr \rangle} .
$$
The box integral at the Landau singularity can be written both as a single term, $D(B,C)$, given by cutting the internal propagators $B$, $C$ and multiplying by the Jacobian factor $\frac{1}{8 i \Delta_{BC}}$, or as a sum over two terms. Each one of these two terms, $D(B,B')$ and $D(B,C')$, corresponds to a particular atom. This second way of writing the singularity is shown in the second row of figures~\ref{Example_of_different_possible_cuts_of_the_box_diagrams_nondual} and~\ref{Example_of_different_possible_cuts_of_the_box_diagrams_dual}. 

In summary, we have seen that the threshold singularity of the box integral can be written as different sums of residues. Each residue is associated with a Feynman diagram decomposition in which propagators having poles inside the integration contour disappear, i.e.\ they are cut. In this manner, the diagram behaves as a tree-level graph with more external on-shell particles. 
The different ways in which the full diagram, or molecule, can be broken into sums of these atoms
are dictated by the dual description of the diagram. In the case considered the choice of $B$ and $C$ to define the integration variables is special in that it leads to just a single atom. This is a result of the fact that all the other momenta ($B'$ and $C'$) are linear combinations of $B$ and $C$ with negative coefficients. 
No matter what singular box integral we are considering, to find the two special momenta in terms of which all the other vectors are written as negative linear combinations we need to search for the triangle which contained in two concave quadrilaterals. In the example considered the vectors $B$ and $C$ form a triangle $\Delta_{aBC}$ which is contained both in the concave quadrilateral defined by the sides $B$, $B'$, $p_b$-incoming, $p_a$-incoming and in the concave quadrilateral defined by $C$, $C'$, $p_b$-outgoing, $p_a$-incoming. 
Looking at the right hand part of figure \ref{Example_of_on_shell_one_loop_box_diagram}, this is not true for any other triangle. For example the triangle $\Delta_{bB' C}$ is contained in the concave quadrilateral having for sides $B$, $B'$, $p_b$-incoming, $p_a$-incoming but it is also contained in the convex quadrilateral defined by sides $C$, $C'$, $p_a$-outgoing, $p_b$-incoming and having $B'$ for diagonal.
Once we find such special vectors, we have found the two propagators with respect to which taking the residues we generate the complete result as a single atom.
On the other hand, we can write the box singularity as a sum over two different residues: we need to cut one among the two special vectors ($B$ or $C$ in this case) and sum over the cuts of the other two sides.
We write below some of the different sums over atoms through which the leading box singularity can be written 
\begin{equation}
\label{some_different_cuts_D(X,Y)_of_the_box_integral}
D_\Box=D(B,C)=D(B,B')+D(B,C')=D(B',C)+D(C',C) .
\end{equation}
The first equality in this expression is given by closing both the $u$ and $v$ contours in \eqref{box_integral_rescaled_B_C_onshell_B'C'_off_shell_expressed_in_terms_of_u_v} in the lower half plane. In this way, we obtain a single term generated by cutting the propagators associated with momenta $B$ and $C$, i.e.\ the special vectors with respect to which all the other momenta are negative linear combinations. The second equality  
corresponds to closing the $u$-contour in \eqref{box_integral_rescaled_B_C_onshell_B'C'_off_shell_expressed_in_terms_of_u_v} in the lower half plane (there is a single pole in the integration path due to the $B$-propagator) and the $v$-contour in the upper half-plane. Since now there are two poles in the integration $v$-contour we have a sum over two different cuts, associated with $B'$ and $C'$. 
On the other hand on the RHS of the third equality in \eqref{some_different_cuts_D(X,Y)_of_the_box_integral} we have performed the integration over $v$ first, closing the integration path in the lower half complex plane. The $v$-contour encloses a single pole of the propagator associated with the particle $C$, and therefore we obtain a single residue corresponding to the cut of the $C$-propagator. Then closing the $u$-contour in the upper half plane we obtain a sum over two residues, one corresponding to cutting the propagator $B'$ and the other corresponding to cutting the propagator $C'$.

In the next sections, we will study networks of Feynman diagrams contributing to some higher-order singularities. In this context it is important to understand how to decompose each molecule of the network cleverly, to make the simplifications between different atoms manifest. 
We will show how special decompositions of diagrams connected by flipping internal propagators sum to zero and the total result is given by a few surviving atoms. We start by studying this phenomenon for $2^{\text{nd}}$-order poles and then move to $3^{\text{rd}}$-order poles.

Even though we identify universal structures common to all the ADE series of affine Toda models, our results are based on a case-by-case study. 
For a fixed $p$ in~\eqref{General_Laurent_expansion_on_the_pole}, we observe that the coefficients $a_p$ of different affine Toda models
are generated by networks sharing the same topology and the same atomic decomposition. However, the reason why this happens is unclear.
It would be interesting to find a universal explanation of how the networks simplify, for example using common properties of the underlying root systems.

\section{Second-order poles}
\label{Chapter_on_second_order_poles}
Second-order poles appear when the bootstrapped S-matrix~\eqref{S_matrix_bootstrap_result_structure} contains the product of two building blocks $\{x-1 \} \{x+1\}$.
These are even order singularities with $N=1$ and $\nu=0$, for which the leading order contribution~\eqref{S_matrix_contributions_on_the_poles_expanded_and_containing_all_the_leading_singularities} to the Laurent expansion about the pole becomes
\begin{equation}
\label{double_singularities_S_matrix_from_bootstrap}
S^{(l)}_{ab}(\theta) = 1+ \Bigl(\frac{\beta^2}{2 h}\Bigr)^2 \frac{1}{(\theta- i \theta_0)^2}.
\end{equation}
This singular part of the S-matrix was computed using perturbation theory in~\cite{Braden:1990wx}. 
However this paper did not discuss why Feynman diagrams connected by flipping internal propagators, that cancel at tree level in $4$-point non-diagonal processes, do not sum to zero at loop level. In this section, following the approach used in~\cite{First_loop_paper_sagex}, we show that this fact is related to how the loop diagrams are cut into atoms. In addition we highlight that cancellations between different atoms in the network of singular diagrams are nevertheless possible, and we show which atoms do not cancel and contribute to the final result.

\begin{figure}
\includegraphics[width=1\linewidth]{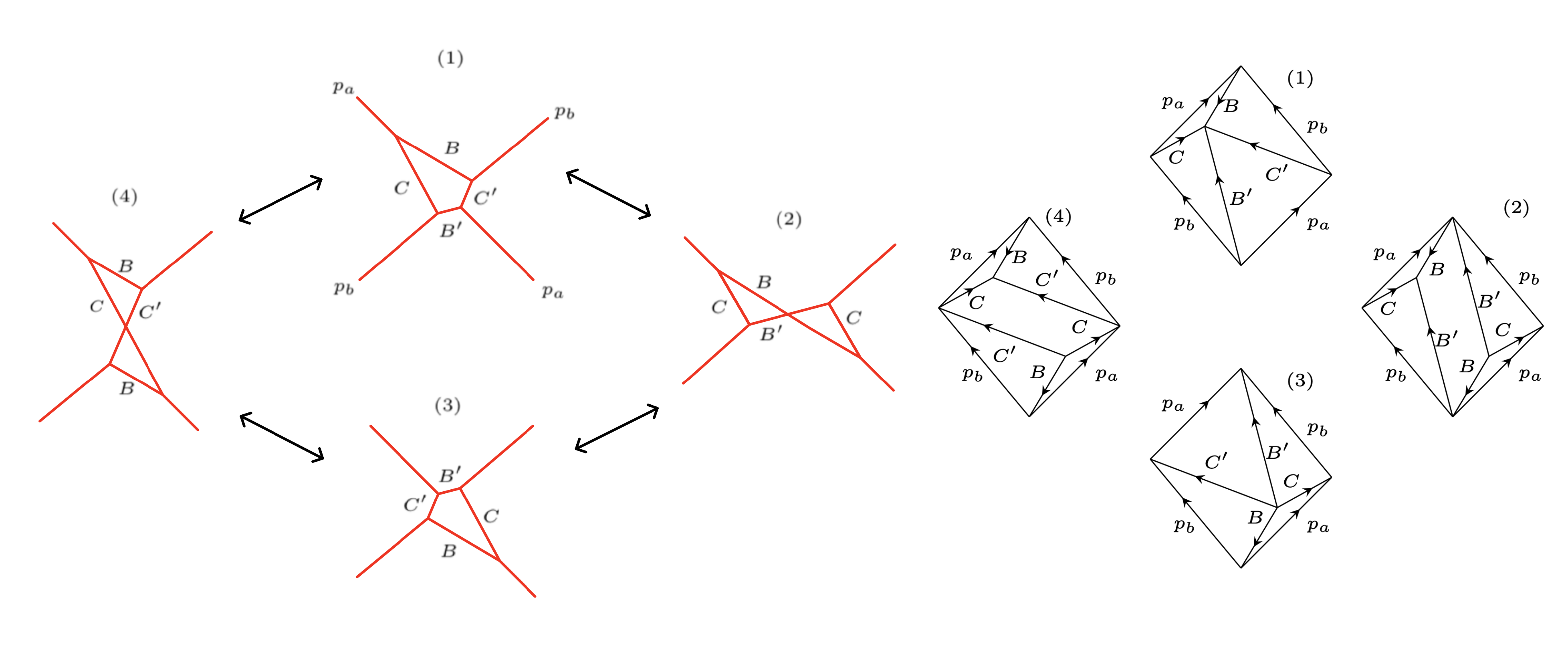}
\caption{Network of on-shell Feynman diagrams (on the left) and their duals (on the right) contributing to a second-order pole.}
\label{allowed_one_loop_box_network_on_shell_and_F_Diagram_description}
\end{figure}

There are in total four Feynman diagrams contributing to the pole. They are connected by flips of type II, according to the convention introduced in~\cite{Braden:1990wx} and reviewed in~\cite{Davide_Patrick_tree_level_paper}.
The full network of on-shell Feynman diagrams, together with their duals, is shown in figure~\ref{allowed_one_loop_box_network_on_shell_and_F_Diagram_description}.
The external shape of each on-shell diagram entering the network is always the same: it is a parallelogram with sides equal to the masses of the scattered particles and angles determined by their rapidities. Internally the parallelogram is tiled by triangles, representing $3$-point vertices, having as sides the masses of the propagating particles. In addition diagrams  $(2)$ and $(4)$ on the RHS of figure \ref{allowed_one_loop_box_network_on_shell_and_F_Diagram_description} have an internal empty space, corresponding to on-shell
propagators that cross each other. 
Starting from any diagram of the network we can generate all the others by flipping internal propagators. For example, to move from diagram $(1)$ to diagram $(2)$ on the RHS of figure~\ref{allowed_one_loop_box_network_on_shell_and_F_Diagram_description} we exchange the positions of the vectors $B$ and $B'$ and we flip $C'$, transforming it into $C$. 
As explained in~\cite{Davide_Patrick_tree_level_paper} this is a flip of type II and it does not change the product of the $3$-point couplings entering in the diagram. This is a requirement to avoid non-diagonal processes at the tree level.
The same type of flip connects the remaining diagrams of the network to one another. Since diagrams $(2)$ and $(4)$ contain products of  $3$-point couplings that are necessarily positive, being the absolute values squared of certain phases, and the product of the $3$-point couplings after each flip does not change sign, we have
\begin{equation*}
\begin{split}
\sigma_{Ba\bar{C}} \sigma_{Cb\bar{B}'} \sigma_{B' \bar{a}\bar{C}'} \sigma_{C' \bar{b}\bar{B}}&\underbrace{=}_{(1) \to (2)} |\sigma_{Ba\bar{C}}|^2 |\sigma_{Cb\bar{B}'}|^2 \underbrace{=}_{(2) \to (3)} \sigma_{Bb\bar{C'}} \sigma_{C' a \bar{B'}}  \sigma_{B' \bar{b}\bar{C}} \sigma_{C\bar{a}\bar{B}}\\
&\underbrace{=}_{(3) \to (4)} |\sigma_{Ba\bar{C}}|^2 |\sigma_{Bb\bar{C}'}|^2=1 .
\end{split}
\end{equation*}
Using this result combined with \eqref{fineal_result_for_the_box_integral_number_1_expressed_with_f_non_explicit_double_cut_BBprime_BCprime}, the first diagram of the network, labelled (1) in figure~\ref{allowed_one_loop_box_network_on_shell_and_F_Diagram_description}, can be written as 
\begin{equation}
\label{final_result_for_the_box_integral_number_1_double_cut_BBp_BCp_full_network_description}
D^{(1)}= D^{(1)}(B,B')+D^{(1)}(B,C')=
-32 i \Bigl( \frac{ \beta}{\sqrt{h}} \Bigr)^4 \frac{\Delta^2_{ab}}{(s-s_0)^2} (\Delta_{B'C}+\Delta_{B'B})
\end{equation}
where, as explained in the previous section, the two pieces in the sum~\eqref{final_result_for_the_box_integral_number_1_double_cut_BBp_BCp_full_network_description} represent two different cuts of the box.

\begin{figure}
\begin{center}
\medskip
\begin{tikzpicture}
\tikzmath{\y = 0.5;}

\draw[] (5.8*\y,1*\y) -- (7.3*\y,2.5*\y);
\draw[] (5.8*\y,1*\y) -- (6.7*\y,1.5*\y);
\draw[] (7.3*\y,-0.8*\y) -- (5.8*\y,1*\y);
\draw[] (7.3*\y,-0.8*\y) -- (6.7*\y,1.5*\y);
\draw[] (7.3*\y,-0.8*\y) -- (8.8*\y,0.7*\y);
\draw[] (7.3*\y,2.5*\y) -- (6.7*\y,1.5*\y);
\draw[] (8.8*\y, 0.7*\y) -- (6.7*\y,1.5*\y);

\draw[] (7.3*\y+1*\y,2.5*\y) -- (6.7*\y+1*\y,1.5*\y);
\draw[] (8.8*\y+1*\y,0.7*\y) -- (7.3*\y+1*\y,2.5*\y);
\draw[] (8.8*\y+1*\y, 0.7*\y) -- (6.7*\y+1*\y,1.5*\y);

\filldraw[black] (10.3*\y,0.8*\y)  node[anchor=west] {\tiny{$+$}};

\draw[blue] (5.8*\y+6*\y,1*\y) -- (7.3*\y+6*\y,2.5*\y);
\draw[blue] (5.8*\y+6*\y,1*\y) -- (6.7*\y+6*\y,1.5*\y);
\draw[blue] (7.3*\y+6*\y,-0.8*\y) -- (5.8*\y+6*\y,1*\y);
\draw[blue] (7.3*\y+6*\y,-0.8*\y) -- (6.7*\y+6*\y,1.5*\y);
\draw[blue] (7.3*\y+6*\y,2.5*\y) -- (6.7*\y+6*\y,1.5*\y);

\draw[blue] (7.3*\y+7*\y,-0.8*\y) -- (8.8*\y+7*\y,0.7*\y);
\draw[blue] (7.3*\y+7*\y,2.5*\y) -- (6.7*\y+7*\y,1.5*\y);
\draw[blue] (8.8*\y+7*\y,0.7*\y) -- (7.3*\y+7*\y,2.5*\y);
\draw[blue] (8.8*\y+7*\y, 0.7*\y) -- (6.7*\y+7*\y,1.5*\y);
\draw[blue] (7.3*\y+7*\y,-0.8*\y) -- (6.7*\y+7*\y,1.5*\y);

\filldraw[] (7*\y,-1.4*\y)  node[anchor=west] {\tiny{$\Delta_{B'C}$}};
\filldraw[blue] (12*\y,-1.4*\y)  node[anchor=west] {\tiny{$\Delta_{BB'}$}};

\filldraw[black] (10.3*\y,2.6*\y)  node[anchor=west] {\tiny{$(1)$}};

\draw[blue] (5.8*\y+8.5*\y+3.5*\y,1*\y-4.5*\y) -- (7.3*\y+8.5*\y+3.5*\y,2.5*\y-4.5*\y);
\draw[blue] (5.8*\y+8.5*\y+3.5*\y,1*\y-4.5*\y) -- (6.7*\y+8.5*\y+3.5*\y,1.5*\y-4.5*\y);
\draw[blue] (7.3*\y+8.5*\y+3.5*\y,2.5*\y-4.5*\y) -- (6.7*\y+8.5*\y+3.5*\y,1.5*\y-4.5*\y);
\draw[blue] (7.3*\y+8.5*\y+3.5*\y,-0.8*\y-4.5*\y) -- (5.8*\y+8.5*\y+3.5*\y,1*\y-4.5*\y);
\draw[blue] (7.3*\y+8.5*\y+3.5*\y,-0.8*\y-4.5*\y) -- (6.7*\y+8.5*\y+3.5*\y,1.5*\y-4.5*\y);

\draw[blue] (7.9*\y+9.5*\y+3.5*\y,0.2*\y-4.5*\y) -- (7.3*\y+9.5*\y+3.5*\y,2.5*\y-4.5*\y);
\draw[blue] (7.9*\y+9.5*\y+3.5*\y,0.2*\y-4.5*\y) -- (7.3*\y+9.5*\y+3.5*\y,-0.8*\y-4.5*\y);
\draw[blue] (7.9*\y+9.5*\y+3.5*\y,0.2*\y-4.5*\y) -- (8.8*\y+9.5*\y+3.5*\y,0.7*\y-4.5*\y);
\draw[blue] (7.3*\y+9.5*\y+3.5*\y,-0.8*\y-4.5*\y) -- (8.8*\y+9.5*\y+3.5*\y,0.7*\y-4.5*\y);
\draw[blue] (8.8*\y+9.5*\y+3.5*\y,0.7*\y-4.5*\y) -- (7.3*\y+9.5*\y+3.5*\y,2.5*\y-4.5*\y);

\filldraw[blue] (15*\y,-3.5*\y)  node[anchor=west] {\tiny{$-\Delta_{BB'}$}};

\filldraw[black] (19.4*\y,-1*\y)  node[anchor=west] {\tiny{$(2)$}};

\draw[red] (5.8*\y,1*\y-10*\y) -- (7.3*\y,2.5*\y-10*\y);
\draw[red] (7.9*\y,0.2*\y-10*\y) -- (5.8*\y,1*\y-10*\y) ;
\draw[red] (7.9*\y,0.2*\y-10*\y) -- (7.3*\y,2.5*\y-10*\y);
\draw[red] (7.9*\y,0.2*\y-10*\y) -- (8.8*\y,0.7*\y-10*\y);
\draw[red] (8.8*\y,0.7*\y-10*\y) -- (7.3*\y,2.5*\y-10*\y);

\draw[red] (7.9*\y,0.2*\y-11*\y) -- (5.8*\y,1*\y-11*\y) ;
\draw[red] (7.3*\y,-0.8*\y-11*\y) -- (5.8*\y,1*\y-11*\y);
\draw[red] (7.9*\y,0.2*\y-11*\y) -- (7.3*\y,-0.8*\y-11*\y);
\draw[red] (7.3*\y,-0.8*\y-11*\y) -- (8.8*\y,0.7*\y-11*\y);
\draw[red] (7.9*\y,0.2*\y-11*\y) -- (8.8*\y,0.7*\y-11*\y);

\filldraw[red] (6.5*\y,-6.5*\y)  node[anchor=west] {\tiny{$\Delta_{CC'}$}};

\filldraw[black] (10.3*\y,-9.8*\y)  node[anchor=west] {\tiny{$+$}};

\draw[] (5.8*\y+6*\y,1*\y-11*\y) -- (7.3*\y+6*\y,2.5*\y-11*\y);
\draw[] (7.9*\y+6*\y,0.2*\y-11*\y) -- (5.8*\y+6*\y,1*\y-11*\y) ;
\draw[] (7.9*\y+6*\y,0.2*\y-11*\y) -- (7.3*\y+6*\y,2.5*\y-11*\y);
\draw[] (7.9*\y+6*\y,0.2*\y-11*\y) -- (5.8*\y+6*\y,1*\y-11*\y) ;
\draw[] (7.3*\y+6*\y,-0.8*\y-11*\y) -- (5.8*\y+6*\y,1*\y-11*\y);
\draw[] (7.9*\y+6*\y,0.2*\y-11*\y) -- (7.3*\y+6*\y,-0.8*\y-11*\y);
\draw[] (7.3*\y+6*\y,-0.8*\y-11*\y) -- (8.8*\y+6*\y,0.7*\y-11*\y);
\draw[] (7.9*\y+6*\y,0.2*\y-11*\y) -- (8.8*\y+6*\y,0.7*\y-11*\y);

\draw[] (7.9*\y+6.5*\y,0.2*\y-10*\y) -- (7.3*\y+6.5*\y,2.5*\y-10*\y);
\draw[] (7.9*\y+6.5*\y,0.2*\y-10*\y) -- (8.8*\y+6.5*\y,0.7*\y-10*\y);
\draw[] (8.8*\y+6.5*\y,0.7*\y-10*\y) -- (7.3*\y+6.5*\y,2.5*\y-10*\y);

\filldraw[] (12.5*\y,-6.5*\y)  node[anchor=west] {\tiny{$\Delta_{BC'}$}};

\filldraw[black] (10.3*\y,-7*\y)  node[anchor=west] {\tiny{$(3)$}};

\draw[red] (5.8*\y-3*\y-3.5*\y,1*\y-4.5*\y) -- (7.3*\y-3*\y-3.5*\y,2.5*\y-4.5*\y);
\draw[red] (5.8*\y-3*\y-3.5*\y,1*\y-4.5*\y) -- (6.7*\y-3*\y-3.5*\y,1.5*\y-4.5*\y);
\draw[red] (7.3*\y-3*\y-3.5*\y,2.5*\y-4.5*\y) -- (6.7*\y-3*\y-3.5*\y,1.5*\y-4.5*\y);
\draw[red] (8.8*\y-3*\y-3.5*\y,0.7*\y-4.5*\y) -- (7.3*\y-3*\y-3.5*\y,2.5*\y-4.5*\y);
\draw[red] (8.8*\y-3*\y-3.5*\y, 0.7*\y-4.5*\y) -- (6.7*\y-3*\y-3.5*\y,1.5*\y-4.5*\y);

\draw[red] (7.9*\y-3*\y-3.5*\y,0.2*\y-5.5*\y) -- (7.3*\y-3*\y-3.5*\y,-0.8*\y-5.5*\y);
\draw[red] (7.9*\y-3*\y-3.5*\y,0.2*\y-5.5*\y) -- (5.8*\y-3*\y-3.5*\y,1*\y-5.5*\y);
\draw[red] (7.9*\y-3*\y-3.5*\y,0.2*\y-5.5*\y) -- (8.8*\y-3*\y-3.5*\y, 0.7*\y-5.5*\y);
\draw[red] (7.3*\y-3*\y-3.5*\y,-0.8*\y-5.5*\y) -- (5.8*\y-3*\y-3.5*\y,1*\y-5.5*\y);
\draw[red] (7.3*\y-3*\y-3.5*\y,-0.8*\y-5.5*\y) -- (8.8*\y-3*\y-3.5*\y,0.7*\y-5.5*\y);

\filldraw[red] (2.5*\y,-3.5*\y)  node[anchor=west] {\tiny{$-\Delta_{CC'}$}};

\filldraw[black] (1.4*\y,-1*\y)  node[anchor=west] {\tiny{$(4)$}};

\end{tikzpicture}

\caption{On-shell decomposition of diagrams into (sums of) products of tree-level graphs. Atoms
differing by the presence of two $4$-point inelastic Feynman diagrams connected by flipping one propagator
cancel in the final sum. Near each diagram, its value contributing to the double pole residue is reported. A overall factor $-32 i \Bigl( \frac{ \beta}{\sqrt{h}} \Bigr)^4 \frac{\Delta^2_{ab}}{(s-s_0)^2}$ common to each diagram has been omitted and needs to be introduced after the sum.}
\label{allowed_one_loop_box_network_geometry_cuts_with_different_colours_on_shell_description}
\end{center}
\end{figure}
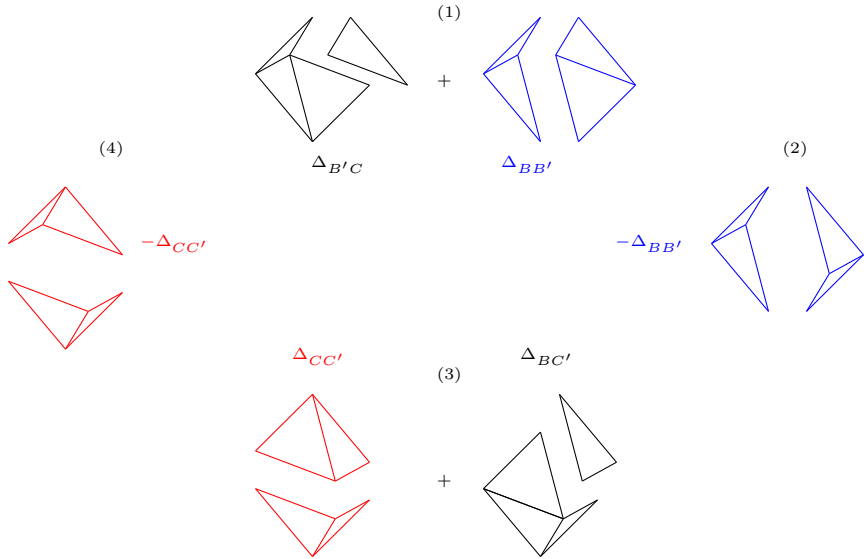

The values associated with the other diagrams are straightforwardly obtained in the same way. Diagram  $(3)$ is similar to diagram  $(1)$; in this case we write its value by summing the atoms $D^{(3)}(C,B')$ and $D^{(3)}(C,C')$. It is given by
\begin{equation*}
D^{(3)}=D^{(3)}(C,B')+D^{(3)}(C,C')=
-32 i \Bigl( \frac{ \beta}{\sqrt{h}} \Bigr)^4 \frac{\Delta^2_{ab}}{(s-s_0)^2} (\Delta_{C'B}+\Delta_{C'C}).
 \end{equation*}
\begin{figure}
\begin{center}
\includegraphics[width=1\linewidth]{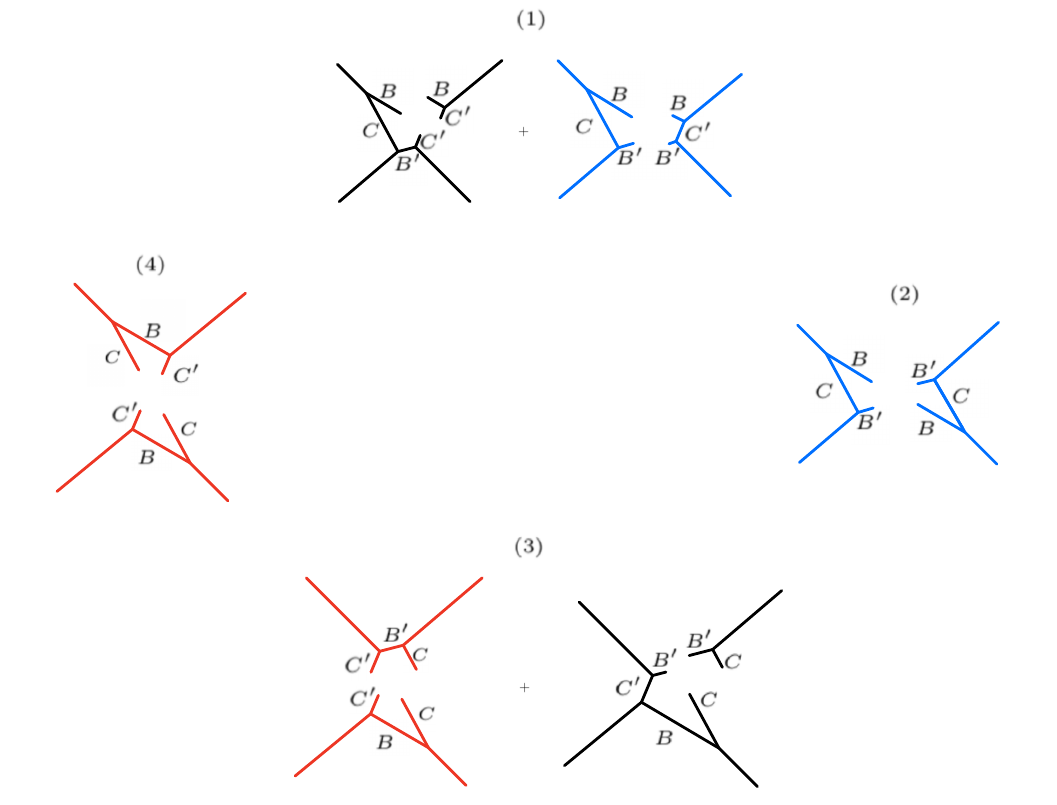}
\caption{Feynman diagram decompositions into sums of products of tree-level graphs.}
\label{Box_network_geometry_cuts_with_different_colours_FD}
\end{center}
\end{figure}
The remaining diagrams $(2)$ and $(4)$ are also simple; their integrals are 
\begin{equation*}
\begin{split}
I^{(2)}&= \frac{1}{(s-s_0)^{2}} \frac{1}{(2\pi)^2 8 i \Delta_{BB'}} \int du dv \ \underbrace{ \frac{1}{u + i \epsilon}}_{B}  \  \underbrace{ \frac{1}{v + i \epsilon}}_{B'} \underbrace{\frac{1}{\bigl(\frac{d C^2}{d s}\Bigl|_{B,B'} -  \frac{\Delta_{CB'}}{\Delta_{BB'}} u-\frac{\Delta_{CB}}{\Delta_{BB'}} v + i \epsilon\bigr)^2}}_{C} , \\
I^{(4)}&= \frac{1}{(s-s_0)^{2}} \frac{1}{(2\pi)^2 8 i \Delta_{CC'}} \int du dv \ \underbrace{ \frac{1}{u + i \epsilon}}_{C}  \  \underbrace{ \frac{1}{v + i \epsilon}}_{C'} \underbrace{\frac{1}{\bigl(\frac{d B^2}{d s}\Bigl|_{C,C'} -  \frac{\Delta_{BC'}}{\Delta_{CC'}} u-\frac{\Delta_{BC}}{\Delta_{CC'}} v + i \epsilon\bigr)^2}}_{B}.
\end{split}
\end{equation*}
Each of these results is given by a single atom, which is $D^{(2)}(B,B')$ in the case of diagram  $(2)$ and $D^{(4)}(C,C')$ in the case of diagram  $(4)$:
\begin{equation*}
\begin{split}
D^{(2)}(B,B')&=-32 i \Bigl( \frac{ \beta}{\sqrt{h}} \Bigr)^4 \frac{\Delta^2_{ab}}{(s-s_0)^2} \times (-\Delta_{B'B}),\\
D^{(4)}(C,C')&=-32 i \Bigl( \frac{ \beta}{\sqrt{h}} \Bigr)^4 \frac{\Delta^2_{ab}}{(s-s_0)^2} \times (-\Delta_{C'C}).
\end{split}
\end{equation*}
The full network of on-shell diagrams, each one decomposed into tree-level atoms, is shown in figure~\ref{Box_network_geometry_cuts_with_different_colours_FD}, while figure \ref{allowed_one_loop_box_network_geometry_cuts_with_different_colours_on_shell_description} shows their dual descriptions.
Atoms that differ by a $4$-point inelastic tree-level diagram in which one propagator is flipped are plotted in the same colour. 
For example, 
the two blue atoms coming from diagram $(1)$ and $(2)$ can be written as
\begin{equation}
\label{cut_BBprime_diagram_1_2_double_pole}
\begin{split}
     &D^{(1)}(B, B')=\frac{1}{8 i \Delta_{B B'}} X_3 \times X_4\\
     &\text{and}\\
     &D^{(2)}(B, B')=\frac{1}{8 i \Delta_{B B'}} X_3 \times X^{(\text{flipped})}_4
\end{split}
\end{equation}
respectively.
The expressions for $X_3$ and $X_4$ are written in~\eqref{expressions_for_X3_X4_X5_X6} while $X^{(\text{flipped})}_4$ is given by
\begin{equation}
    X^{(\text{flipped})}_4 = (-i C_{B \bar{a} \bar{C}})\frac{i}{\frac{d C^2}{d s}(s-s_0)} (-i C_{C \bar{b}\bar{B'}})  \sim (-i C_{B \bar{a} \bar{C}})\frac{i}{C^2 - m^2_{C}} (-i C_{C \bar{b}\bar{B'}}).
\end{equation}
These atoms share a common tree-level diagram ($X_3$) and differ by a pair of flipped tree-level diagrams ($X_4$ and $X_4^{(\text{flipped})}$) in which a particle propagating in the $t$-channel and a particle propagating in the $u$-channel are simultaneously on-shell. A fundamental property, necessary for the absence of inelastic processes at  tree level, is that the singular parts 
of flipped diagrams contributing to two-to-two inelastic processes sum to zero.
The fact that this does hold follows from the area rule~\eqref{Connection_among_three_point_couplings_and_areas} together with certain properties of structure constants and can be easily proven, as was done for example in~\cite{Davide_Patrick_tree_level_paper,Braden:1991vz}.
For this reason, we have
$$
X_4+X^{(\text{flipped})}_4= \text{finite}
$$
and the sum of the two blue atoms does not contribute to the second order singularity.

The effect is that only two among the six  atoms in figure \ref{allowed_one_loop_box_network_geometry_cuts_with_different_colours_on_shell_description} contribute to the final result; they are the two black graphs having values
\begin{equation}
\label{first_surviving_cut_in_second_order_pole_network}
D^{(1)}(B,C')= -32 i \Bigl( \frac{ \beta}{\sqrt{h}} \Bigr)^4 \frac{\Delta^2_{ab}}{(s-s_0)^2} \Delta_{B'C}
\end{equation}
and
\begin{equation}
\label{second_surviving_cut_in_second_order_pole_network}
D^{(3)}(B',C)= -32 i \Bigl( \frac{ \beta}{\sqrt{h}} \Bigr)^4 \frac{\Delta^2_{ab}}{(s-s_0)^2} \Delta_{C'B}.
\end{equation}
Noting that $\Delta_{B' C}+\Delta_{B C'}=\Delta_{ab}$\footnote{It can be checked that, given a point inside a parallelogram and its corresponding subdivision into four triangles, the sum of the areas of the two triangles on the opposite sides of the point is always equal to half the area of the whole parallelogram.} we obtain
\begin{equation}
\label{sum_of_surviving_cuts_in_second_order_pole_network}
D^{(1)}(B,C')+D^{(3)}(B',C)= -32 i \Bigl( \frac{ \beta}{\sqrt{h}} \Bigr)^4 \frac{\Delta^3_{ab}}{(s-s_0)^2}.
\end{equation}
Before comparing this result with the residue of the singularity in~\eqref{double_singularities_S_matrix_from_bootstrap} we need to perform the substitution~\eqref{s_as_function_of_the_rapidity} and multiply by the overall factor in~\eqref{overall_multiplicative_factor_coming_from_momentum_conservation}. If we do so we obtain 
$$
\Bigl( \frac{ \beta^2}{2h} \Bigr)^2 \frac{1}{(s-s_0)^2},
$$
which matches perfectly with the leading order part of the second order singularity in the bootstrapped S-matrix. 
It is worth mentioning that the cut decomposition represented in figure \ref{allowed_one_loop_box_network_geometry_cuts_with_different_colours_on_shell_description} is not unique. There is another legitimate choice, which is decomposing $D^{(1)}$ into $D^{(1)}(C,C')+D^{(1)}(C,B')$ and $D^{(3)}$ into $D^{(3)}(B,B')+D^{(3)}(B,C')$. In this case we would have observed atomic cancellations between diagrams (1) and (4), and between diagrams (2) and (3).

This result has been obtained by noting that some atoms in the network differ from each other by the presence of a pair of $4$-point tree-level inelastic diagrams connected by flipping an internal propagator.
Such atoms cancel in the final sum so that only two atoms remain and contain the information necessary to obtain the S-matrix at the pole position. The problem of evaluating all the Feynman diagrams at their threshold singularity reduces therefore to a decomposition problem, in which the only relevant thing is to understand how the loop needs to be cut. Then, without evaluating every single atom, we know which ones cancel in the sum and we reproduce the final result by evaluating only those that survive.

We conclude the section by pointing out that, as noted in \cite{Patrick_thesis}, in some degenerate cases $B$ and $B'$ may be parallel. In these cases, the area $\Delta_{B B'}$ of the triangle associated with the two aligned vectors is equal to zero and diagram $(2)$ (proportional to $\Delta_{BB'}$) has a null residue at the second order singularity. These are situations in which the atoms coloured blue in figure \ref{allowed_one_loop_box_network_geometry_cuts_with_different_colours_on_shell_description} are null and diagram $(2)$ can simply be omitted. A similar situation occurs when $C$ and $C'$ are parallel: in this case, it is diagram $(3)$ to be zero and the atoms coloured red are separately null.
This observation confirms that the atomic decomposition shown in figure~\ref{allowed_one_loop_box_network_geometry_cuts_with_different_colours_on_shell_description} is general, and also includes such degenerate situations. It is also worth noting that even when atoms coming from non-planar graphs are non-null, they are completely cancelled in the final sum. This turns out also to be true for the $3^\text{rd}$-order pole networks considered in the next section; whether it generalises to higher-order networks is at present unknown.

\section{Third-order poles}
\label{Chapter_on_third_order_poles}

We now turn to the origin of third-order poles in perturbation theory. We restrict our analysis to the $s$-channel since poles in the $t$-channel can be obtained by sending $\theta \to i\pi-\theta$ through crossing symmetry.
Therefore, as shown in figure
\ref{brick_towers_close_each_other},
we consider a general S-matrix containing the following product of building blocks: $\{x-1 \}^2 \{x+1\}$. 
In this case, the Laurent expansion of the bootstrapped S-matrix around the pole $\theta=i\theta_0=i \frac{\pi x}{h}$, at the leading order in $\beta$, is
\begin{equation}
\label{third_order_singularities_S_matrix_from_bootstrap}
S^{(l)}_{ab}(\theta) = 1+ i \Bigl(\frac{\beta^2}{2 h}\Bigr) \frac{1}{(\theta- i \theta_0)} +  \Bigl(\frac{\beta^2}{2 h}\Bigr)^2 \frac{1}{(\theta- i \theta_0)^2} + i \Bigl(\frac{\beta^2}{2 h}\Bigr)^3 \frac{1}{(\theta- i \theta_0)^3},
\end{equation}
where the relation above is obtained by substituting $N=1$ and $\nu=i$ in \eqref{S_matrix_contributions_on_the_poles_expanded_and_containing_all_the_leading_singularities}. In this section we will reproduce the different terms of~\eqref{third_order_singularities_S_matrix_from_bootstrap} using perturbation theory.
The identity is of course the non-interacting part of the S-matrix while the contribution $i \Bigl(\frac{\beta^2}{2 h}\Bigr) \frac{1}{(\theta- i \theta_0)}$ comes from a tree-level diagram in which an intermediate bound state is propagating in the direct channel. Since a detailed study of tree-level
scattering has already been performed in~\cite{Davide_Patrick_tree_level_paper}, we focus here only on the last two contributions in~\eqref{third_order_singularities_S_matrix_from_bootstrap}, having poles of order two and three respectively.

\subsection{Vertex corrections contributing to third-order singularities}
\begin{figure}
\begin{center}
\includegraphics[width=0.7\linewidth]{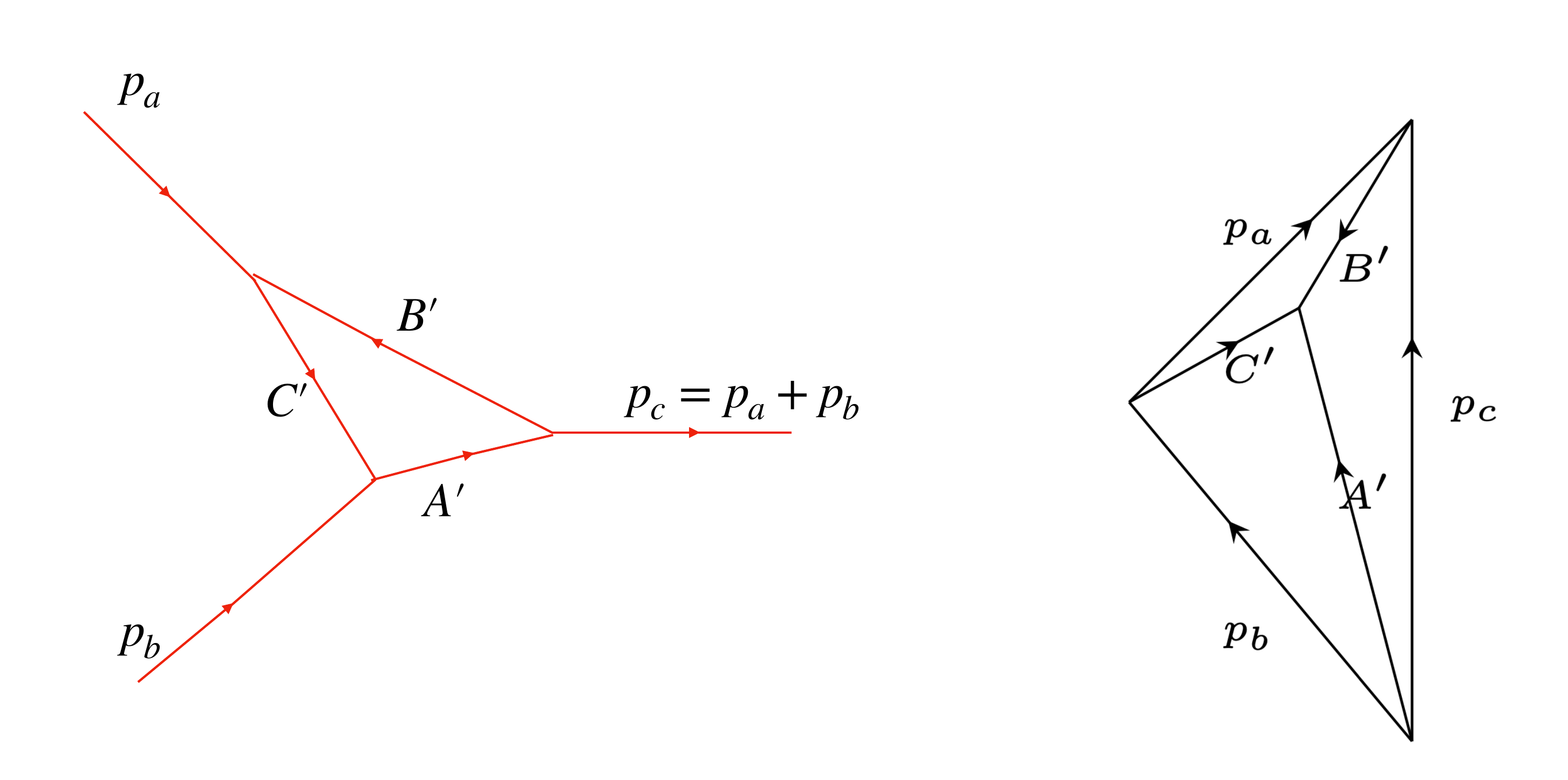}
\end{center}
\caption{Triangle diagram (on the left) and its on-shell representation (on the right).}
\label{triangle_FD_with_geometric_representation}
\end{figure}

Before moving on to the study of the S-matrix we focus on possible singular vertex corrections. These corrections are due to diagrams of the form shown in figure \ref{triangle_FD_with_geometric_representation}, when there exists a value of the loop integration variables $(l_0,l_1)$ at which all the internal propagators are on-shell simultaneously. We focus first on the problem of determining the sign of the product of the $3$-point couplings appearing in the diagram. 
Let us consider the following inelastic process 
\begin{equation}
\label{non_allowed_process_ab_Aprime_Bprime}
    a+b \to A'+ \bar{B}'.
\end{equation}
The process is well reproduced on the RHS of figure \ref{triangle_FD_with_geometric_representation}, where $p_c$ and $C'$ correspond to a copy of particles propagating in the direct and crossed channels in a pair of tree-level Feynman diagrams contributing to the process~\eqref{non_allowed_process_ab_Aprime_Bprime}. Since the two channels are connected by a flip of type III, according to the conventions used in~\cite{Davide_Patrick_tree_level_paper},  we know that the product of the couplings entering the two different channels changes sign after the flip. We have
$$
\sigma_{B'a\bar{C}'} \sigma_{C'b\bar{A}'}=-\sigma_{ab\bar{c}}\sigma_{\bar{A}' c B'} .
$$
Multiplying both the right and the left hand side part of the expression above by $\sigma_{A' \bar{c} \bar{B'} }=\sigma^*_{\bar{A'}c B' }$ we obtain
\begin{equation}
\label{product_of_f_functions_three_point_coupling_number_of_tiles_equal_three}
\sigma_{B'a\bar{C}'} \sigma_{C'b\bar{A}'} \sigma_{A' \bar{c} \bar{B}' }=- \sigma_{ab\bar{c}}.
\end{equation}
This universal formula, already noticed in~\cite{Braden:1990wx},  connects the product of the signs entering a singular vertex correction to the sign of the tree-level vertex. 

Now we derive the pole residue at $(p_a+p_b)^2=m_c^2\equiv s_0$. Referring to figure \ref{triangle_FD_with_geometric_representation} we parametrise the internal loop momenta in the following way
\begin{equation}
\begin{split}
&(A'+l)^2-m_{A'}^2+i \epsilon=2 A' \cdot l + l^2 + i \epsilon \\
&(B'+l)^2-m_{B'}^2+i \epsilon=2 B' \cdot l + l^2 + i \epsilon\\
&(C'+l)^2-m_{C'}^2+i \epsilon=\frac{dC'^2}{ds}\Bigl|_{A', B'} (s-s_0) +2 C' \cdot l + l^2 + i \epsilon\\
\end{split}
\end{equation}
where $A'$ and $B'$ are on-shell at $l=(0,0)$. Using formula~\eqref{general_formula_for_the_pole_integrating_over_tilde_variables} we obtain
\begin{equation}
I_{\triangle'} = \frac{1}{s-s_0} \int \frac{d^2 \tilde{l}}{(2 \pi)^2} \frac{1}{2  A' \cdot \tilde{l} + i \epsilon} \ \frac{1}{2 B' \cdot \tilde{l} + i \epsilon} \ \frac{1}{\frac{d C'^2}{ds}\Bigl|_{A',B'} +2 C' \cdot \tilde{l} + i \epsilon}.
\end{equation}
We adopt the usual choice of integration variables 
\begin{equation}
u = 2 A' \cdot \tilde{l}  \hspace{3mm} , \hspace{3mm}   v = 2  B' \cdot \tilde{l}  .
\end{equation}
The relation in \eqref{writing_D_in_terms_of_A_and_B_general_relation} allows to express $C'$ as a negative linear combination of $A'$ and $B'$ leading to the following expression for the integral
\begin{equation}
I_{\triangle'} = \frac{1}{s-s_0}  \frac{1}{(2 \pi)^2 8 i \Delta_{A'B'}} \int du dv \frac{1}{u+i \epsilon} \ \frac{1}{v+i \epsilon} \ \frac{1}{\frac{d C'^2}{ds}\Bigr|_{A',B'} -\frac{\Delta_{C'B'}}{\Delta_{A'B'}} u -  \frac{\Delta_{C'A'}}{\Delta_{A'B'}} v +i \epsilon} .
\end{equation}
As for the box, the integral can be easily computed by closing both the $u$ and $v$ contours in the lower half plane. Using Cauchy's theorem the result is
\begin{equation}
\label{Vertex_correction_from_derivative_to_triangle_areas}
I_{\triangle'} = \frac{ i}{8  \Delta_{A'B'}}  \frac{1}{\frac{dC'^2}{ds}\Bigr|_{A',B'}} \frac{1}{s-s_0} = \frac{ i}{8 }  \frac{\Delta_{ab} }{\Delta_{B' C'} \Delta_{A' C'}} \frac{1}{s-s_0} .
\end{equation}
The derivative of $C'^2$ respect to $s$ has been taken keeping the lengths of $A'$ and $B'$ fixed at their mass-shell values. The last equality in~\eqref{Vertex_correction_from_derivative_to_triangle_areas} can be checked by using relation \eqref{relation_amonf_squaredistances_of_points_in_a_plane}.
Multiplying~\eqref{Vertex_correction_from_derivative_to_triangle_areas} by the extra vertex and propagator factors, as written in~\eqref{extra_multiplicative_factors_to_be_added_in_the_end}, we obtain
\begin{equation}
\label{singular_term_vertex_correction_third_order_poles_all_factors_added}
\vv_{\triangle'} = \frac{ i}{8 }  \frac{\Delta_{ab} \Delta_{A' B'}}{s-s_0} \Bigl( \frac{4 \beta}{\sqrt{h}} \Bigl)^3 \sigma_{A' \bar{c} \bar{B}'} \sigma_{B'a \bar{C}'} \sigma_{C'b \bar{A}'}\sim - 2   \Bigl(\frac{\beta}{\sqrt{h}} \Bigl)^3 \sigma_{ab\bar{c}}   \frac{\Delta_{A' B'}}{\theta-i\theta_0}
\end{equation}
where in the second equality  we used \eqref{product_of_f_functions_three_point_coupling_number_of_tiles_equal_three} and we wrote $s-s_0$ in terms of $\theta$, the difference between the rapidities of the $a$- and $b$-particles. 

\begin{figure}
\medskip
\begin{center}
\begin{tikzpicture}
\tikzmath{\y=1;}

\draw[directed] (5.8*\y-4*\y+15*\y,0*\y+3.5*\y) -- (7.3*\y-4*\y+15*\y,1.5*\y+3.5*\y);
\draw[directed] (5.8*\y-4*\y+15*\y,0*\y+3.5*\y) -- (6.7*\y-4*\y+15*\y,0.5*\y+3.5*\y);
\draw[directed] (7.3*\y-4*\y+15*\y,1.5*\y+3.5*\y) -- (6.7*\y-4*\y+15*\y,0.5*\y+3.5*\y);
\draw[directed] (7.3*\y-4*\y+15*\y,-1.8*\y+3.5*\y) -- (5.8*\y-4*\y+15*\y,0*\y+3.5*\y);
\draw[directed] (7.3*\y-4*\y+15*\y,-1.8*\y+3.5*\y) -- (6.7*\y-4*\y+15*\y,0.5*\y+3.5*\y);
\draw[directed] (7.3*\y-4*\y+15*\y,-1.8*\y+3.5*\y) -- (7.3*\y-4*\y+15*\y,1.5*\y+3.5*\y);

\filldraw[black] (17*\y,3.6*\y)  node[anchor=west] {\tiny{$C'$}};
\filldraw[black] (17.75*\y,4.3*\y)  node[anchor=west] {\tiny{$B'$}};
\filldraw[black] (17.75*\y,3.35*\y)  node[anchor=west] {\tiny{$A'$}};
\filldraw[black] (17*\y,4.5*\y)  node[anchor=west] {\tiny{$p_a$}};
\filldraw[black] (17*\y,2.5*\y)  node[anchor=west] {\tiny{$p_b$}};
\filldraw[black] (18.3*\y,3.5*\y)  node[anchor=west] {\tiny{$p_c$}};

\filldraw[black] (7.1*\y+10.5*\y,1.3*\y)  node[anchor=west] {\tiny{$D_{\Delta'}$}};

\draw[directed] (5.8*\y+15*\y,0*\y+3.5*\y) -- (7.3*\y+15*\y,1.5*\y+3.5*\y);                  
\draw[directed] (7.3*\y+15*\y,-1.8*\y+3.5*\y) -- (5.8*\y+15*\y,0*\y+3.5*\y);
\draw[directed]  (7.3*\y+15*\y,-1.8*\y+3.5*\y) -- (6.8*\y+15*\y,-0.1*\y+3.5*\y);
\draw[directed] (5.8*\y+15*\y,0*\y+3.5*\y) -- (6.8*\y+15*\y,-0.1*\y+3.5*\y);
\draw[directed] (7.3*\y+15*\y,1.5*\y+3.5*\y) -- (6.8*\y+15*\y,-0.1*\y+3.5*\y);
\draw[directed] (7.3*\y+15*\y,-1.8*\y+3.5*\y) -- (7.3*\y+15*\y,1.5*\y+3.5*\y);

\filldraw[black] (6.2*\y+15*\y,-0.2*\y+3.5*\y)  node[anchor=west] {\tiny{$C$}};
\filldraw[black] (21.6*\y,3.9*\y)  node[anchor=west] {\tiny{$B$}};
\filldraw[black] (21.65*\y,2.85*\y)  node[anchor=west] {\tiny{$A$}};
\filldraw[black] (21*\y,4.5*\y)  node[anchor=west] {\tiny{$p_a$}};
\filldraw[black] (21*\y,2.5*\y)  node[anchor=west] {\tiny{$p_b$}};
\filldraw[black] (22.3*\y,3.5*\y)  node[anchor=west] {\tiny{$p_c$}};

\filldraw[black] (7.1*\y+14.5*\y,1.3*\y)  node[anchor=west] {\tiny{$D_{\Delta}$}};

\draw[directed] (5.8*\y+15*\y+4*\y,0*\y+3.5*\y) -- (7.3*\y+15*\y+4*\y,1.5*\y+3.5*\y);
\draw[directed] (7.3*\y+15*\y+4*\y,-1.8*\y+3.5*\y) -- (5.8*\y+15*\y+4*\y,0*\y+3.5*\y);
\draw[directed] (7.3*\y+15*\y+4*\y,-1.8*\y+3.5*\y) -- (6.9*\y+15*\y+4*\y,-0.7*\y+3.5*\y);
\draw[directed] (7.3*\y+15*\y+4*\y,1.5*\y+3.5*\y) -- (6.9*\y+15*\y+4*\y,-0.7*\y+3.5*\y);
\draw[directed] (5.8*\y+15*\y+4*\y,0*\y+3.5*\y) -- (6.9*\y+15*\y+4*\y,-0.7*\y+3.5*\y);
\draw[directed] (7.3*\y+15*\y+4*\y,-1.8*\y+3.5*\y) -- (7.3*\y+15*\y+4*\y,1.5*\y+3.5*\y);

\filldraw[black] (6.2*\y+15*\y+4*\y,-0.2*\y+3.5*\y)  node[anchor=west] {\tiny{$C''$}};
\filldraw[black] (25.6*\y,3.9*\y)  node[anchor=west] {\tiny{$B''$}};
\filldraw[black] (25.65*\y,2.4*\y)  node[anchor=west] {\tiny{$A''$}};
\filldraw[black] (25*\y,4.5*\y)  node[anchor=west] {\tiny{$p_a$}};
\filldraw[black] (25*\y,2.5*\y)  node[anchor=west] {\tiny{$p_b$}};
\filldraw[black] (26.3*\y,3.5*\y)  node[anchor=west] {\tiny{$p_c$}};

\filldraw[black] (7.1*\y+15*\y+4*\y,1.3*\y)  node[anchor=west] {\tiny{$D_{\Delta''}$}};

\end{tikzpicture}
\end{center}
\caption{Three types of vertex corrections contributing to a third-order pole.}
\label{3_kind_of_triangular_diagrams_in_allowed_processes_contributing_to_3rd_order_poles}
\end{figure}
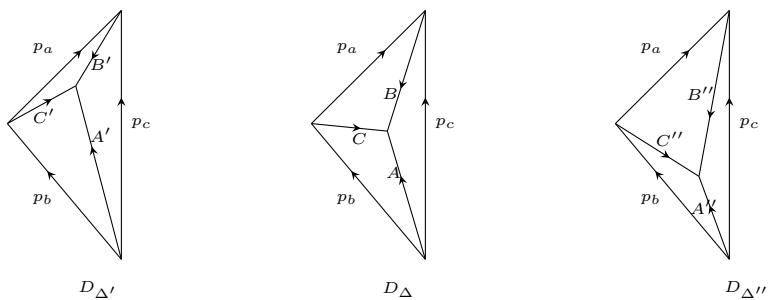

In all the cases analysed we find exactly three vertex corrections entering Feynman diagrams contributing to third-order poles. This fact, already noted in \cite{Braden:1990wx}, still requires an explanation in terms of root system properties. The three different vertex corrections are shown in figure \ref{3_kind_of_triangular_diagrams_in_allowed_processes_contributing_to_3rd_order_poles} following an increasing order with respect to the angle formed between $p_a$ and the internal propagator $C'$, $C$ and $C''$.
Their constituents always satisfy the following universal relations \cite{Braden:1990wx}
\begin{equation}
\label{sum_areas_three_point_couplings_third_orderd_pole_vertex_corrections}
\begin{split}
&\Delta_{B' C'}+\Delta_{B C}+\Delta_{B'' C''}=\Delta_{ab}\\
&\Delta_{A' C'}+\Delta_{A C}+\Delta_{A'' C''}=\Delta_{ab}\\
&\Delta_{A' B'}+\Delta_{A B}+\Delta_{A'' B''}=\Delta_{ab}.
\end{split}
\end{equation}
A general proof of these
identities is also still missing, though we expect they follow from universal properties of the Coxeter geometry associated with root systems. 
Summing the three vertex corrections and combining \eqref{singular_term_vertex_correction_third_order_poles_all_factors_added} with the third line in \eqref{sum_areas_three_point_couplings_third_orderd_pole_vertex_corrections} we obtain a universal formula for the leading order singularity of the vertex contributing to a third-order pole in the S-matrix. It is given by
\begin{equation}
\label{sum_of_the_three_types_of_vertex_corrections}
\vv_{\triangle'}+\vv_\triangle+\vv_{\triangle''} =
- 2  \Bigl(\frac{\beta}{\sqrt{h}} \Bigl)^3 
  \sigma_{ab\bar{c}} \frac{ \Delta_{ab}}{\theta-i \theta_0}= -\frac{ \beta^2}{2h}    \frac{C_{ab \bar{c}}}{\theta-i \theta_0}  .
\end{equation}
In the next sections, we will use the value of the vertex correction to compute the leading order expansion of the S-matrix around the pole position. 

\subsection{One-loop contributions}
\label{Computation_of_one_loop_S_matrix_around_3rd_order_pole}

\begin{figure}
\begin{center}
\includegraphics[width=1\linewidth]{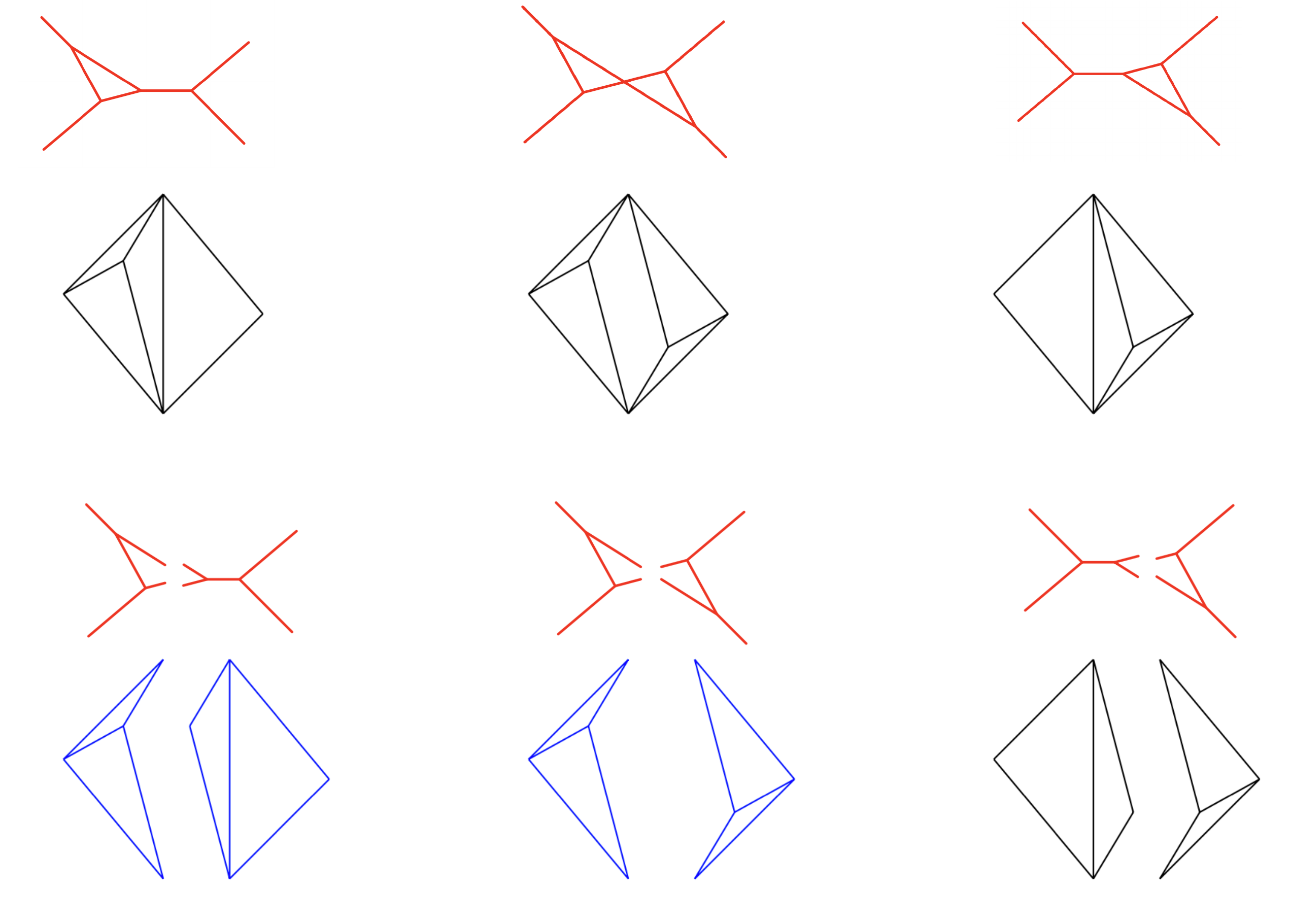}
\end{center}
\caption{Part of the network of one-loop on-shell diagrams contributing to the second order pole in~\eqref{third_order_singularities_S_matrix_from_bootstrap}. In the second line cut diagrams are shown. Diagrams coloured blue cancel in the sum since they differ by a pair of flipped $4$-point tree-level graphs.}
\label{One_loop_third_order_pole_diagram_from_vertex_corrections_partial_tilings}
\end{figure}

We compute one loop contributions first; these contributions are responsible for the coefficient of order $(\theta-i\theta_0)^{-2}$ in~\eqref{third_order_singularities_S_matrix_from_bootstrap}. 
To each of the vertex corrections $\vv_{\triangle'}$, $\vv_{\triangle}$ and $\vv_{\triangle''}$ three singular Feynman diagrams are associated, connected by flipping internal propagators. In figure \ref{One_loop_third_order_pole_diagram_from_vertex_corrections_partial_tilings} we show the three diagrams associated with the vertex correction $\vv_{\triangle'}$. We start considering a one-particle reducible Feynman diagram having on the LHS a vertex correction of type $\vv_{\triangle'}$, and in which a bound state is propagating on its on-shell value. Both $\vv_{\triangle'}$ and the propagating bound state carries a pole of order one so that their product generates an order two singularity, matching the order $\frac{1}{(\theta-\theta_0)^2}$ we intend to analyse. Among the three different diagrams,  two cancel in the sum (the blue ones in figure \ref{One_loop_third_order_pole_diagram_from_vertex_corrections_partial_tilings}). Indeed when we compute the loop integral they split into tree-level graphs two of which differ by flipping one internal propagator. 
The sum of the three graphs in figure \ref{One_loop_third_order_pole_diagram_from_vertex_corrections_partial_tilings} is therefore simply given by a single one-particle reducible graph presenting in order a $3$-point vertex $(-i C_{ab \bar{c}})$, an on-shell bound state propagator $\frac{i}{s-s_0}$ and the one loop vertex correction $\vv_{\triangle'}$. 
Since we need to repeat the same analysis for Feynman diagrams obtained by flipping one-particle reducible graphs containing $\vv_{\triangle}$ and $\vv_{\triangle''}$ the final result for the second-order S-matrix term, after adding the flux factor~\eqref{overall_multiplicative_factor_coming_from_momentum_conservation}, is given by
$$
\frac{1}{8i \Delta_{ab}}(-i C_{ab\bar{c}}) \frac{i}{s-s_0} (\vv_{\triangle'}+\vv_{\triangle}+\vv_{\triangle''}).
$$
If we use \eqref{sum_of_the_three_types_of_vertex_corrections} and we express the pole in terms of the rapidity difference we obtain 
\begin{equation}
\label{one_loop_3rd_order_pole_S_matrix_perturbation_theory}
\frac{1}{32  \Delta^2_{ab}} \frac{ \beta^2}{2h}    \frac{|C_{ab \bar{c}}|^2}{(\theta-i \theta_0)^2}= \Bigl(\frac{ \beta^2}{2h} \Bigr)^2  \frac{1}{(\theta-i \theta_0)^2}.
\end{equation}
This exactly matches the bootstrapped formula~\eqref{third_order_singularities_S_matrix_from_bootstrap} at  order $(\theta-i\theta_0)^{-2}$.

\subsection{Two-loop contributions}

The $3^{\text{rd}}$-order singularity in~\eqref{third_order_singularities_S_matrix_from_bootstrap} comes from a two-loop computation. The complete network of Feynman diagrams, together with their values, contributing to such a result was found in \cite{Braden:1990wx}. However, in that paper, the authors limited themselves to giving the final answer and did not identify the simplification between diagrams composing the network. In this section, we show how the result is generated by summing suitable atoms located at the boundary of the network while the internal part simplifies. Using the sign rules for $3$-point couplings given in~\cite{Davide_Patrick_tree_level_paper} we also explain how to obtain the correct sign of each Feynman diagram composing the network.

\begin{figure}
\includegraphics[width=1\linewidth]{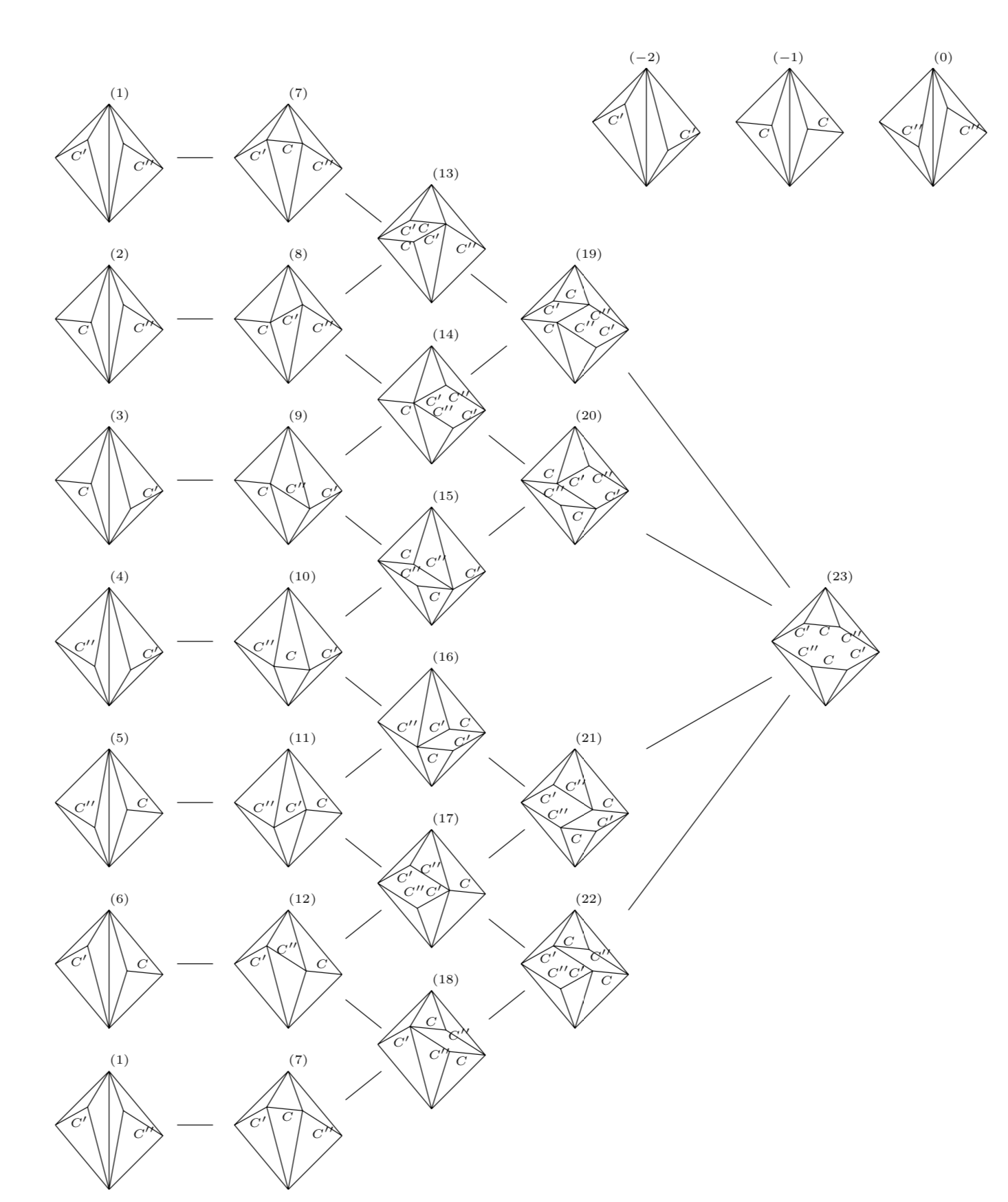}
\caption{Network of Feynman diagrams contributing to the $3^{\text{rd}}$-order pole in~\eqref{third_order_singularities_S_matrix_from_bootstrap}.}
\label{Network_of_onshell_diagrams_contributing_to_the_third_order_poles}
\end{figure}

The full network of singular diagrams is 
shown in figure~\ref{Network_of_onshell_diagrams_contributing_to_the_third_order_poles}, taken from~\cite{Patrick_thesis}.
The network has the topology of a disk. Indeed after six vertical steps along the first column, the diagrams repeat periodically. The first two columns, together with the three isolated diagrams $(-2)$, $(-1)$ and $(0)$ are planar. Diagrams $(13), (14),  \dots , (23)$ in the centre of the disk
are instead non-planar and gaps are added in their tilings to represent propagators that cross. Moreover, graphs in the network connected by a line are related by a flipping move.
Again following the convention of~\cite{Braden:1990wx,Davide_Patrick_tree_level_paper}, the first and second column are connected by a flip of type I, while in all the other situations (both moving from the second to the third column, from the third to the fourth column and from the fourth column to the central diagram $(23)$) the flip entering in play is of type II.

Among the different graphs,  the one-particle reducible ones are particularly simple to compute. These diagrams are obtained by gluing vertex corrections on the opposite sides of an on-shell bound state propagator. They are reported in the first column, containing diagrams $(1) \dots (6)$, and in the isolated sector of the network, containing diagrams $(-2)$, $(-1)$ and $(0)$.   In total there are $9$ one-particle reducible diagrams of this type, $3 \times 3$ since there are three types of vertex corrections (those reported in figure \ref{3_kind_of_triangular_diagrams_in_allowed_processes_contributing_to_3rd_order_poles}).  
Summing all the one-particle reducible two-loop diagrams and multiplying by the flux factor~\eqref{overall_multiplicative_factor_coming_from_momentum_conservation}, on the pole we obtain
\begin{equation}
\label{one_particle_reducible_sum_two_loops_third_order_pole}
\frac{1}{8 i \Delta_{ab}}(\vv_{\triangle'}+\vv_{\triangle}+\vv_{\triangle''}) \frac{i}{s-s_0} (\vv_{\triangle'}+\vv_{\triangle}+\vv_{\triangle''})=-i \Bigl(\frac{ \beta^2}{2h}\Bigr)^3  \frac{1}{(\theta-i\theta_0)^3},
\end{equation} 
where in the last equality we used relations \eqref{s_as_function_of_the_rapidity}, \eqref{sum_of_the_three_types_of_vertex_corrections} together with the area rule \eqref{Connection_among_three_point_couplings_and_areas}.
The result in \eqref{one_particle_reducible_sum_two_loops_third_order_pole} is universal and does not depend on the simply-laced Toda model studied. However, as remarked in \cite{Braden:1990wx}, it carries a minus sign compared to the third-order pole in~\eqref{third_order_singularities_S_matrix_from_bootstrap}. We will show now how the remaining diagrams $(7)$, $(8)$, \dots , $(23)$ conspire to reproduce exactly $2 \times i \Bigl(\frac{ \beta^2}{2h}\Bigr)^3  \frac{1}{(\theta-i\theta_0)^3}$ in such a way that summing this new term to \eqref{one_particle_reducible_sum_two_loops_third_order_pole} we exactly obtain the expected $3^{\text{rd}}$-order singularity of~\eqref{third_order_singularities_S_matrix_from_bootstrap}.

\begin{figure}
\begin{center}
\includegraphics[width=0.7\linewidth]{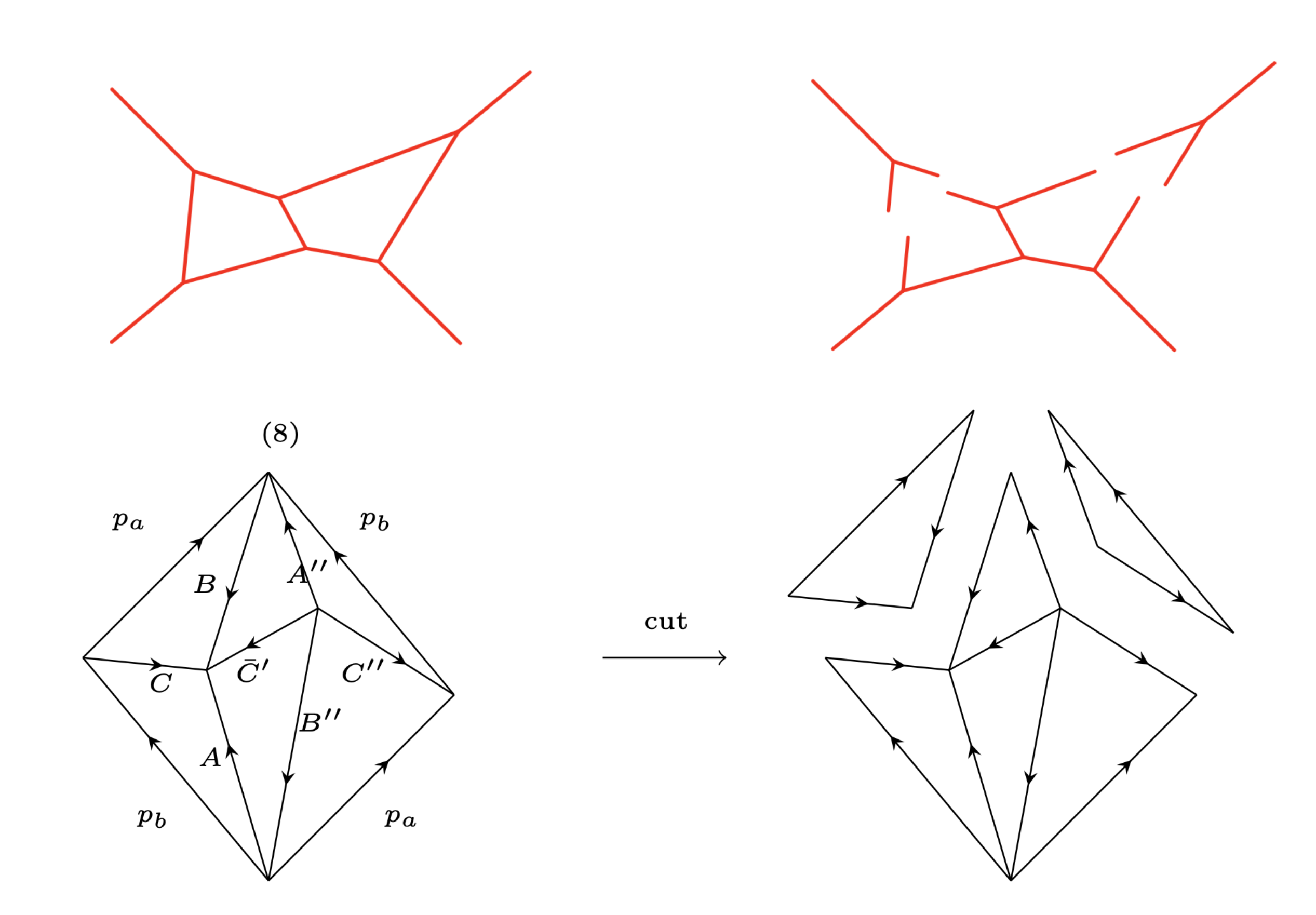}
\end{center}
\caption{Example of double box cut contributing to the final result.}
\label{double_box_contribution_with_relevant_cut}
\end{figure}

The contributions necessary to restore the bootstrapped result come from particular decompositions of diagrams $(7), (8), \ldots, (12)$ into atoms. In the following, we first explain the required cuts, and then show why these cuts are the correct ones.  The needed cuts are obtained by removing respectively the top left and top right vertices or the bottom left and bottom right vertices from diagrams $(7), (8), \ldots, (12)$. The important thing is that once we choose what vertices to remove, those sitting at the top or the bottom of the graph, we need to maintain the same choice for all the diagrams in the second column of the network in figure \ref{Network_of_onshell_diagrams_contributing_to_the_third_order_poles}. We decide to locate the cuts at the top of the diagrams, as shown in figure \ref{double_box_contribution_with_relevant_cut} where graph number $(8)$ is considered. 
The cut is associated with a particular residue that is picked up when we compute the integral and its result, omitting the additional factors in~\eqref{extra_multiplicative_factors_to_be_added_in_the_end}, is
\begin{multline}
    \label{double_box_relevant_cut_associated_to_the_diagram_D_8}
 I^{(8)} \rightsquigarrow \frac{i}{8  \Delta_{BC}} \frac{i}{8  \Delta_{A'' C''}} \frac{1}{\frac{d A^2 }{ds}} \frac{1}{\frac{d B''^2 }{ds}}\frac{1}{\frac{d C'^2 }{ds}}\frac{1}{(s-s_0)^3}\\
 =\frac{i}{8  \Delta_{BC}} \frac{i}{8  \Delta_{A'' C''}}  \frac{\Delta_{ab}}{\Delta_{AC}} \frac{\Delta_{ab}}{\Delta_{B''C''}} \Bigl(\frac{-\Delta_{ab}}{\Delta_{BA''}}\Bigr) \frac{1}{(s-s_0)^3}.
\end{multline}
A curly arrow has been used in~\eqref{double_box_relevant_cut_associated_to_the_diagram_D_8} instead of an equality to indicate that we are neglecting atoms arising from the computation of $I^{(8)}$ that, as will be shown below, cancel after summing over all Feynman diagrams of the network. The two factors $\frac{1}{8  \Delta_{BC}}$ and  $\frac{1}{8  \Delta_{A'' C''}}$, come from the Jacobians of the choice of variables adopted to perform the integration: in our case they are $2 B \cdot l=u_l$ and $2 C \cdot l=v_l$ for the box on the LHS part of the diagram and $2 A'' \cdot k=u_k$ and $2 C'' \cdot k=v_k$ for the box on the RHS. 

Then there are the propagators that have not been cut, and close to the pole are given by 
$\frac{1}{A^2-m^2_A}\sim\frac{1}{\frac{d A^2 }{ds} (s-s_0)}$, 
$\frac{1}{B''^2-m^2_{B''}}\sim\frac{1}{\frac{d B''^2 }{ds} (s-s_0)}$, 
$\frac{1}{C'^2-m^2_{C'}}\sim\frac{1}{\frac{d C'^2 }{ds} (s-s_0)}$. The derivatives in the first line of~\eqref{double_box_relevant_cut_associated_to_the_diagram_D_8} are performed keeping fixed the lengths of the momenta associated to the cut propagators and their values are explicitly written on the second line  \eqref{double_box_relevant_cut_associated_to_the_diagram_D_8}.
At this point we still need to multiply the expression in \eqref{double_box_relevant_cut_associated_to_the_diagram_D_8} by the remaining quantities in \eqref{extra_multiplicative_factors_to_be_added_in_the_end}, and express the pole in terms of the rapidity. To this end we start by computing the product of the signs $\sigma$ entering the $3$-point vertices of diagrams $(7) - (12)$.

The flip used to move from the double-vertex graphs $(1), \ldots , (6)$ in the first column to the double-box diagrams $(7) - (12)$ in the second column of figure \ref{Network_of_onshell_diagrams_contributing_to_the_third_order_poles} is of type I and, as discussed  in~\cite{Davide_Patrick_tree_level_paper}, does not change the sign of the product of the $3$-point couplings.
Therefore, if we use relation \eqref{product_of_f_functions_three_point_coupling_number_of_tiles_equal_three}, the product of the signs entering the different graphs in the second column of the network is always given by
\begin{equation}
\prod \sigma_{ijk}=(-\sigma_{ab\bar{c}}) \times (-\sigma_{\bar{a} \bar{b}c} )= 1 .
\end{equation}
Using this fact and multiplying the expression \eqref{double_box_relevant_cut_associated_to_the_diagram_D_8} by the remaining terms in \eqref{extra_multiplicative_factors_to_be_added_in_the_end} the cut  of diagram $(8)$ can be written in terms of the rapidity difference as
\begin{equation}
\label{Main_cut_of_diagram_number_8}
D^{(8)} \rightsquigarrow - \Bigl( \frac{ \beta}{\sqrt{h}} \Bigr)^6  \frac{\Delta_{A B''}}{(\theta-i \theta_0)^3}.
\end{equation}
The result for this atom is proportional to the area $\Delta_{A B''}$ of the triangle in the central bottom part of the graph.
At this point, we note that diagram  $(8)$ admits an equivalent diagram,  $(11)$, which is diagram $(8)$ rotated by $\pi$. If we decide to cut such a diagram in the same way, i.e. we remove the top left and the top right vertices, the associated atom is 
\begin{equation}
    \label{Main_cut_of_diagram_number_11}
D^{(11)} \rightsquigarrow - \Bigl( \frac{ \beta}{\sqrt{h}} \Bigr)^6   \frac{\Delta_{B A''}}{(\theta-i \theta_0)^3}.
\end{equation}
Noting that $\Delta_{A B''}+\Delta_{B A''}=\Delta_{AB}+\Delta_{A'' B''}$ the sum of~\eqref{Main_cut_of_diagram_number_8} and~\eqref{Main_cut_of_diagram_number_11} is
$$
D^{(8)}+D^{(11)} \rightsquigarrow - \Bigl( \frac{ \beta}{\sqrt{h}} \Bigr)^6  \frac{1}{(\theta-i \theta_0)^3} (\Delta_{AB}+\Delta_{A''B''}).
$$
The same analysis can be repeated for the copy of diagrams $(7)$ and $(10)$, which again are related by a rotation by $\pi$. In this case, we obtain that the sum is proportional to $\Delta_{A'B'}+\Delta_{A''B''}$. Finally, summing the results obtained by cutting the $3$-point vertices on the top of diagrams $(9)$ and $(12)$ generates $\Delta_{AB}+\Delta_{A'B'}$. Therefore, the sum of the six atoms obtained by cutting the diagrams in the second column of the network in the way just explained returns
\begin{equation}
\label{remaining_piece_to_be_summed_to_one_particle_reducible_diagrams_to_obtain_the_final_answer}
\frac{1}{8 i \Delta_{ab}}\sum_{n=7}^{12}D^{(n)} \rightsquigarrow i \Bigl( \frac{ \beta}{\sqrt{2h}} \Bigr)^6 \frac{2(\Delta_{AB}+\Delta_{A'B'}+\Delta_{A''B''})}{\Delta_{ab}}  \frac{1}{(\theta-i \theta_0)^3}=2 i \Bigl( \frac{ \beta}{\sqrt{2h}} \Bigr)^6  \frac{1}{(\theta-i \theta_0)^3},
\end{equation}
In the expression above we divided by the flux factor~\eqref{overall_multiplicative_factor_coming_from_momentum_conservation} and we used the third identity in \eqref{sum_areas_three_point_couplings_third_orderd_pole_vertex_corrections}.
Both formulas \eqref{one_particle_reducible_sum_two_loops_third_order_pole} and \eqref{remaining_piece_to_be_summed_to_one_particle_reducible_diagrams_to_obtain_the_final_answer} are universal: they do not depend by the particular simply-laced Toda theory considered and their sum returns the correct bootstrapped result. Two questions arise at this point. First of all, we did not explain why the atoms that we considered for the double-box integrals reproduce the right answer. Secondly, we still need to explain why we are not considering the non-planar graphs in the centre of the network in figure~\ref{Network_of_onshell_diagrams_contributing_to_the_third_order_poles}. In the remaining part of this section, we will answer these questions.

\begin{figure}
\includegraphics[width=1\linewidth]{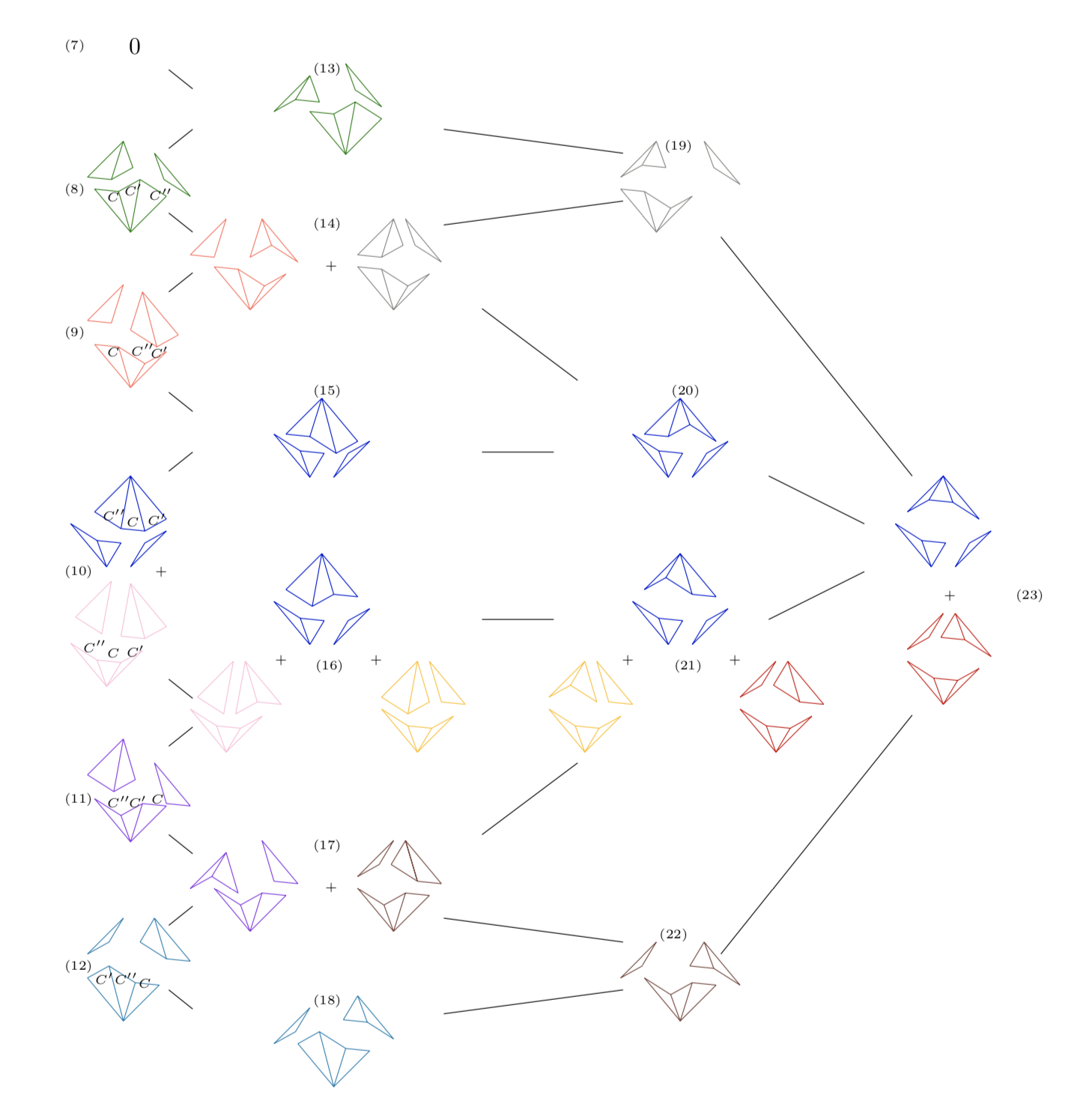}
\caption{Cancellation between atoms in centre of the network. Atoms of the same colour differ by tree-level diagrams contributing to inelastic processes and sum to zero due to the tree-level integrability of the model.}
\label{Bulk_contributions_which_cancel_in_the_third_order_pole_network}
\end{figure}

Let us consider the computation of diagram $(8)$ in more detail. 
Looking at figure \ref{double_box_contribution_with_relevant_cut} we see that such a Feynman diagram is composed of two boxes: one on the RHS containing momenta $\bar{C}'$, $B''$, $C''$ and $A''$ and one on the LHS containing momenta $C$, $A$, $\bar{C}'$ and $B$.  
In the on-shell diagram in figure~\ref{double_box_contribution_with_relevant_cut} we have reversed the direction of the $C'$-vector compared to the vertex correction in figure~\ref{3_kind_of_triangular_diagrams_in_allowed_processes_contributing_to_3rd_order_poles} to keep the same flow of the propagators in each  of the two one-loop boxes. This is consistent if at the same time we change the particle label $C'$ to its antiparticle $\bar{C}'$ as we did in figure~\ref{double_box_contribution_with_relevant_cut}. In other words, we can see the $C'$-propagator as a particle propagating from left to right, or as an antiparticle propagating from right to left. We use this second convention.
\begin{figure}
\includegraphics[width=1\linewidth]{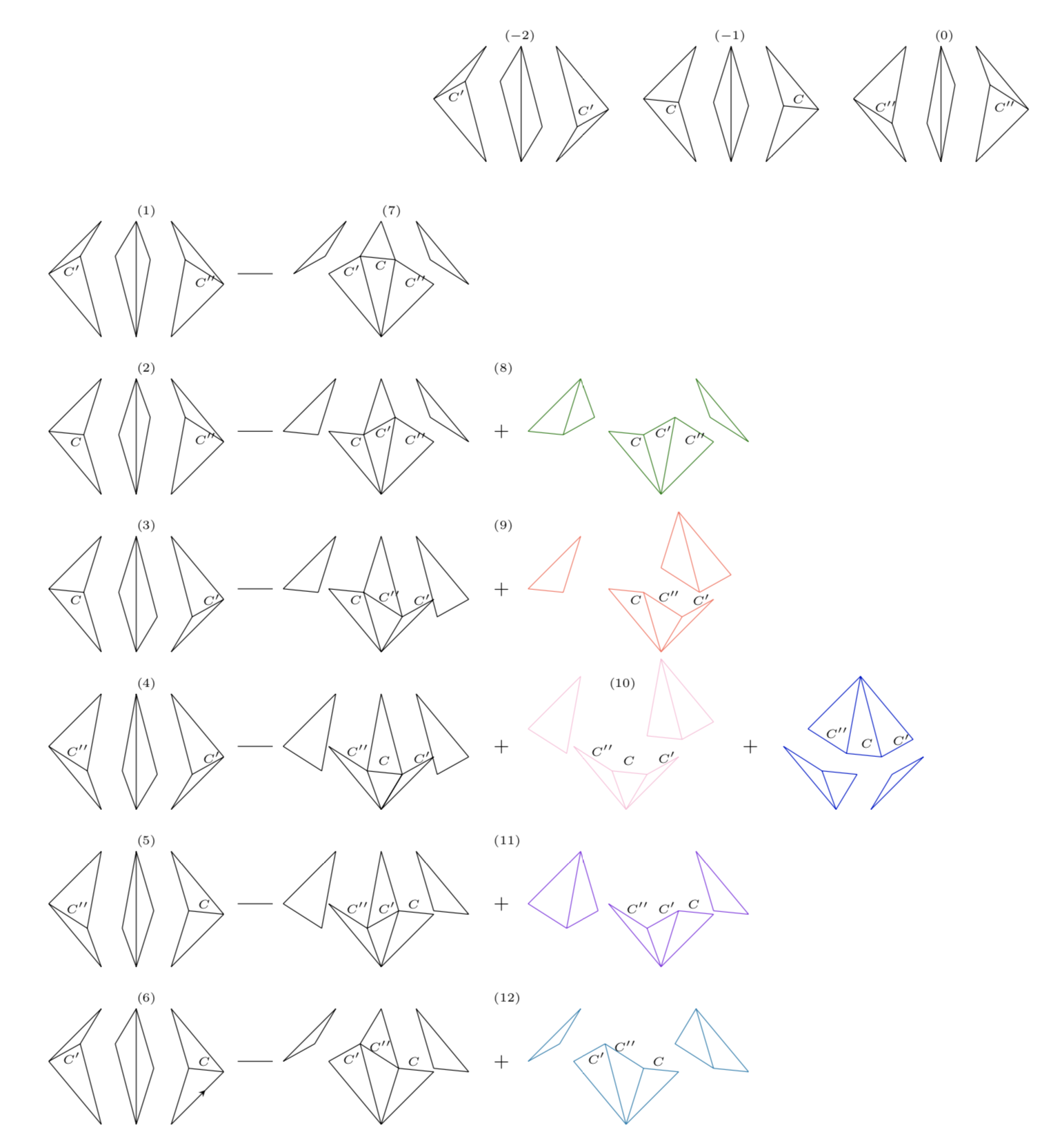}
\caption{Atoms contributing to the third-order pole coefficient of the S-matrix (coloured black) and additional terms (having different colours) which cancel against the internal part of the network. This internal part is depicted in figure \ref{Bulk_contributions_which_cancel_in_the_third_order_pole_network} below.}
\label{Boundary_terms_contributing_to_the_third_order_pole_and_canceling_cuts}
\end{figure}
Each momentum carries then a loop integration variable, say $l$ for the momenta flowing in the LHS box and $k$ for the momenta flowing in the RHS one, with  $\bar{C}'$ containing both $l$ and $k$ since it belongs to both the boxes. If all the LHS momenta are translated by $l$ with propagators $\frac{1}{(C+l)^2-m^2_{C}+i\epsilon}$, $\frac{1}{(A+l)^2-m^2_{A}+i\epsilon}$, $\frac{1}{(B+l)^2-m^2_{B}+i\epsilon}$ and similarly the RHS propagators are translated by $k$, then $\bar{C}'$ is translated by $l+k$.  
The double-box integral can then be performed by computing the single-box integrals one at a time, first the one on the RHS and then the one on the LHS of the graph. If we integrate the loop variable $k$ first, we note that both $B''$ and $\bar{C}'$ can be written as negative linear combinations of $A''$ and $C''$, so the box on the RHS can be broken by cutting the propagators associated to the particles $A''$ and $C''$ which generate a single term as explained in the previous section. Similarly, on the LHS box, we see that the two vectors with respect to which all the others are negative linear combinations are $A$ and $C$. Indeed both $\bar{C}'$ and $B$ can be written as linear combinations of them with negative coefficients. Integrating over $l$ we can write the sum over residues of the LHS box equivalently as a single cut of $(A,C)$ or as a sum of two cuts $(B,C)$ and $(B,C')$ exactly as we did in expression \eqref{some_different_cuts_D(X,Y)_of_the_box_integral}. In this case, we obtain
\begin{equation}
D^{(8)}=D^{(8)}(A,C,A'',C'')=D^{(8)}(B,C,A'',C'')+D^{(8)}(C',C,A'',C'') \,.
\end{equation}
In the first equality, we have written the complete third-order singularity of diagram $(8)$. It corresponds to a single atom in which the propagators $A$, $C$, $A''$ and $C''$ have been cut. However, as in the one-loop case, it is possible to write the result as a sum over different atoms, corresponding to different choices of residues when we use Cauchy's theorem. In this respect 
$D^{(8)}(B,C,A'',C'')$ returns the desired atom contributing to the bootstrapped result while the remaining atom $D^{(8)}(C',C,A'',C'')$ is a piece that will be cancelled by the bulk contributions of the network.
Figure~\ref{Boundary_terms_contributing_to_the_third_order_pole_and_canceling_cuts} shows all the atoms associated with the boundary of the network; the atom $D^{(8)}(C',C,A'',C'')$ just mentioned corresponds to the green cut box on the figure.
The black pieces are the atoms contributing to the final result expected from the bootstrapped S-matrix, while the coloured atoms turn out to cancel with atoms coming from the interior (or bulk) of the network. This cancellation mechanism is shown in 
figure~\ref{Bulk_contributions_which_cancel_in_the_third_order_pole_network}. 
Apart from diagram  $(7)$, which only contributes to the boundary and generates a zero bulk contribution, all the other diagrams $(8), \ldots, (12)$ are split into multiple pieces, some contributing to the bootstrapped answer, the others cancelling diagrams $(13), \dots, (23)$ in the central part of the network. The cancellation mechanism happens in different sectors, which we depicted with different colours in figure \ref{Bulk_contributions_which_cancel_in_the_third_order_pole_network}. There are in total nine pairs of cut diagrams composed of graphs differing by a single flipped four-point tree-level diagram. If we look for example at the green atoms associated with diagrams $(8)$ and $(13)$ in figure \ref{Bulk_contributions_which_cancel_in_the_third_order_pole_network} we see they are equivalent but for a singular $4$-point tree-level graph that is flipped moving from $(8)$ to $(13)$. Since the sum of such a pair of $4$-point tree-level diagrams is zero, the sum of the two green atoms is null. In addition to these pairs of cancelling atoms, there are six atoms (coloured blue) coming from diagrams $(10), (15), (16), (20), (21), (23)$ which cancel in a way not seen before: they are equal with respect to their $3$- and $4$-point tree-level diagrams but contain different $5$-point tree-level diagrams connected by flipping propagators. All these $5$-point diagrams contribute to the singular part of a $5$-point scattering process at the tree level and therefore they have to sum to zero. 

\begin{figure}
\begin{center}
         \includegraphics[width=1.1\linewidth]{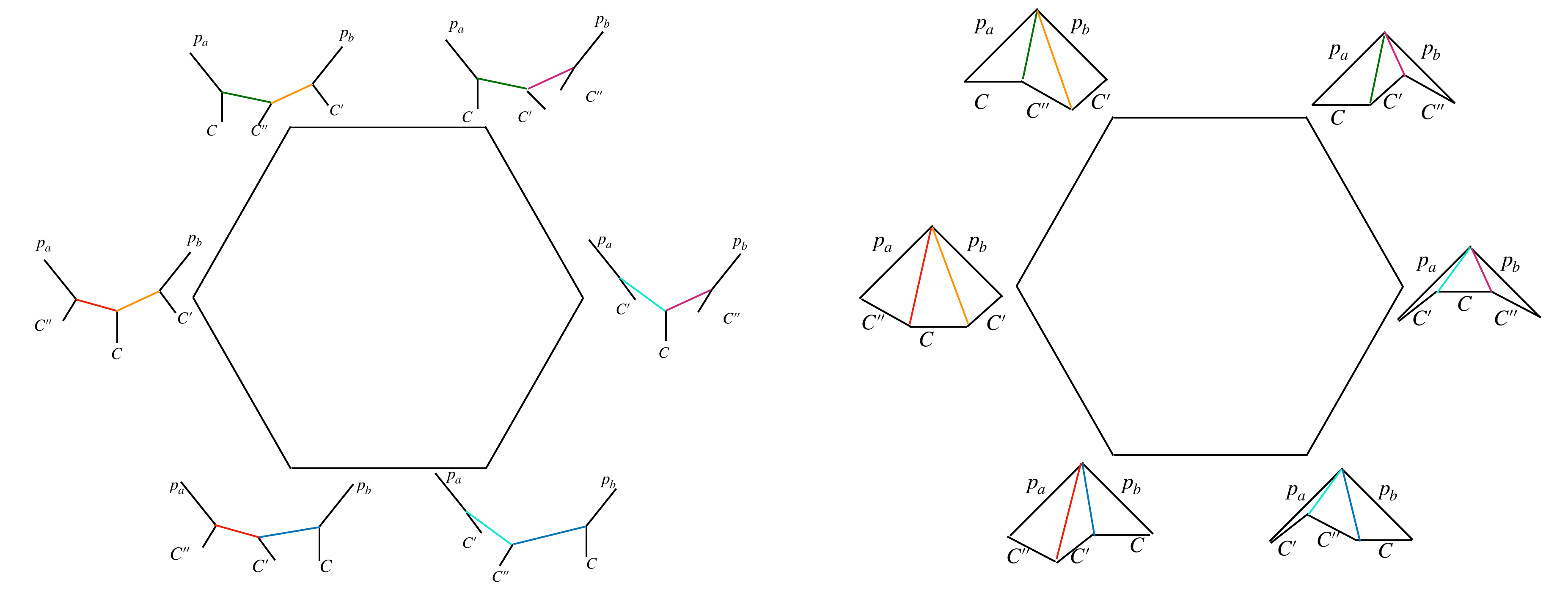}
                 \end{center}
        \caption{The network of simultaneously singular Feynman diagrams contributing to a tree-level $5$-point process (on the left) and their duals (on the right). We use different colours to distinguish different propagators. For the case at hand, $p_a$ and $p_b$ should be identified with the external momenta of the two-loop Landau diagrams under discussion, while $C$, $C'$ and $C''$ are the momenta of the internal loop propagators after the atomic decomposition is realised.}
        \label{fig:5_point_dual_Feyn_diagrams}
\end{figure}

To explain this last point in more detail we follow the discussion in appendix~C of~\cite{Polvara:2022mve} which shows how diagrams contributing to five-point processes can be arranged on the vertices of certain planar polygons, as depicted in figure~\ref{fig:5_point_dual_Feyn_diagrams}, where each side of the polygon corresponds to the flip of a propagator. 
As explained in~\cite{Polvara:2022mve}, for a given tree-level $5$-point diagram we can replace the product of its $3$-point couplings by
\begin{equation}
\prod_{j=1}^3 C^{(3)}_j = \lambda (a_1 - a_2)
\end{equation}
($j$ is a label for the three couplings in the $5$-point diagram) and its two propagators by 
\begin{equation}
P_1= \frac{1}{(s-s_0) (x+a_1)} \hspace{3mm}, \hspace{3mm} P_2= \frac{1}{(s-s_0) (x+a_2)} \,.
\end{equation}
Here the parameter $s=(p_a+p_b)^2$ is the usual Mandelstam variable associated with the momenta of the external particles entering the two-loop Landau diagram and each $s-s_0$ is proportional to the displacement of the square of a momentum flowing in a propagator of an associated tree-level diagram from its on-shell value.
The quantities $(x+a_1)$ and $(x+a_2)$ correspond to the derivatives of the momenta squared of the propagators with respect to $s$, as shown for example in~\eqref{B'_C'_box_integral_expansion_around_s0_example_to_explain_the_method}. If we label this diagram by $D^{(\text{I})}$ then we can write
\begin{equation}
D^{(\text{I})}=\frac{\lambda}{(s-s_0)^2}  \frac{a_1 - a_2}{(x+a_1) (x+a_2)} \,.
\end{equation}
So far nothing special happened: we just decided to parameterise the product of the couplings and the derivatives of the squared of the propagator momenta in terms of $a_1$, $a_2$, $\lambda$ and $x$. 
Once this parameterization is chosen for $D^{(\text{I})}$, a diagram $D^{(\text{II})}$ connected to $D^{(\text{I})}$ by flipping the propagator $P_1$ (and sharing with $D^{(\text{I})}$ the common propagator $P_2$) will be given by
\begin{equation}
D^{(\text{II})}=\frac{\lambda}{(s-s_0)^2}  \frac{a_2 - a_3}{(x+a_2) (x+a_3)} \,.
\end{equation}
The reason for the structure of $D^{(\text{II})}$ is the following: in the expression above we parameterised the new propagator after the flip by
\begin{equation}
P_3= \frac{1}{(s-s_0) (x+a_3)} \,.
\end{equation}
With this parameterization, the product of the $3$-point couplings entering diagram $D^{(\text{II})}$ has to be equal to $\lambda(a_2 - a_3)$. Indeed, if we consider the limit in which $P_2$ diverges more quickly than $P_1$ and $P_3$ (i.e. for $x \sim - a_2$) then the residue of  $D^{(\text{I})}+D^{(\text{II})}$ at $x=a_2$ has to be null: this residue is indeed proportional to the sum of two flipped diagrams contributing to the same two-to-two tree-level inelastic process.
Repeating the same argument for all the diagrams composing the hexagon in figure~\ref{fig:5_point_dual_Feyn_diagrams} we obtain that the tree-level $5$-point amplitude at the leading-order singularity is given by
\begin{equation}
\label{leading_singular_part_5_point_tree_level_process}
\begin{split}
M_5^{(0)} &= \frac{\lambda}{(s-s_0)^2} \Bigl( \frac{a_1 - a_2}{(x+a_1) (x+a_2)}+ \frac{a_2 - a_3}{(x+a_2) (x+a_3)}+ \dots + \frac{a_6 - a_1}{(x+a_6) (x+a_1)} \Bigl)\\
&= \frac{\lambda}{(s-s_0)^2} \Bigl( -\frac{1}{x+a_1}+\frac{1}{x+a_2}-\frac{1}{x+a_2}+\frac{1}{x+a_3}-\dots - \frac{1}{x+a_6}+\frac{1}{x+a_1} \Bigl)=0 \,.
\end{split}
\end{equation}
Diagrams that are connected by a flipping move share a common propagator and the sum of all diagrams vanishes.
We remark that in the sum of the blue atoms in the network in figure~\ref{Bulk_contributions_which_cancel_in_the_third_order_pole_network} $x$ is a fixed number set by the direction we follow when we approach the double pole generated by the five-point diagrams. However, the sum in~\eqref{leading_singular_part_5_point_tree_level_process} is null no matter the value of $x$.

Even though the residue obtained through the Landau analysis is expected from the bootstrap, the way that it comes about is striking. The pole structure of the bootstrapped S-matrix emerges in perturbation theory from an underlying simplification 
which is explained using tree-level properties of the model. 
A subset of the atoms for the original Feynman diagrams yields the expected answer, through contributions entirely located at the boundary of the network. However, how the diagrams need to be cut into atoms to observe such a simplification is still an empirical fact, not always evident from the on-shell structure of the Feynman diagrams. The most difficult situations occur when we try to split the diagram into more than two atoms, as happens for diagrams $(10)$, $(16)$ and $(21)$. In the computation, a fundamental role is played by the fusing angles associated with the on-shell geometry of the Feynman diagrams that determine which cuts are allowed and which are forbidden. These angles are determined by the root system associated with the theory and a  deeper understanding of our observations is likely to be found by using some universal properties of this underlying Coxeter geometry.

\section{Some universal features of higher-order singularities}
\label{Universal_properties_of_higher_order_singularities}

In this section, we present general features of higher-order singularities starting by reviewing some empirical properties of the geometry of the tiled parallelograms, as first discovered in~\cite{Braden:1990wx}. 
We show the relevance of these properties in the computation of certain Laurent coefficients of the S-matrix expansion, extending the discussion started in~\cite{First_loop_paper_sagex} also to odd-order singularities.

In the region of purely imaginary rapidities, we can always associate to a $n$-point scattering process a dual $n$-gon having as sides the masses of the interacting particles. Following~\cite{Braden:1990wx}, we identify the \textit{depth} of such a polygon with the number of elementary triangles necessary to tile it completely. By elementary triangles we mean triangles which cannot be tiled into smaller constituents, so in particular they have unit depth. For example, the parallelogram corresponding to the second-order pole in figure \ref{allowed_one_loop_box_network_on_shell_and_F_Diagram_description} has a depth equal to $4$ since it can be tiled by four elementary triangles. Similarly, the fusing triangle in figure \ref{3_kind_of_triangular_diagrams_in_allowed_processes_contributing_to_3rd_order_poles} has a depth $3$ since it can be tiled in three smaller triangles having unit depth. From this definition it is clear that the depth is an additive quantity and any figure composed by $\mathcal{N}$ constituents has a depth equal to
\begin{equation}
\label{total_depth_definition}
d=\sum_{i=1}^\mathcal{N} d_i
\end{equation}
where $d_i$ is the depth of each constituent. This is the same as the definition introduced in~\cite{Braden:1990wx}.
If all the constituents we are looking at are elementary triangles then all of them have unit depth and $d=\mathcal{N}$. 

Under some circumstances, it may be interesting to look not at the complete tilings of a polygon but only at some of its partial tilings. An example is provided by figure \ref{One_loop_third_order_pole_diagram_from_vertex_corrections_partial_tilings} where to compute the second-order coefficient of the Laurent expansion of the S-matrix around a third-order singularity we had to tile only partially the parallelogram, contrarily to what happens in the network in figure~\ref{Network_of_onshell_diagrams_contributing_to_the_third_order_poles} where the full tiles need to be taken into account. This example easily generalises as follows: we say that a tiling of a polygon of depth $d$ by $\tilde d<d$
triangles, where some of these triangles must therefore be non-elementary, has \emph{partial depth} equal to $\tilde d$.
We use this concept
to explain the different coefficients $a_p$ of the Laurent expansion~\eqref{General_Laurent_expansion_on_the_pole}.
A parallelogram of depth $d=2P$ is associated with any pole of order $P$ of the S-matrix. Then the coefficient $a_P$ in~\eqref{General_Laurent_expansion_on_the_pole} is generated by summing Feynman diagrams corresponding to complete tilings of the parallelogram.
All the lower order coefficients $a_p$ with $p<P$ are instead obtained by summing over tilings of partial depth $2p$. This is exactly what we did to obtain the coefficients $a_2$ and $a_3$ of the expansion around the third-order pole: we summed over tilings of partial depth $4$, and depth $6$, respectively 
(see figures \ref{One_loop_third_order_pole_diagram_from_vertex_corrections_partial_tilings} and \ref{Network_of_onshell_diagrams_contributing_to_the_third_order_poles}). 
This is not surprising if we think that we are summing over Feynman diagrams with only $3$-point vertices\footnote{Diagrams with higher-point vertices can contribute to the sub-leading coefficients $b_p$ in~\eqref{General_Laurent_expansion_on_the_pole} but they never contribute to the coefficients $a_p$.}. If we label the number of vertices by $V$, the number of loops by $L$ and the number of propagators by $I$ the topology of such diagrams imposes the relation $V=2I-4L$ which is exactly equal to $2p$, two times the order of the pole as shown in \eqref{general_formula_for_the_pole_integrating_over_tilde_variables}. Since the number of vertices is exactly equal to the number of constituents composing the tiling, and therefore to the partial (total) depth considered, this agrees with our previous claim that the partial (total) depths of the tilings contributing to the different coefficients $a_p$ are equal to $2p$.

We now distinguish two different situations depending on whether the order $P$ of the maximal singular contribution is even or odd.

\textbf{\underline{Even-order poles.}}
If the maximal singular contribution at the pole is even, say $P=2N$ (i.e. the bootstrapped S-matrix contains two bricks of the form $\{x-1\}^N \{x+1\}^N $), then the parallelogram corresponding to the process has a depth $4N$ and it turns out that there exist $N$ copies of the network in figure~\ref{allowed_one_loop_box_network_on_shell_and_F_Diagram_description} in which the internal constituents can be further tiled. In~\cite{First_loop_paper_sagex} this observation allowed for the computation of the coefficients $a_2$ and $b_1$ of the S-matrix expansion. We will focus here only on the coefficient $a_2$. The derivation is the following. 
Each of the $N$ copies of the network is composed of tilings of partial depth $4$ and can be computed separately. 
It carries a contribution~\eqref{sum_of_surviving_cuts_in_second_order_pole_network} so that the result given by the sum over all the $N$ copies is simply $N$ times the result of a single network:
\begin{equation}
-32 i N \Bigl( \frac{ \beta}{\sqrt{h}} \Bigr)^4 \frac{\Delta^3_{ab}}{(s-s_0)^2}.
\end{equation}
After performing the substitution~\eqref{s_as_function_of_the_rapidity} and multiplying by~\eqref{overall_multiplicative_factor_coming_from_momentum_conservation} we obtain 
\begin{equation}
\label{fourth_order_singularities_S_matrix_from_perturbation_theroy}
N \Bigl(\frac{\beta^2}{2 h}\Bigr)^2 \frac{1}{(\theta- i \theta_0)^2} .
\end{equation}
This result is in agreement with the bootstrapped S-matrix~\eqref{S_matrix_contributions_on_the_poles_expanded_and_containing_all_the_leading_singularities} where the $a_2$ coefficient of the expansion around the $2N^{\text{th}}$-order pole is the binomial factor $\binom{N}{1}$.
\begin{figure}

\begin{center}
\begin{tikzpicture}
\tikzmath{\y=0.5;}

\fill[bostonuniversityred] (-2*\y,7*\y+0.4*\y) -- (-2*\y+0.4*\y,7*\y+0.4*\y) -- (-2*\y+0.4*\y,7*\y+0*\y) -- (-2*\y+0*\y,7*\y+0*\y);
\filldraw[] (-2*\y+0.4*\y,7*\y+0.15*\y)  node[anchor=west] {\small{$: m_1$}};
\fill[orange] (-2*\y,6*\y+0.4*\y) -- (-2*\y+0.4*\y,6*\y+0.4*\y) -- (-2*\y+0.4*\y,6*\y+0*\y) -- (-2*\y+0*\y,6*\y+0*\y);
\filldraw[] (-2*\y+0.4*\y,6*\y+0.15*\y)  node[anchor=west] {\small{$: m_2$}};
\fill[] (-2*\y,5*\y+0.4*\y) -- (-2*\y+0.4*\y,5*\y+0.4*\y) -- (-2*\y+0.4*\y,5*\y+0*\y) -- (-2*\y+0*\y,5*\y+0*\y);
\filldraw[black] (-2*\y+0.4*\y,5*\y+0.15*\y)  node[anchor=west] {\small{$: m_3$}};
\fill[ao(english)] (6*\y,7*\y+0.4*\y) -- (6*\y+0.4*\y,7*\y+0.4*\y) -- (6*\y+0.4*\y,7*\y+0*\y) -- (6*\y+0*\y,7*\y+0*\y);
\filldraw[] (6*\y+0.4*\y,7*\y+0.15*\y)  node[anchor=west] {\small{$: m_4$}};
\fill[blue] (6*\y,6*\y+0.4*\y) -- (6*\y+0.4*\y,6*\y+0.4*\y) -- (6*\y+0.4*\y,6*\y+0*\y) -- (6*\y+0*\y,6*\y+0*\y);
\filldraw[] (6*\y+0.4*\y,6*\y+0.15*\y)  node[anchor=west] {\small{$: m_5$}};
\fill[brown] (6*\y,5*\y+0.4*\y) -- (6*\y+0.4*\y,5*\y+0.4*\y) -- (6*\y+0.4*\y,5*\y+0*\y) -- (6*\y+0*\y,5*\y+0*\y);
\filldraw[] (6*\y+0.4*\y,5*\y+0.15*\y)  node[anchor=west] {\small{$: m_6$}};
\fill[blue-violet] (12*\y,7*\y+0.4*\y) -- (12*\y+0.4*\y,7*\y+0.4*\y) -- (12*\y+0.4*\y,7*\y+0*\y) -- (12*\y+0*\y,7*\y+0*\y);
\filldraw[] (12*\y+0.4*\y,7*\y+0.15*\y)  node[anchor=west] {\small{$: m_7$}};
\fill[capri] (12*\y,6*\y+0.4*\y) -- (12*\y+0.4*\y,6*\y+0.4*\y) -- (12*\y+0.4*\y,6*\y+0*\y) -- (12*\y+0*\y,6*\y+0*\y);
\filldraw[] (12*\y+0.4*\y,6*\y+0.15*\y)  node[anchor=west] {\small{$: m_8$}};

\draw[bostonuniversityred] (-0.522642*\y-1.868096*\y,2.45884*\y-1.682042*\y) -- (-3.09017*\y,0*\y);

\draw[] (-0.522642*\y-1.868096*\y,2.45884*\y-1.682042*\y) -- (-3.09017*\y+-0.522642*\y,0*\y+2.45884*\y);

\draw[ao(english)] (0*\y+0*\y, 0*\y+0*\y) -- (-0.522642*\y+0*\y,2.45884*\y+0*\y);
\draw[ao(english)] (-3.09017*\y+-0.522642*\y,0*\y+2.45884*\y) -- (-3.09017*\y+0*\y,0*\y+0*\y);
\draw[ao(english)] (-0.522642*\y-1.868096*\y,2.45884*\y-1.682042*\y) -- (-0.522642*\y+0*\y,2.45884*\y+0*\y);
\draw[ao(english)] (-0.522642*\y-1.868096*\y,2.45884*\y-1.682042*\y) -- (0*\y+0*\y, 0*\y+0*\y);

\draw[blue] (0*\y+0*\y, 0*\y+0*\y) -- (-3.09017*\y+0*\y,0*\y+0*\y);
\draw[blue] (0*\y+-0.522642*\y, 0*\y+2.45884*\y) -- (-3.09017*\y+-0.522642*\y,0*\y+2.45884*\y);

\filldraw[] (-1.8*\y,1*\y)  node[anchor=west] {\small{$3$}};
\filldraw[] (-3*\y,2*\y)  node[anchor=west] {\small{$3$}};

\draw[bostonuniversityred] (-0.522642*\y-1.868096*\y+4.5*\y,2.45884*\y-1.682042*\y-3*\y) -- (-3.09017*\y+4.5*\y,0*\y-3*\y);
\draw[bostonuniversityred] (-0.522642*\y+4.5*\y,2.45884*\y+0*\y-3*\y) -- (-0.699432*\y+-0.522642*\y+4.5*\y,-0.776798*\y+2.45884*\y-3*\y);

\draw[] (-0.522642*\y-1.868096*\y+4.5*\y,2.45884*\y-1.682042*\y-3*\y) -- (-3.09017*\y+-0.522642*\y+4.5*\y,0*\y+2.45884*\y-3*\y);
\draw[] (-0.699432*\y+-0.522642*\y+4.5*\y,-0.776798*\y+2.45884*\y-3*\y) -- (0*\y+4.5*\y, -3*\y);

\draw[ao(english)] (+4.5*\y, -3*\y) -- (-0.522642*\y+4.5*\y,2.45884*\y-3*\y);
\draw[ao(english)] (-3.09017*\y+-0.522642*\y+4.5*\y,0*\y+2.45884*\y-3*\y) -- (-3.09017*\y+4.5*\y,-3*\y);
\draw[ao(english)] (-0.522642*\y-1.868096*\y+4.5*\y,2.45884*\y-1.682042*\y-3*\y) -- (+4.5*\y, -3*\y);
\draw[ao(english)] (-0.699432*\y+-0.522642*\y+4.5*\y,-0.776798*\y+2.45884*\y-3*\y) -- (-3.09017*\y+-0.522642*\y+4.5*\y,0*\y+2.45884*\y-3*\y);

\draw[blue] (+4.5*\y, -3*\y) -- (-3.09017*\y+4.5*\y,-3*\y);
\draw[blue] (0*\y+-0.522642*\y+4.5*\y, 0*\y+2.45884*\y-3*\y) -- (-3.09017*\y+-0.522642*\y+4.5*\y,0*\y+2.45884*\y-3*\y);

\filldraw[] (2*\y,-1.5*\y)  node[anchor=west] {\small{$4$}};


\draw[bostonuniversityred] (-0.522642*\y+0*\y,2.45884*\y+0*\y-6*\y) -- (-0.699432*\y+-0.522642*\y,-0.776798*\y+2.45884*\y-6*\y);

\draw[] (-0.699432*\y+-0.522642*\y,-0.776798*\y+2.45884*\y-6*\y) -- (0*\y, -6*\y);

\draw[ao(english)] (-0.699432*\y+-0.522642*\y,-0.776798*\y+2.45884*\y-6*\y) -- (-3.09017*\y+-0.522642*\y,0*\y+2.45884*\y-6*\y);
\draw[ao(english)] (0*\y, -6*\y) -- (-0.522642*\y,2.45884*\y+0*\y-6*\y);
\draw[ao(english)] (-3.09017*\y+-0.522642*\y,0*\y+2.45884*\y-6*\y) -- (-3.09017*\y,-6*\y);
\draw[ao(english)] (-3.09017*\y,-6*\y) -- (-0.699432*\y+-0.522642*\y,-0.776798*\y+2.45884*\y-6*\y);

\draw[blue] (0*\y, -6*\y) -- (-3.09017*\y,-6*\y);
\draw[blue] (0*\y+-0.522642*\y, 0*\y+2.45884*\y-6*\y) -- (-3.09017*\y+-0.522642*\y,0*\y+2.45884*\y-6*\y);

\filldraw[] (-1.8*\y,-5*\y)  node[anchor=west] {\small{$3$}};
\filldraw[] (-3*\y,-4.5*\y)  node[anchor=west] {\small{$3$}};


\draw[bostonuniversityred] (-0.522642*\y+0*\y-4.5*\y,2.45884*\y+0*\y-3*\y) -- (-0.699432*\y+-0.522642*\y-4.5*\y,-0.776798*\y+2.45884*\y-3*\y);
\draw[bostonuniversityred] (-0.522642*\y-1.868096*\y-4.5*\y,2.45884*\y-1.682042*\y-3*\y) -- (-3.09017*\y-4.5*\y,0*\y-3*\y);

\draw[] (-0.699432*\y+-0.522642*\y-4.5*\y,-0.776798*\y+2.45884*\y-3*\y) -- (-4.5*\y, -3*\y);
\draw[] (-0.522642*\y-1.868096*\y-4.5*\y,2.45884*\y-1.682042*\y-3*\y) -- (-3.09017*\y+-0.522642*\y-4.5*\y,0*\y+2.45884*\y-3*\y);

\draw[ao(english)] (-4.5*\y, -3*\y) -- (-0.522642*\y-4.5*\y,2.45884*\y+0*\y-3*\y);
\draw[ao(english)] (-3.09017*\y+-0.522642*\y-4.5*\y,0*\y+2.45884*\y-3*\y) -- (-3.09017*\y-4.5*\y,-3*\y);
\draw[ao(english)] (-3.09017*\y-4.5*\y,-3*\y) -- (-0.699432*\y+-0.522642*\y-4.5*\y,-0.776798*\y+2.45884*\y-3*\y);
\draw[ao(english)] (-0.522642*\y-1.868096*\y-4.5*\y,2.45884*\y-1.682042*\y-3*\y) -- (-0.522642*\y-4.5*\y,2.45884*\y-3*\y);

\draw[blue] (-4.5*\y, -3*\y) -- (-3.09017*\y-4.5*\y,-3*\y);
\draw[blue] (0*\y+-0.522642*\y-4.5*\y, 0*\y+2.45884*\y-3*\y) -- (-3.09017*\y+-0.522642*\y-4.5*\y,0*\y+2.45884*\y-3*\y);

\filldraw[] (-6.3*\y,-2.2*\y)  node[anchor=west] {\small{$3$}};
\filldraw[] (-6.7*\y,-1*\y)  node[anchor=west] {\small{$3$}};


\draw[bostonuniversityred] (-3.09017*\y+-0.522642*\y+15*\y, 0*\y+2.45884*\y) -- (-2.39074*\y+-0.522642*\y+15*\y,-0.776798*\y+2.45884*\y);

\draw[orange] (-2.39074*\y+-0.522642*\y+15*\y,-0.776798*\y+2.45884*\y) -- (-3.09017*\y+15*\y,0*\y+0*\y);

\draw[ao(english)] (+15*\y, 0*\y+0*\y) -- (-0.522642*\y+15*\y,2.45884*\y+0*\y);
\draw[ao(english)] (0*\y+-0.522642*\y+15*\y, 0*\y+2.45884*\y) -- (-2.39074*\y+-0.522642*\y+15*\y,-0.776798*\y+2.45884*\y);
\draw[ao(english)] (-3.09017*\y+-0.522642*\y+15*\y,0*\y+2.45884*\y) -- (-3.09017*\y+15*\y,0*\y+0*\y);

\draw[blue] (+15*\y, 0*\y+0*\y) -- (-3.09017*\y+15*\y,0*\y+0*\y);
\draw[blue] (0*\y+-0.522642*\y+15*\y, 0*\y+2.45884*\y) -- (-3.09017*\y+-0.522642*\y+15*\y,0*\y+2.45884*\y);

\draw[brown] (-2.39074*\y+-0.522642*\y+15*\y,-0.776798*\y+2.45884*\y) -- (+15*\y, 0*\y+0*\y);

\filldraw[] (13.4*\y,1.6*\y)  node[anchor=west] {\small{$3$}};
\filldraw[] (12*\y,1*\y)  node[anchor=west] {\small{$3$}};


\draw[bostonuniversityred] (-3.09017*\y+-0.522642*\y+19.5*\y, 0*\y+2.45884*\y-3*\y) -- (-2.39074*\y+-0.522642*\y+19.5*\y,-0.776798*\y+2.45884*\y-3*\y);
\draw[bostonuniversityred] (-3.09017*\y+2.39074*\y+19.5*\y, 0.776798*\y-3*\y) -- (+19.5*\y,0*\y-3*\y);

\draw[orange] (-2.39074*\y+-0.522642*\y+19.5*\y,-0.776798*\y+2.45884*\y-3*\y) -- (-3.09017*\y+19.5*\y,0*\y+0*\y-3*\y);
\draw[orange] (-3.09017*\y+2.39074*\y+19.5*\y, 0.776798*\y-3*\y) -- (0*\y+-0.522642*\y+19.5*\y, 0*\y+2.45884*\y-3*\y);

\draw[ao(english)] (+19.5*\y, -3*\y) -- (-0.522642*\y+19.5*\y,2.45884*\y-3*\y);
\draw[ao(english)] (-3.09017*\y+-0.522642*\y+19.5*\y,+2.45884*\y-3*\y) -- (-3.09017*\y+19.5*\y,-3*\y);

\draw[blue] (+19.5*\y, -3*\y) -- (-3.09017*\y+19.5*\y,-3*\y);
\draw[blue] (0*\y+-0.522642*\y+19.5*\y, +2.45884*\y-3*\y) -- (-3.09017*\y+-0.522642*\y+19.5*\y,+2.45884*\y-3*\y);

\draw[brown] (-2.39074*\y+-0.522642*\y+19.5*\y,-0.776798*\y+2.45884*\y-3*\y) -- (+19.5*\y, -3*\y);
\draw[brown] (-3.09017*\y+2.39074*\y+19.5*\y, 0.776798*\y-3*\y) -- (-3.09017*\y+-0.522642*\y+19.5*\y,+2.45884*\y-3*\y);

\filldraw[] (17*\y,-2.3*\y)  node[anchor=west] {\small{$3$}};
\filldraw[] (17.3*\y,-0.9*\y)  node[anchor=west] {\small{$3$}};


\draw[bostonuniversityred] (-3.09017*\y+2.39074*\y+15*\y, 0.776798*\y-6*\y) -- (+15*\y,0*\y-6*\y);

\draw[orange] (-3.09017*\y+2.39074*\y+15*\y, 0.776798*\y-6*\y) -- (0*\y+-0.522642*\y+15*\y, 0*\y+2.45884*\y-6*\y);

\draw[ao(english)] (+15*\y, -6*\y) -- (-0.522642*\y+15*\y,2.45884*\y-6*\y);
\draw[ao(english)] (-3.09017*\y+-0.522642*\y+15*\y,+2.45884*\y-6*\y) -- (-3.09017*\y+15*\y,-6*\y);
\draw[ao(english)] (-3.09017*\y+2.39074*\y+15*\y, 0.776798*\y-6*\y) -- (-3.09017*\y+15*\y,-6*\y);

\draw[blue] (+15*\y, -6*\y) -- (-3.09017*\y+15*\y,-6*\y);
\draw[blue] (0*\y+-0.522642*\y+15*\y, +2.45884*\y-6*\y) -- (-3.09017*\y+-0.522642*\y+15*\y,+2.45884*\y-6*\y);

\draw[brown] (-3.09017*\y+2.39074*\y+15*\y, 0.776798*\y-6*\y) -- (-3.09017*\y+-0.522642*\y+15*\y,+2.45884*\y-6*\y);

\filldraw[] (13*\y,-4*\y)  node[anchor=west] {\small{$3$}};
\filldraw[] (12*\y,-5*\y)  node[anchor=west] {\small{$3$}};


\draw[bostonuniversityred] (-3.09017*\y+-0.522642*\y+10.5*\y, 0*\y+2.45884*\y-3*\y) -- (-2.39074*\y+-0.522642*\y+10.5*\y,-0.776798*\y+2.45884*\y-3*\y);
\draw[bostonuniversityred] (-3.09017*\y+2.39074*\y+10.5*\y, 0.776798*\y-3*\y) -- (+10.5*\y,0*\y-3*\y);

\draw[orange] (-3.09017*\y+2.39074*\y+10.5*\y, 0.776798*\y-3*\y) -- (0*\y+-0.522642*\y+10.5*\y, 0*\y+2.45884*\y-3*\y);
\draw[orange] (-2.39074*\y+-0.522642*\y+10.5*\y,-0.776798*\y+2.45884*\y-3*\y) -- (-3.09017*\y+10.5*\y,0*\y+0*\y-3*\y);

\draw[ao(english)] (+10.5*\y, -3*\y) -- (-0.522642*\y+10.5*\y,2.45884*\y-3*\y);
\draw[ao(english)] (-3.09017*\y+-0.522642*\y+10.5*\y,+2.45884*\y-3*\y) -- (-3.09017*\y+10.5*\y,-3*\y);
\draw[ao(english)] (-3.09017*\y+2.39074*\y+10.5*\y, 0.776798*\y-3*\y) -- (-3.09017*\y+10.5*\y,-3*\y);
\draw[ao(english)] (0*\y+-0.522642*\y+10.5*\y, 0*\y+2.45884*\y-3*\y) -- (-2.39074*\y+-0.522642*\y+10.5*\y,-0.776798*\y+2.45884*\y-3*\y);

\draw[blue] (+10.5*\y, -3*\y) -- (-3.09017*\y+10.5*\y,-3*\y);
\draw[blue] (0*\y+-0.522642*\y+10.5*\y, +2.45884*\y-3*\y) -- (-3.09017*\y+-0.522642*\y+10.5*\y,+2.45884*\y-3*\y);

\filldraw[] (8*\y,-1.8*\y)  node[anchor=west] {\small{$4$}};

\end{tikzpicture}
\end{center}
\caption{Pair of networks contributing to the $a_2$ coefficient (see \eqref{General_Laurent_expansion_on_the_pole}) of the Laurent expansion of the S-matrix around a $4^{\text{th}}$-order pole. The example has been taken from the $e_8^{(1)}$ affine Toda model and its masses are reported with different colours following the order $m_1<m_2<\dots<m_8$. Inside the different tiles (apart from those with unit depth) their depth is reported.}
\label{4th_order_pole_tile_level_one_networks}
\end{figure}
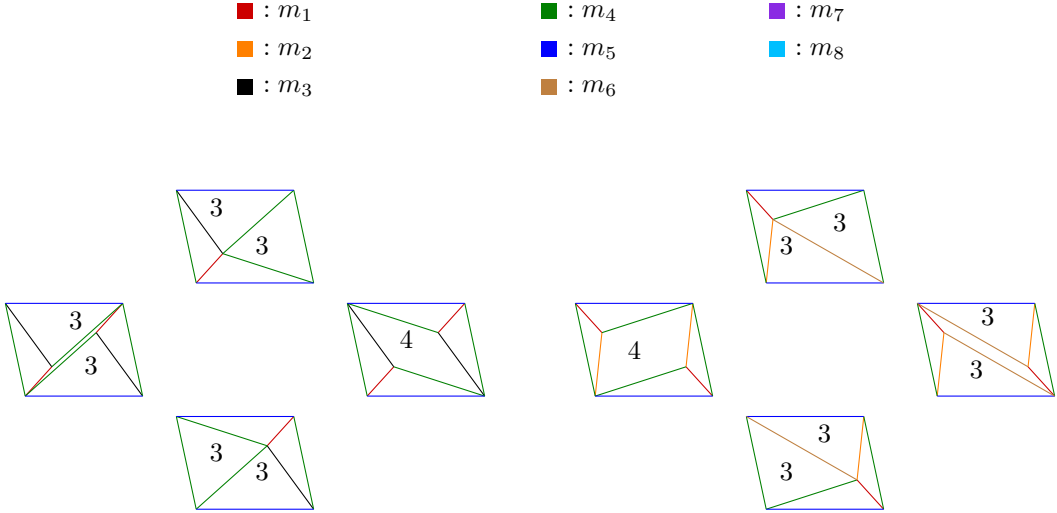

As an example in figure~\ref{4th_order_pole_tile_level_one_networks}
we show the two networks contributing to the second-order coefficient around a $4^{\text{th}}$-order pole. 
In the image, we see that the two networks at this depth level are disjoint and are not connected by any flip. Each one contributes separately to the final result. This is no longer the case if we start searching for tilings with a depth higher than $4$, which in this case can be $6$ or $8$. The latter case corresponds to the search for Feynman diagrams contributing to the $a_4$ coefficient at the pole. The number of such diagrams is huge. The graphs containing two vertices of depth $3$ in figure \ref{4th_order_pole_tile_level_one_networks} contribute with $9$ Feynman diagrams ($3 \times 3$ since each vertex of depth $3$ counts $3$ possible corrections of the form in figure~\ref{3_kind_of_triangular_diagrams_in_allowed_processes_contributing_to_3rd_order_poles}) while the graphs containing a depth-four parallelogram contribute with $4$ diagrams (those coming from the second-order pole network). In total there are $62$ diagrams contributing to the $a_4$ coefficient to which we should add all the graphs obtained by them through the flipping move. The fact that such a big number of diagrams should reproduce the simple result that we observe in the bootstrapped S-matrix suggests that probably, as it happens for third-order poles, the computation is reproduced by only a small number of cuts. It would be very interesting to investigate the space of Feynman diagrams contributing to such a simple result. 
On the other hand, there exist also Feynman diagrams with partial depth equal to six. These are two-loop diagrams contributing to the $a_3$ coefficient of the Laurent expansion around the pole. We expect a zero result for this number. Indeed it is evident from the expression \eqref{S_matrix_contributions_on_the_poles_expanded_and_containing_all_the_leading_singularities} that all the odd-order coefficients of the expansion around a generic even-order singularity are zero. Once again we do not have a motivation for why this happens, but it is probably due to an underlying simplification that can be obtained by cutting the different diagrams inside a network. A universal explanation, based on diagrammatic computations, of the absence of all the odd-order terms in the expansion~\eqref{S_matrix_contributions_on_the_poles_expanded_and_containing_all_the_leading_singularities} around a generic even-order pole would be very interesting.

\textbf{\underline{Odd-order poles.}}
If $P$ in~\eqref{General_Laurent_expansion_on_the_pole} is odd, then we have an odd-order singularity in the S-matrix. In this case, the parallelogram associated with the diagonal process can be split into two triangles each one of depth $P$. 
In the known examples coming from affine Toda field theories, $P$ can be any odd number from 1 up to and including 11. 

Based on empirical observations we note that each triangle of depth $P=2N+1$ can be tiled by three subtriangles in $3N$ ways, as was remarked in~\cite{Braden:1990wx}. We
further note that these $3N$ tilings can be grouped into $N$ triplets 
$$
(\Delta^{(1)},\Delta^{(2)},\Delta^{(3)}), \ldots,(\Delta^{(3N-2)},\Delta^{(3N-1)},\Delta^{(3N)}) \,,
$$
such that the relations in~\eqref{sum_areas_three_point_couplings_third_orderd_pole_vertex_corrections}  apply separately to each of these $N$ triplets.
If we associate the labels $\Delta_a$, $\Delta_b$ and $\Delta_c$ to the triangles constructed over the sides $p_a$, $p_b$ and $p_c$ respectively,
then we have 
\begin{equation}
\label{relations_between_triangle_areas_of_the_3q_tilings}
\sum_{i=1}^3 \Delta_a^{(3n-3+i)}=\sum_{i=1}^3 \Delta_b^{(3n-3+i)}=\sum_{i=1}^3 \Delta_c^{(3n-3+i)}=\Delta_{ab}
\end{equation}
for each triplet $n=1,\dots,N$. 
At the moment we do not have a general way to specify these triplets apart from the observation that they can be chosen so as to satisfy~\eqref{relations_between_triangle_areas_of_the_3q_tilings}.
This observation is stronger than the one highlighted in~\cite{Braden:1990wx}, where it was noted that the sum over \emph{all} the triangles composing the $3N$ different tilings respects
$$
\sum_{i=1}^{3N} \Delta_a^{(i)}=\sum_{i=1}^{3N} \Delta_b^{(i)}=\sum_{i=1}^{3N} \Delta_c^{(i)}=N \Delta_{ab}.
$$
The sign rule~\eqref{product_of_f_functions_three_point_coupling_number_of_tiles_equal_three} is applied to each one of the $3N$ different tilings. Labelling by $(\sigma_a^{(i)},\sigma_b^{(i)},\sigma_c^{(i)})$ the three phases entering the $3$-point couplings constructed over the sides $a$, $b$ and $c$ in the $i$-th tiling, with $i=1,\ldots,3N$, 
 we have for each $i$
\begin{equation}
\label{relations_between_f_terms_of_the_3q_tilings}
\sigma_a^{(i)} \sigma_b^{(i)} \sigma_c^{(i)}=- \sigma_{ab\bar{c}}.
\end{equation}
At this point, it is easy to explain the origin of the coefficient $a_2= \binom{N}{1}$ in the S-matrix expansion~\eqref{S_matrix_contributions_on_the_poles_expanded_and_containing_all_the_leading_singularities}. It comes by performing $N$ times the computation reproduced in section~\ref{Computation_of_one_loop_S_matrix_around_3rd_order_pole}, one for each triplet of vertex corrections; the final result is therefore given by multiplying the result in~\eqref{one_loop_3rd_order_pole_S_matrix_perturbation_theory} by $N$. 

As for even-order singularities, to study the higher-order coefficients of the Laurent expansion $a_3, a_4, \ldots$ we need to keep into account depths of higher order. The $N$ triplets of vertex corrections, that are disconnected at one loop, by considering nested tilings start to become connected by flipping internal propagators. We show an example of vertex corrections contributing to a $5^{\text{th}}$-order pole in the next subsection.

We conclude this discussion of the properties of higher-order poles by noting that the relation~\eqref{relations_between_f_terms_of_the_3q_tilings} can be iterated, a fact that was already noted in~\cite{Braden:1990wx}. 
We have already explained that a triangle of total depth $2N+1$ can be tiled in three subtriangles exactly in $3N$ ways. Then the triangles composing each one of these $3N$ tilings can be further tiled in triangles and so on. By looking to more and more nested tilings we end up with a configuration composed of $2N+1$ pieces.
By iterating relation~\eqref{relations_between_f_terms_of_the_3q_tilings}, the product of all the phases for each of these maximal depth nested tilings is given by 
\begin{equation}
\label{Multiplication_of_f_in_arbitrary_nested_vertex}
\prod_{k=1}^{2N+1} \sigma^{(j)}_{k} = (-1)^N \sigma_{ab \bar{c}}
\end{equation}
where $j=1,\dots,M$ labels the different nested tilings. As remarked in~\cite{Braden:1990wx}, the number $M$ of such tilings (which are themselves a subset of all possible tilings) grows rapidly with $N$.

\subsection{Two-loop vertex corrections and 5th-order poles}

We compute the maximal leading order singularity of a $3$-point vertex having depth $5$, and contributing therefore to a $5$th-order pole in the S-matrix. Also in this case we show that the result is universal and does not depend on the affine Toda theory considered.

The triangle in question can be divided into three sub-triangles in $6$ different ways. We identify two triplets of tilings at one loop, $[(1,\bullet),(2,\bullet),(3,\bullet)]$ and $[(4,\bullet),(5,\bullet),(6,\bullet)]$, satisfying separately relation~\eqref{relations_between_triangle_areas_of_the_3q_tilings} (with $N=2$). In this case, we have
\begin{equation}
\label{sum_of_main_triangles_constructed_over_c_in_a_depth_5_vertex}
\Delta_c^{(1,\bullet)}+\Delta_c^{(2,\bullet)}+\Delta_c^{(3,\bullet)}=\Delta_{ab} \hspace{5mm} \text{and} \hspace{5mm} \Delta_c^{(4,\bullet)}+\Delta_c^{(5,\bullet)}+\Delta_c^{(6,\bullet)}=\Delta_{ab}.
\end{equation}
We used the subscript  $c$ to indicate that we are summing, within the same triplet, all the triangles constructed on the side $c$. The same relation is valid summing the triangles constructed on the $a$- and $b$-side as we pointed out in~\eqref{relations_between_triangle_areas_of_the_3q_tilings}.

Each of these six different tilings is composed of two triangles having a depth of $1$ and one triangle having a depth of $3$; the latter can therefore be tiled in three different ways.
The total depth,  given by summing all the elementary constituents of a single tiling, is then $5$.
To take into account such a nested level of tiles we use a second index, that is absent in~\eqref{sum_of_main_triangles_constructed_over_c_in_a_depth_5_vertex} and substituted with a bullet. Since for a given one-loop tiling we identify $3$ additional nested tilings
this second index takes values in $\{1,2,3\}$. Then the pair of numbers $(j,k)$, running from $1$ to $6$ and $1$ to $3$ respectively, label the complete tiling.
For example, the pair $(1,2)$ indicates the tiling number one at the depth level $3$, and (among the three different possibilities to further divide one of its constituents) the sub-tiling number two at the depth level $5$. We did not identify a preferred order to arrange these tilings; we follow the enumeration shown in figure~\ref{2_loops_vertex_corrections_with_flip_symbols_below_the_figures}. Three non-planar vertex corrections $(7,1)$, $(7,2)$ and $(7,3)$ are also present, but they completely cancel with certain atoms generated by cutting the planar diagrams, as shown in figure~\ref{2_loops_vertex_corrections_with_flip_symbols_below_the_figures}.
Apart from these non-planar diagrams, the tilings are arranged in two blocks of $9$ diagrams each. The six different rows label the main tiling, while the three columns provide a label for the nested tilings. We then use the convention of labelling by $\Delta_c^{(j,\bullet)}$ the triangle constructed over the side $c$ in the one-loop configuration $j$  while we label by $\Delta_c^{(j, k)}$ the triangle constructed over the side $c$ in the nested tiling $(j,k)$. To give an example, $\Delta_c^{(3,\bullet)}$ corresponds to the triangle $\Delta_{c e'' m''}$ reported in any of the configurations $(3,1)$, $(3,2)$ and $(3,3)$ in figure~\ref{2_loops_vertex_corrections_with_flip_symbols_below_the_figures} while $\Delta_c^{(3,1)}$ corresponds to $\Delta_{c d'' o}$ that is one of the components of the configuration $(3,1)$.

\afterpage{%
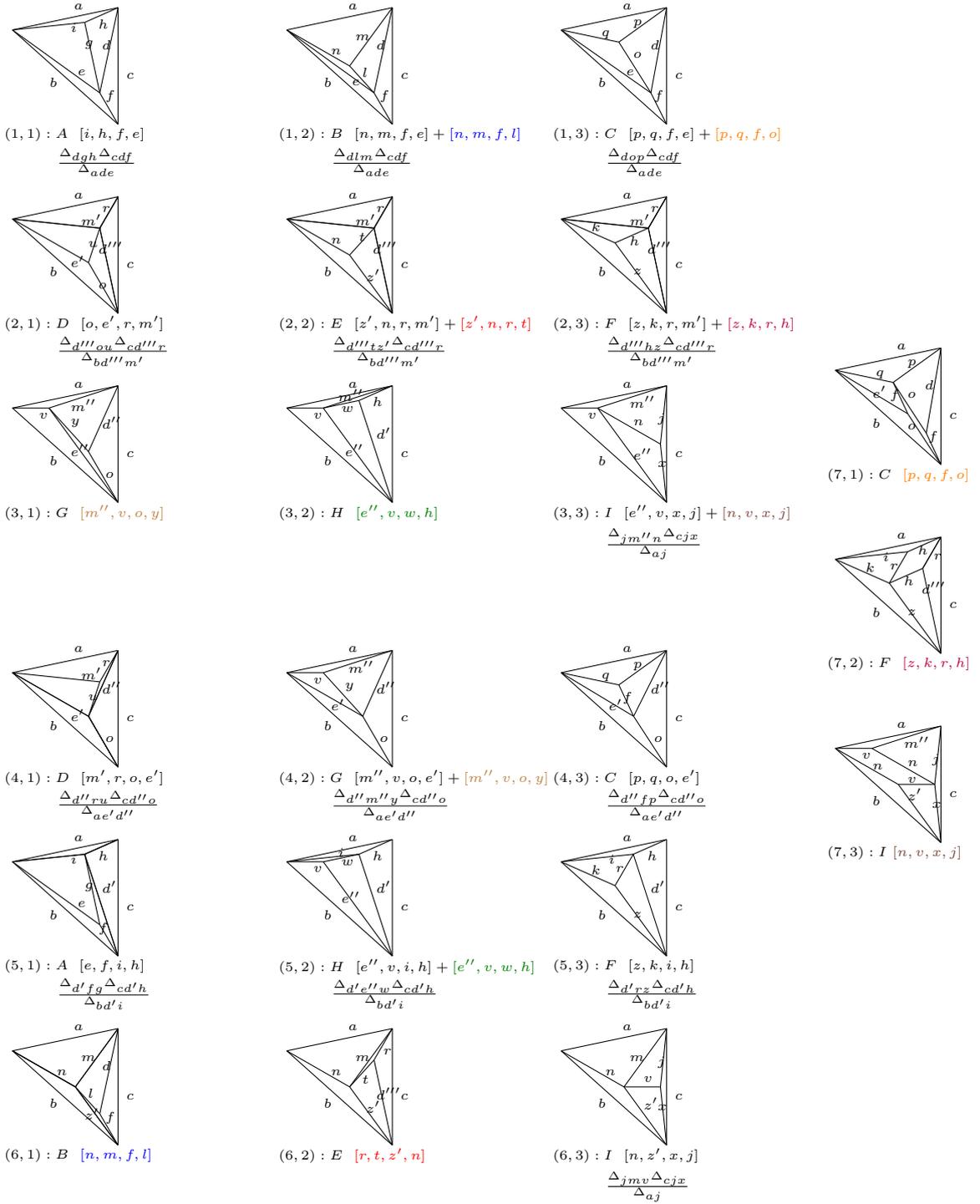
\begin{figure}[p]
\begin{center}
\begin{tikzpicture}
\tikzmath{\y=0.55; \x=0.72; \z=0.6;}

\filldraw[black] (0*\y+0*\y-3.5*\y, 0*\y+0*\y-0.3*\y)  node[anchor=west] {\tiny{$(1,1): A \ \  [i,h,f,e]$}};
\filldraw[black] (0*\y+0*\y-2*\y, 0*\y+0*\y-1.1*\y)  node[anchor=west] {\tiny{$\frac{\Delta_{dgh}\Delta_{cdf}}{\Delta_{ade}}$}};

\draw[] (0*\y+0*\y, 0*\y+0*\y) -- (0*\y+0*\y,3.36408*\y+0*\y);
\draw[] (0*\y+0*\y, 0*\y+0*\y) -- (-3.02264*\y+0*\y,2.7216*\y+0*\y);
\draw[] (-3.02264*\y+0*\y, 2.7216*\y+0*\y) -- (0*\y+0*\y,3.36408*\y+0*\y);
\draw[] (0*\y+0*\y, 0*\y+0*\y) -- (-0.522642*\y+0*\y,0.905243*\y+0*\y);
\draw[] (0*\y+0*\y, 3.36408*\y+0*\y) -- (-0.522642*\y+0*\y,0.905243*\y+0*\y);
\draw[] (-3.02264*\y+0*\y, 2.7216*\y+0*\y) -- (-0.522642*\y+0*\y,0.905243*\y+0*\y);
\draw[] (-0.522642*\y+0*\y, 0.905243*\y+0*\y) -- (-0.954915*\y+0*\y,2.93893*\y+0*\y);
\draw[] (0*\y+0*\y, 3.36408*\y+0*\y) -- (-0.954915*\y+0*\y,2.93893*\y+0*\y);
\draw[] (-3.02264*\y+0*\y, 2.7216*\y+0*\y) -- (-0.954915*\y+0*\y,2.93893*\y+0*\y);

\filldraw[black] (2*\y+0*\y-3.5*\y, 3.7*\y+0*\y-0.3*\y)  node[anchor=west] {\tiny{$a$}};
\filldraw[black] (1.3*\y+0*\y-3.5*\y, 1.5*\y+0*\y-0.3*\y)  node[anchor=west] {\tiny{$b$}};
\filldraw[black] (3.5*\y+0*\y-3.5*\y, 1.7*\y+0*\y-0.3*\y)  node[anchor=west] {\tiny{$c$}};
\filldraw[black] (2.8*\y+0*\y-3.5*\y, 2.6*\y+0*\y-0.3*\y)  node[anchor=west] {\tiny{$d$}};
\filldraw[black] (2.1*\y+0*\y-3.5*\y, 1.8*\y+0*\y-0.3*\y)  node[anchor=west] {\tiny{$e$}};
\filldraw[black] (2.9*\y+0*\y-3.5*\y, 1.1*\y+0*\y-0.3*\y)  node[anchor=west] {\tiny{$f$}};
\filldraw[black] (2.3*\y+0*\y-3.5*\y, 2.6*\y+0*\y-0.3*\y)  node[anchor=west] {\tiny{$g$}};
\filldraw[black] (2.7*\y+0*\y-3.5*\y, 3.2*\y+0*\y-0.3*\y)  node[anchor=west] {\tiny{$h$}};
\filldraw[black] (1.9*\y+0*\y-3.5*\y, 3.1*\y+0*\y-0.3*\y)  node[anchor=west] {\tiny{$i$}};

\textcolor{blue}{This is a sample text in blue.}
\filldraw[black] (0*\y+6*\x-3.5*\y, 0*\y+0*\y-0.3*\y)  node[anchor=west] {\tiny{$(1,2): B \ \ [n,m,f,e]+ \textcolor{blue}{[n,m,f,l]}$}};
\filldraw[black] (0*\y+6*\x-2*\y, 0*\y+0*\y-1.1*\y)  node[anchor=west] {\tiny{$\frac{\Delta_{dlm}\Delta_{cdf}}{\Delta_{ade}}$}};
\draw[] (0*\y+6*\x, 0*\y+0*\y) -- (0*\y+6*\x,3.36408*\y+0*\y);
\draw[] (0*\y+6*\x, 0*\y+0*\y) -- (-3.02264*\y+6*\x,2.7216*\y+0*\y);
\draw[] (-3.02264*\y+6*\x, 2.7216*\y+0*\y) -- (0*\y+6*\x,3.36408*\y+0*\y);
\draw[] (0*\y+6*\x, 0*\y+0*\y) -- (-0.522642*\y+6*\x,0.905243*\y+0*\y);
\draw[] (0*\y+6*\x, 3.36408*\y+0*\y) -- (-0.522642*\y+6*\x,0.905243*\y+0*\y);
\draw[] (-3.02264*\y+6*\x, 2.7216*\y+0*\y) -- (-0.522642*\y+6*\x,0.905243*\y+0*\y);
\draw[] (-0.522642*\y+6*\x, 0.905243*\y+0*\y) -- (-1.22207*\y+6*\x,1.68204*\y+0*\y);
\draw[] (0*\y+6*\x, 3.36408*\y+0*\y) -- (-1.22207*\y+6*\x,1.68204*\y+0*\y);
\draw[] (-3.02264*\y+6*\x, 2.7216*\y+0*\y) -- (-1.22207*\y+6*\x,1.68204*\y+0*\y);

\filldraw[black] (2*\y+6*\x-3.5*\y, 3.7*\y+0*\y-0.3*\y)  node[anchor=west] {\tiny{$a$}};
\filldraw[black] (1.3*\y+6*\x-3.5*\y, 1.5*\y+0*\y-0.3*\y)  node[anchor=west] {\tiny{$b$}};
\filldraw[black] (3.5*\y+6*\x-3.5*\y, 1.7*\y+0*\y-0.3*\y)  node[anchor=west] {\tiny{$c$}};
\filldraw[black] (2.8*\y+6*\x-3.5*\y, 2.6*\y+0*\y-0.3*\y)  node[anchor=west] {\tiny{$d$}};
\filldraw[black] (2.1*\y+6*\x-3.5*\y, 1.5*\y+0*\y-0.3*\y)  node[anchor=west] {\tiny{$e$}};
\filldraw[black] (2.9*\y+6*\x-3.5*\y, 1.1*\y+0*\y-0.3*\y)  node[anchor=west] {\tiny{$f$}};
\filldraw[black] (2.4*\y+6*\x-3.5*\y, 1.8*\y+0*\y-0.3*\y)  node[anchor=west] {\tiny{$l$}};
\filldraw[black] (2.2*\y+6*\x-3.5*\y, 2.8*\y+0*\y-0.3*\y)  node[anchor=west] {\tiny{$m$}};
\filldraw[black] (1.5*\y+6*\x-3.5*\y, 2.4*\y+0*\y-0.3*\y)  node[anchor=west] {\tiny{$n$}};

\filldraw[black] (0*\y+12*\x-3.5*\y, 0*\y+0*\y-0.3*\y)  node[anchor=west] {\tiny{$(1,3): C  \ \ [p,q,f,e]+ \textcolor{orange}{[p,q,f,o]}$}};
\filldraw[black] (0*\y+12*\x-2*\y, 0*\y+0*\y-1.1*\y)  node[anchor=west] {\tiny{$\frac{\Delta_{dop}\Delta_{cdf}}{\Delta_{ade}}$}};
\draw[] (0*\y+12*\x, 0*\y+0*\y) -- (0*\y+12*\x,3.36408*\y+0*\y);
\draw[] (0*\y+12*\x, 0*\y+0*\y) -- (-3.02264*\y+12*\x,2.7216*\y+0*\y);
\draw[] (-3.02264*\y+12*\x, 2.7216*\y+0*\y) -- (0*\y+12*\x,3.36408*\y+0*\y);
\draw[] (0*\y+12*\x, 0*\y+0*\y) -- (-0.45*\y+12*\x,0.905243*\y+0*\y);
\draw[] (0*\y+12*\x, 3.36408*\y+0*\y) -- (-0.45*\y+12*\x,0.905243*\y+0*\y);
\draw[] (-3.02264*\y+12*\x, 2.7216*\y+0*\y) -- (-0.45*\y+12*\x,0.905243*\y+0*\y);
\draw[] (-0.45*\y+12*\x, 0.905243*\y+0*\y) -- (-1.3683*\y+12*\x,2.36996*\y+0*\y);
\draw[] (0*\y+12*\x, 3.36408*\y+0*\y) -- (-1.3683*\y+12*\x,2.36996*\y+0*\y);
\draw[] (-3.02264*\y+12*\x, 2.7216*\y+0*\y) -- (-1.3683*\y+12*\x,2.36996*\y+0*\y);

\filldraw[black] (2*\y+12*\x-3.5*\y, 3.7*\y+0*\y-0.3*\y)  node[anchor=west] {\tiny{$a$}};
\filldraw[black] (1.3*\y+12*\x-3.5*\y, 1.5*\y+0*\y-0.3*\y)  node[anchor=west] {\tiny{$b$}};
\filldraw[black] (3.5*\y+12*\x-3.5*\y, 1.7*\y+0*\y-0.3*\y)  node[anchor=west] {\tiny{$c$}};
\filldraw[black] (2.8*\y+12*\x-3.5*\y, 2.6*\y+0*\y-0.3*\y)  node[anchor=west] {\tiny{$d$}};
\filldraw[black] (2.1*\y+12*\x-3.5*\y, 1.8*\y+0*\y-0.3*\y)  node[anchor=west] {\tiny{$e$}};
\filldraw[black] (2.9*\y+12*\x-3.5*\y, 1.1*\y+0*\y-0.3*\y)  node[anchor=west] {\tiny{$f$}};
\filldraw[black] (2.3*\y+12*\x-3.5*\y, 2.3*\y+0*\y-0.3*\y)  node[anchor=west] {\tiny{$o$}};
\filldraw[black] (2.3*\y+12*\x-3.5*\y, 3.2*\y+0*\y-0.3*\y)  node[anchor=west] {\tiny{$p$}};
\filldraw[black] (1.4*\y+12*\x-3.5*\y, 2.9*\y+0*\y-0.3*\y)  node[anchor=west] {\tiny{$q$}};

\filldraw[black] (0*\y+0*\y-3.5*\y, 0*\y+-5*\z-0.3*\y)  node[anchor=west] {\tiny{$(2,1): D \ \  [o,e', r,m']$}};
  \filldraw[black] (0*\y+0*\y-2*\y, 0*\y+-5*\z-1.1*\y)  node[anchor=west] {\tiny{$\frac{\Delta_{d'''ou} \Delta_{cd'''r} }{\Delta_{bd'''m'}}$}};

\draw[] (0*\y+0*\y, 0*\y+-5*\z) -- (0*\y+0*\y,3.36408*\y+-5*\z);
\draw[] (0*\y+0*\y, 0*\y+-5*\z) -- (-3.02264*\y+0*\y,2.7216*\y+-5*\z);
\draw[] (-3.02264*\y+0*\y, 2.7216*\y+-5*\z) -- (0*\y+0*\y,3.36408*\y+-5*\z);
\draw[] (0*\y+0*\y, 0*\y+-5*\z) -- (-0.522642*\y+0*\y,2.45884*\y+-5*\z);
\draw[] (0*\y+0*\y, 3.36408*\y+-5*\z) -- (-0.522642*\y+0*\y,2.45884*\y+-5*\z);
\draw[] (-3.02264*\y+0*\y, 2.7216*\y+-5*\z) -- (-0.522642*\y+0*\y,2.45884*\y+-5*\z);
\draw[] (-0.522642*\y+0*\y, 2.45884*\y+-5*\z) -- (-0.522642*\y+0*\y,2.45884*\y+-5*\z);
\draw[] (0*\y+0*\y, 3.36408*\y+-5*\z) -- (-0.522642*\y+0*\y,2.45884*\y+-5*\z);
\draw[] (-3.02264*\y+0*\y, 2.7216*\y+-5*\z) -- (-0.522642*\y+0*\y,2.45884*\y+-5*\z);
\draw[] (-0.522642*\y+0*\y, 2.45884*\y+-5*\z) -- (-0.845653*\y+0*\y,1.46471*\y+-5*\z);
\draw[] (-3.02264*\y+0*\y, 2.7216*\y+-5*\z) -- (-0.845653*\y+0*\y,1.46471*\y+-5*\z);
\draw[] (0*\y+0*\y, 0*\y+-5*\z) -- (-0.845653*\y+0*\y,1.46471*\y+-5*\z);
\draw[] (-0.522642*\y+0*\y, 2.45884*\y+-5*\z) -- (-0.522642*\y+0*\y,2.45884*\y+-5*\z);
\draw[] (0*\y+0*\y, 0*\y+-5*\z) -- (-0.522642*\y+0*\y,2.45884*\y+-5*\z);
\draw[] (0*\y+0*\y, 3.36408*\y+-5*\z) -- (-0.522642*\y+0*\y,2.45884*\y+-5*\z);

\filldraw[black] (2*\y+0*\y-3.5*\y, 3.7*\y-5*\z-0.3*\y)  node[anchor=west] {\tiny{$a$}};
\filldraw[black] (1.3*\y+0*\y-3.5*\y, 1.5*\y-5*\z-0.3*\y)  node[anchor=west] {\tiny{$b$}};
\filldraw[black] (3.5*\y+0*\y-3.5*\y, 1.7*\y-5*\z-0.3*\y)  node[anchor=west] {\tiny{$c$}};
\filldraw[black] (2.7*\y+0*\y-3.5*\y, 2.2*\y-5*\z-0.3*\y)  node[anchor=west] {\tiny{$d'''$}};
\filldraw[black] (1.9*\y+0*\y-3.5*\y, 1.8*\y-5*\z-0.3*\y)  node[anchor=west] {\tiny{$e'$}};
\filldraw[black] (2.7*\y+0*\y-3.5*\y, 1.1*\y-5*\z-0.3*\y)  node[anchor=west] {\tiny{$o$}};
\filldraw[black] (2.8*\y+0*\y-3.5*\y, 3.3*\y-5*\z-0.3*\y)  node[anchor=west] {\tiny{$r$}};
\filldraw[black] (2.2*\y+0*\y-3.5*\y, 3*\y-5*\z-0.3*\y)  node[anchor=west] {\tiny{$m'$}};
\filldraw[black] (2.4*\y+0*\y-3.5*\y, 2.3*\y-5*\z-0.3*\y)  node[anchor=west] {\tiny{$u$}};

\filldraw[black] (0*\y+6*\x-3.5*\y, 0*\y+-5*\z-0.3*\y)  node[anchor=west] {\tiny{$(2,2): E \ \ [z',n, r, m']+  \textcolor{red}{[z', n, r, t]} $}};
  \filldraw[black] (0*\y+6*\x-2*\y, 0*\y+-5*\z-1.1*\y)  node[anchor=west] {\tiny{$\frac{\Delta_{d''' t z' } \Delta_{c d'''r} }{\Delta_{bd'''m'}}$}};
\draw[] (0*\y+6*\x, 0*\y+-5*\z) -- (0*\y+6*\x,3.36408*\y+-5*\z);
\draw[] (0*\y+6*\x, 0*\y+-5*\z) -- (-3.02264*\y+6*\x,2.7216*\y+-5*\z);
\draw[] (-3.02264*\y+6*\x, 2.7216*\y+-5*\z) -- (+6*\x,3.36408*\y+-5*\z);
\draw[] (0*\y+6*\x, 0*\y+-5*\z) -- (-0.522642*\y+6*\x,2.45884*\y+-5*\z);
\draw[] (0*\y+6*\x, 3.36408*\y+-5*\z) -- (-0.522642*\y+6*\x,2.45884*\y+-5*\z);
\draw[] (-3.02264*\y+6*\x, 2.7216*\y+-5*\z) -- (-0.522642*\y+6*\x,2.45884*\y+-5*\z);
\draw[] (-0.522642*\y+6*\x, 2.45884*\y+-5*\z) -- (-0.522642*\y+6*\x,2.45884*\y+-5*\z);
\draw[] (0*\y+6*\x, 3.36408*\y+-5*\z) -- (-0.522642*\y+6*\x,2.45884*\y+-5*\z);
\draw[] (-3.02264*\y+6*\x, 2.7216*\y+-5*\z) -- (-0.522642*\y+6*\x,2.45884*\y+-5*\z);
\draw[] (-0.522642*\y+6*\x, 2.45884*\y+-5*\z) -- (-1.22207*\y+6*\x,1.68204*\y+-5*\z);
\draw[] (-3.02264*\y+6*\x, 2.7216*\y+-5*\z) -- (-1.22207*\y+6*\x,1.68204*\y+-5*\z);
\draw[] (0*\y+6*\x, 0*\y+-5*\z) -- (-1.22207*\y+6*\x,1.68204*\y+-5*\z);
\draw[] (-0.522642*\y+6*\x, 2.45884*\y+-5*\z) -- (-0.522642*\y+6*\x,2.45884*\y+-5*\z);
\draw[] (0*\y+6*\x, 0*\y+-5*\z) -- (-0.522642*\y+6*\x,2.45884*\y+-5*\z);
\draw[] (0*\y+6*\x, 3.36408*\y+-5*\z) -- (-0.522642*\y+6*\x,2.45884*\y+-5*\z);

\filldraw[black] (2*\y+6*\x-3.5*\y, 3.7*\y-5*\z-0.3*\y)  node[anchor=west] {\tiny{$a$}};
\filldraw[black] (1.3*\y+6*\x-3.5*\y, 1.5*\y-5*\z-0.3*\y)  node[anchor=west] {\tiny{$b$}};
\filldraw[black] (3.5*\y+6*\x-3.5*\y, 1.7*\y-5*\z-0.3*\y)  node[anchor=west] {\tiny{$c$}};
\filldraw[black] (2.7*\y+6*\x-3.5*\y, 2.2*\y-5*\z-0.3*\y)  node[anchor=west] {\tiny{$d'''$}};
\filldraw[black] (2.8*\y+6*\x-3.5*\y, 3.3*\y-5*\z-0.3*\y)  node[anchor=west] {\tiny{$r$}};
\filldraw[black] (2.2*\y+6*\x-3.5*\y, 3*\y-5*\z-0.3*\y)  node[anchor=west] {\tiny{$m'$}};
\filldraw[black] (2.5*\y+6*\x-3.5*\y, 1.4*\y-5*\z-0.3*\y)  node[anchor=west] {\tiny{$z'$}};
\filldraw[black] (1.5*\y+6*\x-3.5*\y, 2.4*\y-5*\z-0.3*\y)  node[anchor=west] {\tiny{$n$}};
\filldraw[black] (2.3*\y+6*\x-3.5*\y, 2.5*\y-5*\z-0.3*\y)  node[anchor=west] {\tiny{$t$}};

\filldraw[black] (0*\y+12*\x-3.5*\y, 0*\y+-5*\z-0.3*\y)  node[anchor=west] {\tiny{$(2,3): F \ \ [z, k, r, m']+\textcolor{purple}{[z, k, r, h]}$}};
  \filldraw[black] (0*\y+12*\x-2*\y, 0*\y+-5*\z-1.1*\y)  node[anchor=west] {\tiny{$\frac{\Delta_{d''' h z} \Delta_{c d'''r} }{\Delta_{bd'''m'}}$}};
\draw[] (0*\y+12*\x, 0*\y+-5*\z) -- (0*\y+12*\x,3.36408*\y+-5*\z);
\draw[] (0*\y+12*\x, 0*\y+-5*\z) -- (-3.02264*\y+12*\x,2.7216*\y+-5*\z);
\draw[] (-3.02264*\y+12*\x, 2.7216*\y+-5*\z) -- (0*\y+12*\x,3.36408*\y+-5*\z);
\draw[] (0*\y+12*\x, 0*\y+-5*\z) -- (-0.522642*\y+12*\x,2.45884*\y+-5*\z);
\draw[] (0*\y+12*\x, 3.36408*\y+-5*\z) -- (-0.522642*\y+12*\x,2.45884*\y+-5*\z);
\draw[] (-3.02264*\y+12*\x, 2.7216*\y+-5*\z) -- (-0.522642*\y+12*\x,2.45884*\y+-5*\z);
\draw[] (-0.522642*\y+12*\x, 2.45884*\y+-5*\z) -- (-0.522642*\y+12*\x,2.45884*\y+-5*\z);
\draw[] (0*\y+12*\x, 3.36408*\y+-5*\z) -- (-0.522642*\y+12*\x,2.45884*\y+-5*\z);
\draw[] (-3.02264*\y+12*\x, 2.7216*\y+-5*\z) -- (-0.522642*\y+12*\x,2.45884*\y+-5*\z);
\draw[] (-0.522642*\y+12*\x, 2.45884*\y+-5*\z) -- (-1.47756*\y+12*\x,2.03368*\y+-5*\z);
\draw[] (-3.02264*\y+12*\x, 2.7216*\y+-5*\z) -- (-1.47756*\y+12*\x,2.03368*\y+-5*\z);
\draw[] (0*\y+12*\x, 0*\y+-5*\z) -- (-1.47756*\y+12*\x,2.03368*\y+-5*\z);
\draw[] (-0.522642*\y+12*\x, 2.45884*\y+-5*\z) -- (-0.522642*\y+12*\x,2.45884*\y+-5*\z);
\draw[] (0*\y+12*\x, 0*\y+-5*\z) -- (-0.522642*\y+12*\x,2.45884*\y+-5*\z);
\draw[] (0*\y+12*\x, 3.36408*\y+-5*\z) -- (-0.522642*\y+12*\x,2.45884*\y+-5*\z);

\filldraw[black] (2*\y+12*\x-3.5*\y, 3.7*\y-5*\z-0.3*\y)  node[anchor=west] {\tiny{$a$}};
\filldraw[black] (1.3*\y+12*\x-3.5*\y, 1.5*\y-5*\z-0.3*\y)  node[anchor=west] {\tiny{$b$}};
\filldraw[black] (3.5*\y+12*\x-3.5*\y, 1.7*\y-5*\z-0.3*\y)  node[anchor=west] {\tiny{$c$}};
\filldraw[black] (2.7*\y+12*\x-3.5*\y, 2.2*\y-5*\z-0.3*\y)  node[anchor=west] {\tiny{$d'''$}};
\filldraw[black] (2.8*\y+12*\x-3.5*\y, 3.3*\y-5*\z-0.3*\y)  node[anchor=west] {\tiny{$r$}};
\filldraw[black] (2.2*\y+12*\x-3.5*\y, 3*\y-5*\z-0.3*\y)  node[anchor=west] {\tiny{$m'$}};
\filldraw[black] (1.1*\y+12*\x-3.5*\y, 2.8*\y-5*\z-0.3*\y)  node[anchor=west] {\tiny{$k$}};
\filldraw[black] (2.2*\y+12*\x-3.5*\y, 2.4*\y-5*\z-0.3*\y)  node[anchor=west] {\tiny{$h$}};
\filldraw[black] (2.3*\y+12*\x-3.5*\y, 1.5*\y-5*\z-0.3*\y)  node[anchor=west] {\tiny{$z$}};

\filldraw[black] (0*\y+0*\y-3.5*\y, 0*\y+-10*\z-0.3*\y)  node[anchor=west] {\tiny{$(3,1): G \ \ \textcolor{brown}{[m'', v, o, y]}$}};
\draw[] (0*\y+0*\y, 0*\y+-10*\z) -- (0*\y+0*\y,3.36408*\y+-10*\z);
\draw[] (0*\y+0*\y, 0*\y+-10*\z) -- (-3.02264*\y+0*\y,2.7216*\y+-10*\z);
\draw[] (-3.02264*\y+0*\y, 2.7216*\y+-10*\z) -- (0*\y+0*\y,3.36408*\y+-10*\z);
\draw[] (0*\y+0*\y, 0*\y+-10*\z) -- (-1.97736*\y+0*\y,2.7216*\y+-10*\z);
\draw[] (0*\y+0*\y, 3.36408*\y+-10*\z) -- (-1.97736*\y+0*\y,2.7216*\y+-10*\z);
\draw[] (-3.02264*\y+0*\y, 2.7216*\y+-10*\z) -- (-1.97736*\y+0*\y,2.7216*\y+-10*\z);
\draw[] (-1.97736*\y+0*\y, 2.7216*\y+-10*\z) -- (-0.845653*\y+0*\y,1.46471*\y+-10*\z);
\draw[] (0*\y+0*\y, 0*\y+-10*\z) -- (-0.845653*\y+0*\y,1.46471*\y+-10*\z);
\draw[] (0*\y+0*\y, 3.36408*\y+-10*\z) -- (-0.845653*\y+0*\y,1.46471*\y+-10*\z);

\filldraw[black] (2*\y+0*\y-3.5*\y, 3.7*\y-10*\z-0.3*\y)  node[anchor=west] {\tiny{$a$}};
\filldraw[black] (1.3*\y+0*\y-3.5*\y, 1.5*\y-10*\z-0.3*\y)  node[anchor=west] {\tiny{$b$}};
\filldraw[black] (3.5*\y+0*\y-3.5*\y, 1.7*\y-10*\z-0.3*\y)  node[anchor=west] {\tiny{$c$}};
\filldraw[black] (1.9*\y+0*\y-3.5*\y, 3.1*\y-10*\z-0.3*\y)  node[anchor=west] {\tiny{$m''$}};
\filldraw[black] (1.9*\y+0*\y-3.5*\y, 2.6*\y-10*\z-0.3*\y)  node[anchor=west] {\tiny{$y$}};
\filldraw[black] (2.8*\y+0*\y-3.5*\y, 2.6*\y-10*\z-0.3*\y)  node[anchor=west] {\tiny{$d''$}};
\filldraw[black] (1.9*\y+0*\y-3.5*\y, 1.8*\y-10*\z-0.3*\y)  node[anchor=west] {\tiny{$e''$}};
\filldraw[black] (2.9*\y+0*\y-3.5*\y, 1.1*\y-10*\z-0.3*\y)  node[anchor=west] {\tiny{$o$}};
\filldraw[black] (1*\y+0*\y-3.5*\y, 2.8*\y-10*\z-0.3*\y)  node[anchor=west] {\tiny{$v$}};

\filldraw[black] (0*\y+6*\x-3.5*\y, 0*\y+-10*\z-0.3*\y)  node[anchor=west] {\tiny{$(3,2): H \ \ \textcolor{ao(english)}{[e'', v, w, h]} $}};
\draw[] (0*\y+6*\x, 0*\y+-10*\z) -- (0*\y+6*\x,3.36408*\y+-10*\z);
\draw[] (0*\y+6*\x, 0*\y+-10*\z) -- (-3.02264*\y+6*\x,2.7216*\y+-10*\z);
\draw[] (-3.02264*\y+6*\x, 2.7216*\y+-10*\z) -- (0*\y+6*\x,3.36408*\y+-10*\z);
\draw[] (0*\y+6*\x, 0*\y+-10*\z) -- (-1.97736*\y+6*\x,2.7216*\y+-10*\z);
\draw[] (0*\y+6*\x, 3.36408*\y+-10*\z) -- (-1.97736*\y+6*\x,2.7216*\y+-10*\z);
\draw[] (-3.02264*\y+6*\x, 2.7216*\y+-10*\z) -- (-1.97736*\y+6*\x,2.7216*\y+-10*\z);
\draw[] (-1.97736*\y+6*\x, 2.7216*\y+-10*\z) -- (-0.954915*\y+6*\x,2.93893*\y+-10*\z);
\draw[] (0*\y+6*\x, 0*\y+-10*\z) -- (-0.954915*\y+6*\x,2.93893*\y+-10*\z);
\draw[] (0*\y+6*\x, 3.36408*\y+-10*\z) -- (-0.954915*\y+6*\x,2.93893*\y+-10*\z);

\filldraw[black] (2*\y+6*\x-3.5*\y, 3.7*\y-10*\z-0.3*\y)  node[anchor=west] {\tiny{$a$}};
\filldraw[black] (1.3*\y+6*\x-3.5*\y, 1.5*\y-10*\z-0.3*\y)  node[anchor=west] {\tiny{$b$}};
\filldraw[black] (3.5*\y+6*\x-3.5*\y, 1.7*\y-10*\z-0.3*\y)  node[anchor=west] {\tiny{$c$}};
\filldraw[black] (2.8*\y+6*\x-3.5*\y, 2.3*\y-10*\z-0.3*\y)  node[anchor=west] {\tiny{$d'$}};
\filldraw[black] (2.7*\y+6*\x-3.5*\y, 3.2*\y-10*\z-0.3*\y)  node[anchor=west] {\tiny{$h$}};
\filldraw[black] (1.7*\y+6*\x-3.5*\y, 3.4*\y-10*\z-0.3*\y)  node[anchor=west] {\tiny{$m''$}};
\filldraw[black] (1*\y+6*\x-3.5*\y, 2.8*\y-10*\z-0.3*\y)  node[anchor=west] {\tiny{$v$}};
\filldraw[black] (1.9*\y+6*\x-3.5*\y, 1.8*\y-10*\z-0.3*\y)  node[anchor=west] {\tiny{$e''$}};
\filldraw[black] (1.8*\y+6*\x-3.5*\y, 3*\y-10*\z-0.3*\y)  node[anchor=west] {\tiny{$w$}};

\filldraw[black] (0*\y+12*\x-3.5*\y, 0*\y+-10*\z-0.3*\y)  node[anchor=west] {\tiny{$(3,3): I \ \ [e'', v, x, j]+\textcolor{bole}{[n, v, x, j]}$}};
  \filldraw[black] (0*\y+12*\x-2*\y, 0*\y+-10*\z-1.2*\y)  node[anchor=west] {\tiny{$\frac{\Delta_{j m''n}\Delta_{c j x}}{\Delta_{aj}}$}};
\draw[] (0*\y+12*\x, 0*\y+-10*\z) -- (0*\y+12*\x,3.36408*\y+-10*\z);
\draw[] (0*\y+12*\x, 0*\y+-10*\z) -- (-3.02264*\y+12*\x,2.7216*\y+-10*\z);
\draw[] (-3.02264*\y+12*\x, 2.7216*\y+-10*\z) -- (0*\y+12*\x,3.36408*\y+-10*\z);
\draw[] (0*\y+12*\x, 0*\y+-10*\z) -- (-1.97736*\y+12*\x,2.7216*\y+-10*\z);
\draw[] (0*\y+12*\x, 3.36408*\y+-10*\z) -- (-1.97736*\y+12*\x,2.7216*\y+-10*\z);
\draw[] (-3.02264*\y+12*\x, 2.7216*\y+-10*\z) -- (-1.97736*\y+12*\x,2.7216*\y+-10*\z);
\draw[] (-1.97736*\y+12*\x, 2.7216*\y+-10*\z) -- (-0.17679*\y+12*\x,1.68204*\y+-10*\z);
\draw[] (0*\y+12*\x, 0*\y+-10*\z) -- (-0.17679*\y+12*\x,1.68204*\y+-10*\z);
\draw[] (0*\y+12*\x, 3.36408*\y+-10*\z) -- (-0.17679*\y+12*\x,1.68204*\y+-10*\z);

\filldraw[black] (2*\y+12*\x-3.5*\y, 3.7*\y-10*\z-0.3*\y)  node[anchor=west] {\tiny{$a$}};
\filldraw[black] (1.3*\y+12*\x-3.5*\y, 1.5*\y-10*\z-0.3*\y)  node[anchor=west] {\tiny{$b$}};
\filldraw[black] (3.5*\y+12*\x-3.5*\y, 1.7*\y-10*\z-0.3*\y)  node[anchor=west] {\tiny{$c$}};
\filldraw[black] (2.3*\y+12*\x-3.5*\y, 1.7*\y-10*\z-0.3*\y)  node[anchor=west] {\tiny{$e''$}};
\filldraw[black] (2.2*\y+12*\x-3.5*\y, 3.2*\y-10*\z-0.3*\y)  node[anchor=west] {\tiny{$m''$}};
\filldraw[black] (2.3*\y+12*\x-3.5*\y, 2.6*\y-10*\z-0.3*\y)  node[anchor=west] {\tiny{$n$}};
\filldraw[black] (1*\y+12*\x-3.5*\y, 2.8*\y-10*\z-0.3*\y)  node[anchor=west] {\tiny{$v$}};
\filldraw[black] (3*\y+12*\x-3.5*\y, 2.7*\y-10*\z-0.3*\y)  node[anchor=west] {\tiny{$j$}};
\filldraw[black] (3*\y+12*\x-3.5*\y, 1.4*\y-10*\z-0.3*\y)  node[anchor=west] {\tiny{$x$}};

\filldraw[black] (0*\y+0*\y-3.5*\y, 0*\y+-17*\z-0.3*\y)  node[anchor=west] {\tiny{$(4,1): D \ \ [m', r, o, e']$}};
  \filldraw[black] (0*\y+0*\y-2*\y, 0*\y-17*\z-1.1*\y)  node[anchor=west] {\tiny{$\frac{\Delta_{d''ru}\Delta_{cd''o}}{\Delta_{ae'd''}}$}};

\draw[] (0*\y+0*\y, 0*\y+-17*\z) -- (0*\y+0*\y,3.36408*\y+-17*\z);
\draw[] (0*\y+0*\y, 0*\y+-17*\z) -- (-3.02264*\y+0*\y,2.7216*\y+-17*\z);
\draw[] (-3.02264*\y+0*\y, 2.7216*\y+-17*\z) -- (0*\y+0*\y,3.36408*\y+-17*\z);
\draw[] (0*\y+0*\y, 0*\y+-17*\z) -- (-0.845653*\y+0*\y,1.46471*\y+-17*\z);
\draw[] (0*\y+0*\y, 3.36408*\y+-17*\z) -- (-0.845653*\y+0*\y,1.46471*\y+-17*\z);
\draw[] (-3.02264*\y+0*\y, 2.7216*\y+-17*\z) -- (-0.845653*\y+0*\y,1.46471*\y+-17*\z);
\draw[] (-0.845653*\y+0*\y, 1.46471*\y+-17*\z) -- (-0.522642*\y+0*\y,2.45884*\y+-17*\z);
\draw[] (0*\y+0*\y, 3.36408*\y+-17*\z) -- (-0.522642*\y+0*\y,2.45884*\y+-17*\z);
\draw[] (-3.02264*\y+0*\y, 2.7216*\y+-17*\z) -- (-0.522642*\y+0*\y,2.45884*\y+-17*\z);
\draw[] (-0.845653*\y+0*\y, 1.46471*\y+-17*\z) -- (-0.845653*\y+0*\y,1.46471*\y+-17*\z);
\draw[] (-3.02264*\y+0*\y, 2.7216*\y+-17*\z) -- (-0.845653*\y+0*\y,1.46471*\y+-17*\z);
\draw[] (0*\y+0*\y, 0*\y+-17*\z) -- (-0.845653*\y+0*\y,1.46471*\y+-17*\z);
\draw[] (-0.845653*\y+0*\y, 1.46471*\y+-17*\z) -- (-0.845653*\y+0*\y,1.46471*\y+-17*\z);
\draw[] (0*\y+0*\y, 0*\y+-17*\z) -- (-0.845653*\y+0*\y,1.46471*\y+-17*\z);
\draw[] (0*\y+0*\y, 3.36408*\y+-17*\z) -- (-0.845653*\y+0*\y,1.46471*\y+-17*\z);

\filldraw[black] (2*\y+0*\y-3.5*\y, 3.7*\y-17*\z-0.3*\y)  node[anchor=west] {\tiny{$a$}};
\filldraw[black] (1.3*\y+0*\y-3.5*\y, 1.5*\y-17*\z-0.3*\y)  node[anchor=west] {\tiny{$b$}};
\filldraw[black] (3.5*\y+0*\y-3.5*\y, 1.7*\y-17*\z-0.3*\y)  node[anchor=west] {\tiny{$c$}};
\filldraw[black] (2.8*\y+0*\y-3.5*\y, 2.6*\y-17*\z-0.3*\y)  node[anchor=west] {\tiny{$d''$}};
\filldraw[black] (1.9*\y+0*\y-3.5*\y, 1.8*\y-17*\z-0.3*\y)  node[anchor=west] {\tiny{$e'$}};
\filldraw[black] (2.9*\y+0*\y-3.5*\y, 1.1*\y-17*\z-0.3*\y)  node[anchor=west] {\tiny{$o$}};
\filldraw[black] (2.8*\y+0*\y-3.5*\y, 3.3*\y-17*\z-0.3*\y)  node[anchor=west] {\tiny{$r$}};
\filldraw[black] (2.2*\y+0*\y-3.5*\y, 3*\y-17*\z-0.3*\y)  node[anchor=west] {\tiny{$m'$}};
\filldraw[black] (2.4*\y+0*\y-3.5*\y, 2.3*\y-17*\z-0.3*\y)  node[anchor=west] {\tiny{$u$}};

\filldraw[black] (0*\y+6*\x-3.5*\y, 0*\y+-17*\z-0.3*\y)  node[anchor=west] {\tiny{$(4,2): G \ \  [m'',v,o,e']+\textcolor{brown}{[m'', v, o, y]}$}};
    \filldraw[black] (0*\y+6*\x-2*\y, 0*\y-17*\z-1.1*\y)  node[anchor=west] {\tiny{$\frac{\Delta_{d'' m''y}\Delta_{cd''o}}{\Delta_{ae'd''}}$}};

\draw[] (0*\y+6*\x, 0*\y+-17*\z) -- (0*\y+6*\x,3.36408*\y+-17*\z);
\draw[] (0*\y+6*\x, 0*\y+-17*\z) -- (-3.02264*\y+6*\x,2.7216*\y+-17*\z);
\draw[] (-3.02264*\y+6*\x, 2.7216*\y+-17*\z) -- (0*\y+6*\x,3.36408*\y+-17*\z);
\draw[] (0*\y+6*\x, 0*\y+-17*\z) -- (-0.845653*\y+6*\x,1.46471*\y+-17*\z);
\draw[] (0*\y+6*\x, 3.36408*\y+-17*\z) -- (-0.845653*\y+6*\x,1.46471*\y+-17*\z);
\draw[] (-0.845653*\y+6*\x, 1.46471*\y+-17*\z) -- (-1.97736*\y+6*\x,2.7216*\y+-17*\z);
\draw[] (0*\y+6*\x, 3.36408*\y+-17*\z) -- (-1.97736*\y+6*\x,2.7216*\y+-17*\z);
\draw[] (-3.02264*\y+6*\x, 2.7216*\y+-17*\z) -- (-1.97736*\y+6*\x,2.7216*\y+-17*\z);
\draw[] (-3.02264*\y+6*\x, 2.7216*\y+-17*\z) -- (-0.845653*\y+6*\x,1.46471*\y+-17*\z);

\filldraw[black] (2*\y+6*\x-3.5*\y, 3.7*\y-17*\z-0.3*\y)  node[anchor=west] {\tiny{$a$}};
\filldraw[black] (1.3*\y+6*\x-3.5*\y, 1.5*\y-17*\z-0.3*\y)  node[anchor=west] {\tiny{$b$}};
\filldraw[black] (3.5*\y+6*\x-3.5*\y, 1.7*\y-17*\z-0.3*\y)  node[anchor=west] {\tiny{$c$}};
\filldraw[black] (2.8*\y+6*\x-3.5*\y, 2.6*\y-17*\z-0.3*\y)  node[anchor=west] {\tiny{$d''$}};
\filldraw[black] (1.5*\y+6*\x-3.5*\y, 2.1*\y-17*\z-0.3*\y)  node[anchor=west] {\tiny{$e'$}};
\filldraw[black] (2.9*\y+6*\x-3.5*\y, 1.1*\y-17*\z-0.3*\y)  node[anchor=west] {\tiny{$o$}};
\filldraw[black] (1.9*\y+6*\x-3.5*\y, 2.6*\y-17*\z-0.3*\y)  node[anchor=west] {\tiny{$y$}};
\filldraw[black] (1*\y+6*\x-3.5*\y, 2.8*\y-17*\z-0.3*\y)  node[anchor=west] {\tiny{$v$}};
\filldraw[black] (2*\y+6*\x-3.5*\y, 3.2*\y-17*\z-0.3*\y)  node[anchor=west] {\tiny{$m''$}};

\filldraw[black] (0*\y+12*\x-3.5*\y, 0*\y+-17*\z-0.3*\y)  node[anchor=west] {\tiny{$(4,3): C \ \ [p,q,o,e']$}};
    \filldraw[black] (0*\y+12*\x-2*\y, 0*\y-17*\z-1.1*\y)  node[anchor=west] {\tiny{$\frac{\Delta_{d'' f p}\Delta_{cd''o}}{\Delta_{ae'd''}}$}};
\draw[] (0*\y+12*\x, 0*\y+-17*\z) -- (0*\y+12*\x,3.36408*\y+-17*\z);
\draw[] (0*\y+12*\x, 0*\y+-17*\z) -- (-3.02264*\y+12*\x,2.7216*\y+-17*\z);
\draw[] (-3.02264*\y+12*\x, 2.7216*\y+-17*\z) -- (0*\y+12*\x,3.36408*\y+-17*\z);
\draw[] (-0.945653*\y+12*\x, 1.46471*\y+-17*\z) -- (-1.3683*\y+12*\x,2.36996*\y+-17*\z);
\draw[] (0*\y+12*\x, 3.36408*\y+-17*\z) -- (-1.3683*\y+12*\x,2.36996*\y+-17*\z);
\draw[] (-3.02264*\y+12*\x, 2.7216*\y+-17*\z) -- (-1.3683*\y+12*\x,2.36996*\y+-17*\z);
\draw[] (-3.02264*\y+12*\x, 2.7216*\y+-17*\z) -- (-0.945653*\y+12*\x,1.46471*\y+-17*\z);
\draw[] (0*\y+12*\x, 0*\y+-17*\z) -- (-0.945653*\y+12*\x,1.46471*\y+-17*\z);
\draw[] (0*\y+12*\x, 3.36408*\y+-17*\z) -- (-0.945653*\y+12*\x,1.46471*\y+-17*\z);

\filldraw[black] (2*\y+12*\x-3.5*\y, 3.7*\y-17*\z-0.3*\y)  node[anchor=west] {\tiny{$a$}};
\filldraw[black] (1.3*\y+12*\x-3.5*\y, 1.5*\y-17*\z-0.3*\y)  node[anchor=west] {\tiny{$b$}};
\filldraw[black] (3.5*\y+12*\x-3.5*\y, 1.7*\y-17*\z-0.3*\y)  node[anchor=west] {\tiny{$c$}};
\filldraw[black] (2.8*\y+12*\x-3.5*\y, 2.6*\y-17*\z-0.3*\y)  node[anchor=west] {\tiny{$d''$}};
\filldraw[black] (1.6*\y+12*\x-3.5*\y, 2.1*\y-17*\z-0.3*\y)  node[anchor=west] {\tiny{$e'$}};
\filldraw[black] (2.9*\y+12*\x-3.5*\y, 1.1*\y-17*\z-0.3*\y)  node[anchor=west] {\tiny{$o$}};
\filldraw[black] (2*\y+12*\x-3.5*\y, 2.3*\y-17*\z-0.3*\y)  node[anchor=west] {\tiny{$f$}};
\filldraw[black] (2.3*\y+12*\x-3.5*\y, 3.2*\y-17*\z-0.3*\y)  node[anchor=west] {\tiny{$p$}};
\filldraw[black] (1.4*\y+12*\x-3.5*\y, 2.9*\y-17*\z-0.3*\y)  node[anchor=west] {\tiny{$q$}};

\filldraw[black] (0*\y+0*\y-3.5*\y, 0*\y+-22*\z-0.3*\y)  node[anchor=west] {\tiny{$(5,1): A \ \ [e,f,i,h]$}};
  \filldraw[black] (0*\y+0*\y-2*\y, 0*\y-22*\z-1.1*\y)  node[anchor=west] {\tiny{$\frac{\Delta_{d'fg}\Delta_{cd'h}}{\Delta_{bd' i}}$}};
\draw[] (0*\y+0*\y, 0*\y+-22*\z) -- (0*\y+0*\y,3.36408*\y-22*\z);
\draw[] (0*\y+0*\y, 0*\y-22*\z) -- (-3.02264*\y+0*\y,2.7216*\y-22*\z);
\draw[] (-3.02264*\y+0*\y, 2.7216*\y-22*\z) -- (0*\y+0*\y,3.36408*\y-22*\z);
\draw[] (0*\y+0*\y, 0*\y-22*\z) -- (-0.954915*\y+0*\y,2.93893*\y-22*\z);
\draw[] (0*\y+0*\y, 3.36408*\y-22*\z) -- (-0.954915*\y+0*\y,2.93893*\y-22*\z);
\draw[] (-3.02264*\y+0*\y, 2.7216*\y-22*\z) -- (-0.954915*\y+0*\y,2.93893*\y-22*\z);
\draw[] (-0.954915*\y+0*\y, 2.93893*\y-22*\z) -- (-0.954915*\y+0*\y,2.93893*\y-22*\z);
\draw[] (0*\y+0*\y, 3.36408*\y-22*\z) -- (-0.954915*\y+0*\y,2.93893*\y-22*\z);
\draw[] (-3.02264*\y+0*\y, 2.7216*\y-22*\z) -- (-0.954915*\y+0*\y,2.93893*\y-22*\z);
\draw[] (-0.954915*\y+0*\y, 2.93893*\y-22*\z) -- (-0.522642*\y+0*\y,0.905243*\y-22*\z);
\draw[] (-3.02264*\y+0*\y, 2.7216*\y-22*\z) -- (-0.522642*\y+0*\y,0.905243*\y-22*\z);
\draw[] (0*\y+0*\y, 0*\y-22*\z) -- (-0.522642*\y+0*\y,0.905243*\y-22*\z);
\draw[] (-0.954915*\y+0*\y, 2.93893*\y-22*\z) -- (-0.954915*\y+0*\y,2.93893*\y-22*\z);
\draw[] (0*\y+0*\y, 0*\y-22*\z) -- (-0.954915*\y+0*\y,2.93893*\y-22*\z);
\draw[] (0*\y+0*\y, 3.36408*\y-22*\z) -- (-0.954915*\y+0*\y,2.93893*\y-22*\z);

\filldraw[black] (2*\y+0*\y-3.5*\y, 3.7*\y-22*\z-0.3*\y)  node[anchor=west] {\tiny{$a$}};
\filldraw[black] (1.3*\y+0*\y-3.5*\y, 1.5*\y-22*\z-0.3*\y)  node[anchor=west] {\tiny{$b$}};
\filldraw[black] (3.5*\y+0*\y-3.5*\y, 1.7*\y-22*\z-0.3*\y)  node[anchor=west] {\tiny{$c$}};
\filldraw[black] (2.8*\y+0*\y-3.5*\y, 2.3*\y-22*\z-0.3*\y)  node[anchor=west] {\tiny{$d'$}};
\filldraw[black] (2.1*\y+0*\y-3.5*\y, 1.8*\y-22*\z-0.3*\y)  node[anchor=west] {\tiny{$e$}};
\filldraw[black] (2.7*\y+0*\y-3.5*\y, 1.1*\y-22*\z-0.3*\y)  node[anchor=west] {\tiny{$f$}};
\filldraw[black] (2.3*\y+0*\y-3.5*\y, 2.3*\y-22*\z-0.3*\y)  node[anchor=west] {\tiny{$g$}};
\filldraw[black] (2.7*\y+0*\y-3.5*\y, 3.2*\y-22*\z-0.3*\y)  node[anchor=west] {\tiny{$h$}};
\filldraw[black] (1.9*\y+0*\y-3.5*\y, 3.1*\y-22*\z-0.3*\y)  node[anchor=west] {\tiny{$i$}};

\filldraw[black] (0*\y+6*\x-3.5*\y, 0*\y+-22*\z-0.3*\y)  node[anchor=west] {\tiny{$(5,2): H \ \ [e'',v, i, h] + \textcolor{ao(english)}{[e'', v, w, h]} $}};
    \filldraw[black] (0*\y+6*\x-2*\y, 0*\y-22*\z-1.1*\y)  node[anchor=west] {\tiny{$\frac{\Delta_{d'e'' w}\Delta_{cd'h}}{\Delta_{bd' i}}$}};
\draw[] (0*\y+6*\x, 0*\y+-22*\z) -- (0*\y+6*\x,3.36408*\y+-22*\z);
\draw[] (0*\y+6*\x, 0*\y+-22*\z) -- (-3.02264*\y+6*\x,2.7216*\y+-22*\z);
\draw[] (-3.02264*\y+6*\x, 2.7216*\y+-22*\z) -- (0*\y+6*\x,3.36408*\y+-22*\z);
\draw[] (0*\y+6*\x, 0*\y+-22*\z) -- (-0.954915*\y+6*\x,2.93893*\y+-22*\z);
\draw[] (0*\y+6*\x, 3.36408*\y+-22*\z) -- (-0.954915*\y+6*\x,2.93893*\y+-22*\z);
\draw[] (-3.02264*\y+6*\x, 2.7216*\y+-22*\z) -- (-0.954915*\y+6*\x,2.93893*\y+-22*\z);
\draw[] (-0.954915*\y+6*\x, 2.93893*\y+-22*\z) -- (-1.97736*\y+6*\x,2.7216*\y+-22*\z);
\draw[] (-3.02264*\y+6*\x, 2.7216*\y+-22*\z) -- (-1.97736*\y+6*\x,2.7216*\y+-22*\z);
\draw[] (0*\y+6*\x, 0*\y+-22*\z) -- (-1.97736*\y+6*\x,2.7216*\y+-22*\z);

\filldraw[black] (2*\y+6*\x-3.5*\y, 3.7*\y-22*\z-0.3*\y)  node[anchor=west] {\tiny{$a$}};
\filldraw[black] (1.3*\y+6*\x-3.5*\y, 1.5*\y-22*\z-0.3*\y)  node[anchor=west] {\tiny{$b$}};
\filldraw[black] (3.5*\y+6*\x-3.5*\y, 1.7*\y-22*\z-0.3*\y)  node[anchor=west] {\tiny{$c$}};
\filldraw[black] (2.8*\y+6*\x-3.5*\y, 2.3*\y-22*\z-0.3*\y)  node[anchor=west] {\tiny{$d'$}};
\filldraw[black] (2.7*\y+6*\x-3.5*\y, 3.2*\y-22*\z-0.3*\y)  node[anchor=west] {\tiny{$h$}};
\filldraw[black] (1.7*\y+6*\x-3.5*\y, 3.3*\y-22*\z-0.3*\y)  node[anchor=west] {\tiny{$i$}};
\filldraw[black] (1*\y+6*\x-3.5*\y, 2.8*\y-22*\z-0.3*\y)  node[anchor=west] {\tiny{$v$}};
\filldraw[black] (1.8*\y+6*\x-3.5*\y, 3*\y-22*\z-0.3*\y)  node[anchor=west] {\tiny{$w$}};
\filldraw[black] (1.8*\y+6*\x-3.5*\y, 2*\y-22*\z-0.3*\y)  node[anchor=west] {\tiny{$e''$}};

\filldraw[black] (0*\y+12*\x-3.5*\y, 0*\y+-22*\z-0.3*\y)  node[anchor=west] {\tiny{$(5,3): F \ \ [z,k,i,h]$}};
    \filldraw[black] (0*\y+12*\x-2*\y, 0*\y-22*\z-1.1*\y)  node[anchor=west] {\tiny{$\frac{\Delta_{d'r z}\Delta_{cd'h}}{\Delta_{bd' i}}$}};
\draw[] (0*\y+12*\x, 0*\y+-22*\z) -- (0*\y+12*\x,3.36408*\y+-22*\z);
\draw[] (0*\y+12*\x, 0*\y+-22*\z) -- (-3.02264*\y+12*\x,2.7216*\y+-22*\z);
\draw[] (-3.02264*\y+12*\x, 2.7216*\y+-22*\z) -- (0*\y+12*\x,3.36408*\y+-22*\z);
\draw[] (0*\y+12*\x, 0*\y+-22*\z) -- (-0.954915*\y+12*\x,2.93893*\y+-22*\z);
\draw[] (0*\y+12*\x, 3.36408*\y+-22*\z) -- (-0.954915*\y+12*\x,2.93893*\y+-22*\z);
\draw[] (-3.02264*\y+12*\x, 2.7216*\y+-22*\z) -- (-0.954915*\y+12*\x,2.93893*\y+-22*\z);
\draw[] (-0.954915*\y+12*\x, 2.93893*\y+-22*\z) -- (-1.47756*\y+12*\x,2.03368*\y+-22*\z);
\draw[] (-3.02264*\y+12*\x, 2.7216*\y+-22*\z) -- (-1.47756*\y+12*\x,2.03368*\y+-22*\z);
\draw[] (0*\y+12*\x, 0*\y+-22*\z) -- (-1.47756*\y+12*\x,2.03368*\y+-22*\z);

\filldraw[black] (2*\y+12*\x-3.5*\y, 3.7*\y-22*\z-0.3*\y)  node[anchor=west] {\tiny{$a$}};
\filldraw[black] (1.3*\y+12*\x-3.5*\y, 1.5*\y-22*\z-0.3*\y)  node[anchor=west] {\tiny{$b$}};
\filldraw[black] (3.5*\y+12*\x-3.5*\y, 1.7*\y-22*\z-0.3*\y)  node[anchor=west] {\tiny{$c$}};
\filldraw[black] (2.8*\y+12*\x-3.5*\y, 2.3*\y-22*\z-0.3*\y)  node[anchor=west] {\tiny{$d'$}};
\filldraw[black] (2.7*\y+12*\x-3.5*\y, 3.2*\y-22*\z-0.3*\y)  node[anchor=west] {\tiny{$h$}};
\filldraw[black] (1.6*\y+12*\x-3.5*\y, 3.1*\y-22*\z-0.3*\y)  node[anchor=west] {\tiny{$i$}};
\filldraw[black] (1.1*\y+12*\x-3.5*\y, 2.8*\y-22*\z-0.3*\y)  node[anchor=west] {\tiny{$k$}};
\filldraw[black] (1.8*\y+12*\x-3.5*\y, 2.8*\y-22*\z-0.3*\y)  node[anchor=west] {\tiny{$r$}};
\filldraw[black] (2.3*\y+12*\x-3.5*\y, 1.5*\y-22*\z-0.3*\y)  node[anchor=west] {\tiny{$z$}};

\filldraw[black] (0*\y+0*\y-3.5*\y, 0*\y+-27*\z-0.3*\y)  node[anchor=west] {\tiny{$(6,1): B \ \ \textcolor{blue}{[n,m,f,l]}$}};
\draw[] (0*\y+0*\y, 0*\y-27*\z) -- (0*\y+0*\y,3.36408*\y-27*\z);
\draw[] (0*\y+0*\y, 0*\y-27*\z) -- (-3.02264*\y+0*\y,2.7216*\y-27*\z);
\draw[] (-3.02264*\y+0*\y, 2.7216*\y-27*\z) -- (0*\y+0*\y,3.36408*\y-27*\z);
\draw[] (0*\y+0*\y, 0*\y-27*\z) -- (-1.22207*\y+0*\y,1.68204*\y-27*\z);
\draw[] (0*\y+0*\y, 3.36408*\y-27*\z) -- (-1.22207*\y+0*\y,1.68204*\y-27*\z);
\draw[] (-3.02264*\y+0*\y, 2.7216*\y-27*\z) -- (-1.22207*\y+0*\y,1.68204*\y-27*\z);
\draw[] (-1.22207*\y+0*\y, 1.68204*\y-27*\z) -- (-1.22207*\y+0*\y,1.68204*\y-27*\z);
\draw[] (0*\y+0*\y, 3.36408*\y-27*\z) -- (-1.22207*\y+0*\y,1.68204*\y-27*\z);
\draw[] (-3.02264*\y+0*\y, 2.7216*\y-27*\z) -- (-1.22207*\y+0*\y,1.68204*\y-27*\z);
\draw[] (-1.22207*\y+0*\y, 1.68204*\y-27*\z) -- (-1.22207*\y+0*\y,1.68204*\y-27*\z);
\draw[] (-3.02264*\y+0*\y, 2.7216*\y-27*\z) -- (-1.22207*\y+0*\y,1.68204*\y-27*\z);
\draw[] (0*\y+0*\y, 0*\y-27*\z) -- (-1.22207*\y+0*\y,1.68204*\y-27*\z);
\draw[] (-1.22207*\y+0*\y, 1.68204*\y-27*\z) -- (-0.522642*\y+0*\y,0.905243*\y-27*\z);
\draw[] (0*\y+0*\y, 0*\y-27*\z) -- (-0.522642*\y+0*\y,0.905243*\y-27*\z);
\draw[] (0*\y+0*\y, 3.36408*\y-27*\z) -- (-0.522642*\y+0*\y,0.905243*\y-27*\z);

\filldraw[black] (2*\y+0*\y-3.5*\y, 3.7*\y-27*\z-0.3*\y)  node[anchor=west] {\tiny{$a$}};
\filldraw[black] (1.3*\y+0*\y-3.5*\y, 1.5*\y-27*\z-0.3*\y)  node[anchor=west] {\tiny{$b$}};
\filldraw[black] (3.5*\y+0*\y-3.5*\y, 1.7*\y-27*\z-0.3*\y)  node[anchor=west] {\tiny{$c$}};
\filldraw[black] (2.8*\y+0*\y-3.5*\y, 2.6*\y-27*\z-0.3*\y)  node[anchor=west] {\tiny{$d$}};
\filldraw[black] (2.3*\y+0*\y-3.5*\y, 1.2*\y-27*\z-0.3*\y)  node[anchor=west] {\tiny{$z'$}};
\filldraw[black] (2.9*\y+0*\y-3.5*\y, 1.1*\y-27*\z-0.3*\y)  node[anchor=west] {\tiny{$f$}};
\filldraw[black] (2.4*\y+0*\y-3.5*\y, 1.8*\y-27*\z-0.3*\y)  node[anchor=west] {\tiny{$l$}};
\filldraw[black] (2.2*\y+0*\y-3.5*\y, 2.8*\y-27*\z-0.3*\y)  node[anchor=west] {\tiny{$m$}};
\filldraw[black] (1.5*\y+0*\y-3.5*\y, 2.4*\y-27*\z-0.3*\y)  node[anchor=west] {\tiny{$n$}};

\filldraw[black] (0*\y+6*\x-3.5*\y, 0*\y-27*\z-0.3*\y)  node[anchor=west] {\tiny{$(6,2): E \ \ \textcolor{red}{[r,t,z',n]}$}};
\draw[] (0*\y+6*\x, 0*\y-27*\z) -- (0*\y+6*\x,3.36408*\y-27*\z);
\draw[] (0*\y+6*\x, 0*\y-27*\z) -- (-3.02264*\y+6*\x,2.7216*\y-27*\z);
\draw[] (-3.02264*\y+6*\x, 2.7216*\y-27*\z) -- (0*\y+6*\x,3.36408*\y-27*\z);
\draw[] (0*\y+6*\x, 0*\y-27*\z) -- (-1.22207*\y+6*\x,1.68204*\y-27*\z);
\draw[] (0*\y+6*\x, 3.36408*\y-27*\z) -- (-1.22207*\y+6*\x,1.68204*\y-27*\z);
\draw[] (-3.02264*\y+6*\x, 2.7216*\y-27*\z) -- (-1.22207*\y+6*\x,1.68204*\y-27*\z);
\draw[] (-1.22207*\y+6*\x, 1.68204*\y-27*\z) -- (-0.522642*\y+6*\x,2.45884*\y-27*\z);
\draw[] (0*\y+6*\x, 0*\y-27*\z) -- (-0.522642*\y+6*\x,2.45884*\y-27*\z);
\draw[] (0*\y+6*\x, 3.36408*\y-27*\z) -- (-0.522642*\y+6*\x,2.45884*\y-27*\z);

\filldraw[black] (2*\y+6*\x-3.5*\y, 3.7*\y-27*\z-0.3*\y)  node[anchor=west] {\tiny{$a$}};
\filldraw[black] (1.3*\y+6*\x-3.5*\y, 1.5*\y-27*\z-0.3*\y)  node[anchor=west] {\tiny{$b$}};
\filldraw[black] (3.5*\y+6*\x-3.5*\y, 1.7*\y-27*\z-0.3*\y)  node[anchor=west] {\tiny{$c$}};
\filldraw[black] (2.5*\y+6*\x-3.5*\y, 1.4*\y-27*\z-0.3*\y)  node[anchor=west] {\tiny{$z'$}};
\filldraw[black] (2.2*\y+6*\x-3.5*\y, 2.8*\y-27*\z-0.3*\y)  node[anchor=west] {\tiny{$m$}};
\filldraw[black] (1.5*\y+6*\x-3.5*\y, 2.4*\y-27*\z-0.3*\y)  node[anchor=west] {\tiny{$n$}};
\filldraw[black] (2.4*\y+6*\x-3.5*\y, 2.2*\y-27*\z-0.3*\y)  node[anchor=west] {\tiny{$t$}};
\filldraw[black] (3*\y+6*\x-3.5*\y, 3*\y-27*\z-0.3*\y)  node[anchor=west] {\tiny{$r$}};
\filldraw[black] (2.8*\y+6*\x-3.5*\y, 1.8*\y-27*\z-0.3*\y)  node[anchor=west] {\tiny{$d'''$}};

\filldraw[black] (0*\y+12*\x-3.5*\y, 0*\y-27*\z-0.3*\y)  node[anchor=west] {\tiny{$(6,3): I \ \ [n,z',x,j]$}};
    \filldraw[black] (0*\y+12*\x-2*\y, 0*\y-27*\z-1.2*\y)  node[anchor=west] {\tiny{$\frac{\Delta_{j m v}\Delta_{cjx}}{\Delta_{aj}}$}};
\draw[] (0*\y+12*\x, 0*\y-27*\z) -- (0*\y+12*\x,3.36408*\y-27*\z);
\draw[] (0*\y+12*\x, 0*\y-27*\z) -- (-3.02264*\y+12*\x,2.7216*\y-27*\z);
\draw[] (-3.02264*\y+12*\x, 2.7216*\y-27*\z) -- (0*\y+12*\x,3.36408*\y-27*\z);
\draw[] (0*\y+12*\x, 0*\y-27*\z) -- (-1.22207*\y+12*\x,1.68204*\y-27*\z);
\draw[] (0*\y+12*\x, 3.36408*\y-27*\z) -- (-1.22207*\y+12*\x,1.68204*\y-27*\z);
\draw[] (-3.02264*\y+12*\x, 2.7216*\y-27*\z) -- (-1.22207*\y+12*\x,1.68204*\y-27*\z);
\draw[] (-1.22207*\y+12*\x, 1.68204*\y-27*\z) -- (-0.17679*\y+12*\x,1.68204*\y-27*\z);
\draw[] (0*\y+12*\x, 0*\y-27*\z) -- (-0.17679*\y+12*\x,1.68204*\y-27*\z);
\draw[] (0*\y+12*\x, 3.36408*\y-27*\z) -- (-0.17679*\y+12*\x,1.68204*\y-27*\z);

\filldraw[black] (2*\y+12*\x-3.5*\y, 3.7*\y-27*\z-0.3*\y)  node[anchor=west] {\tiny{$a$}};
\filldraw[black] (1.3*\y+12*\x-3.5*\y, 1.5*\y-27*\z-0.3*\y)  node[anchor=west] {\tiny{$b$}};
\filldraw[black] (3.5*\y+12*\x-3.5*\y, 1.7*\y-27*\z-0.3*\y)  node[anchor=west] {\tiny{$c$}};
\filldraw[black] (2.6*\y+12*\x-3.5*\y, 1.5*\y-27*\z-0.3*\y)  node[anchor=west] {\tiny{$z'$}};
\filldraw[black] (2.2*\y+12*\x-3.5*\y, 2.8*\y-27*\z-0.3*\y)  node[anchor=west] {\tiny{$m$}};
\filldraw[black] (1.5*\y+12*\x-3.5*\y, 2.4*\y-27*\z-0.3*\y)  node[anchor=west] {\tiny{$n$}};
\filldraw[black] (2.6*\y+12*\x-3.5*\y, 2.2*\y-27*\z-0.3*\y)  node[anchor=west] {\tiny{$v$}};
\filldraw[black] (3*\y+12*\x-3.5*\y, 2.7*\y-27*\z-0.3*\y)  node[anchor=west] {\tiny{$j$}};
\filldraw[black] (3*\y+12*\x-3.5*\y, 1.4*\y-27*\z-0.3*\y)  node[anchor=west] {\tiny{$x$}};



\filldraw[black] (0*\y+18*\x-3.5*\y, 0*\y-0.3*\y-9*\z)  node[anchor=west] {\tiny{$(7,1): C \ \ \textcolor{orange}{[p,q,f,o]}$}};
\draw[] (0*\y+18*\x, 0*\y-9*\z) -- (0*\y+18*\x,3.36408*\y-9*\z);
\draw[] (0*\y+18*\x, 0*\y-9*\z) -- (-3.02264*\y+18*\x,2.7216*\y-9*\z);
\draw[] (-3.02264*\y+18*\x, 2.7216*\y-9*\z) -- (0*\y+18*\x,3.36408*\y-9*\z);
\draw[] (0*\y+18*\x, 0*\y-9*\z) -- (-0.42*\y+18*\x,0.905243*\y-9*\z);
\draw[] (-1.3683*\y+0.42*\y+18*\x, +2.36996*\y-0.905243*\y-9*\z) -- (-1.3683*\y+18*\x,+2.36996*\y-9*\z);
\draw[] (-0.42*\y+18*\x, 0.905243*\y-9*\z) -- (-1.3683*\y+18*\x,2.36996*\y-9*\z);
\draw[] (-1.3683*\y+0.42*\y+18*\x, +2.36996*\y-0.905243*\y-9*\z) -- (0*\y+18*\x, 0*\y-9*\z);
\draw[] (-1.3683*\y+0.42*\y+18*\x, +2.36996*\y-0.905243*\y-9*\z) -- (-3.02264*\y+18*\x,2.7216*\y-9*\z);
\draw[] (0*\y+18*\x, 3.36408*\y-9*\z) -- (-0.42*\y+18*\x,0.905243*\y-9*\z);
\draw[] (0*\y+18*\x, 3.36408*\y-9*\z) -- (-1.3683*\y+18*\x,2.36996*\y-9*\z);
\draw[] (-3.02264*\y+18*\x, 2.7216*\y-9*\z) -- (-1.3683*\y+18*\x,2.36996*\y-9*\z);

\filldraw[black] (2*\y+18*\x-3.5*\y, 3.7*\y+0*\y-0.3*\y-9*\z)  node[anchor=west] {\tiny{$a$}};
\filldraw[black] (1.3*\y+18*\x-3.5*\y, 1.5*\y+0*\y-0.3*\y-9*\z)  node[anchor=west] {\tiny{$b$}};
\filldraw[black] (3.5*\y+18*\x-3.5*\y, 1.7*\y+0*\y-0.3*\y-9*\z)  node[anchor=west] {\tiny{$c$}};
\filldraw[black] (2.8*\y+18*\x-3.5*\y, 2.6*\y+0*\y-0.3*\y-9*\z)  node[anchor=west] {\tiny{$d$}};
\filldraw[black] (1.3*\y+18*\x-3.5*\y, 2.4*\y+0*\y-0.3*\y-9*\z)  node[anchor=west] {\tiny{$e'$}};
\filldraw[black] (2.9*\y+18*\x-3.5*\y, 1.1*\y+0*\y-0.3*\y-9*\z)  node[anchor=west] {\tiny{$f$}};
\filldraw[black] (2.3*\y+18*\x-3.5*\y, 2.3*\y+0*\y-0.3*\y-9*\z)  node[anchor=west] {\tiny{$o$}};
\filldraw[black] (2.3*\y+18*\x-3.5*\y, 1.4*\y+0*\y-0.3*\y-9*\z)  node[anchor=west] {\tiny{$o$}};
\filldraw[black] (1.8*\y+18*\x-3.5*\y, 2.3*\y+0*\y-0.3*\y-9*\z)  node[anchor=west] {\tiny{$f$}};
\filldraw[black] (2.3*\y+18*\x-3.5*\y, 3.2*\y+0*\y-0.3*\y-9*\z)  node[anchor=west] {\tiny{$p$}};
\filldraw[black] (1.4*\y+18*\x-3.5*\y, 2.9*\y+0*\y-0.3*\y-9*\z)  node[anchor=west] {\tiny{$q$}};

\filldraw[black] (0*\y+18*\x-3.5*\y, 0*\y+-14*\z-0.3*\y)  node[anchor=west] {\tiny{$(7,2): F \ \ \textcolor{purple}{[z, k, r, h]}$}};
\draw[] (0*\y+18*\x, 0*\y+-14*\z) -- (0*\y+18*\x,3.36408*\y+-14*\z);
\draw[] (0*\y+18*\x, 0*\y+-14*\z) -- (-3.02264*\y+18*\x,2.7216*\y+-14*\z);
\draw[] (-3.02264*\y+18*\x, 2.7216*\y+-14*\z) -- (0*\y+18*\x,3.36408*\y+-14*\z);
\draw[] (0*\y+18*\x, 0*\y+-14*\z) -- (-0.522642*\y+18*\x,2.45884*\y+-14*\z);
\draw[] (0*\y+18*\x, 3.36408*\y+-14*\z) -- (-0.522642*\y+18*\x,2.45884*\y+-14*\z);
\draw[] (-0.522642*\y+18*\x, 2.45884*\y+-14*\z) -- (-1.47756*\y+18*\x,2.03368*\y+-14*\z);
\draw[] (-3.02264*\y+18*\x, 2.7216*\y+-14*\z) -- (-1.47756*\y+18*\x,2.03368*\y+-14*\z);
\draw[] (0*\y+18*\x, 0*\y+-14*\z) -- (-1.47756*\y+18*\x,2.03368*\y+-14*\z);
\draw[] (0*\y+18*\x, 3.36408*\y+-14*\z) -- (-0.522642*\y+18*\x,2.45884*\y+-14*\z);

\draw[] (0*\y+18*\x, 3.36408*\y+-14*\z) -- (-0.954915*\y+18*\x,2.93893*\y+-14*\z);
\draw[] (-0.954915*\y+18*\x, 2.93893*\y+-14*\z) -- (-1.47756*\y+18*\x,2.03368*\y+-14*\z);
\draw[] (-3.02264*\y+18*\x,2.7216*\y+-14*\z) -- (-0.954915*\y+18*\x, 2.93893*\y+-14*\z);

\filldraw[black] (2*\y+18*\x-3.5*\y, 3.7*\y-14*\z-0.3*\y)  node[anchor=west] {\tiny{$a$}};
\filldraw[black] (1.3*\y+18*\x-3.5*\y, 1.5*\y-14*\z-0.3*\y)  node[anchor=west] {\tiny{$b$}};
\filldraw[black] (3.5*\y+18*\x-3.5*\y, 1.7*\y-14*\z-0.3*\y)  node[anchor=west] {\tiny{$c$}};
\filldraw[black] (2.7*\y+18*\x-3.5*\y, 2.2*\y-14*\z-0.3*\y)  node[anchor=west] {\tiny{$d'''$}};
\filldraw[black] (3.05*\y+18*\x-3.5*\y, 3.2*\y-14*\z-0.4*\y)  node[anchor=west] {\tiny{$r$}};
\filldraw[black] (1.1*\y+18*\x-3.5*\y, 2.8*\y-14*\z-0.3*\y)  node[anchor=west] {\tiny{$k$}};
\filldraw[black] (2.2*\y+18*\x-3.5*\y, 2.4*\y-14*\z-0.3*\y)  node[anchor=west] {\tiny{$h$}};
\filldraw[black] (2.3*\y+18*\x-3.5*\y, 1.5*\y-14*\z-0.3*\y)  node[anchor=west] {\tiny{$z$}};
\filldraw[black] (2.6*\y+18*\x-3.5*\y, 3.2*\y-14*\z-0.2*\y)  node[anchor=west] {\tiny{$h$}};
\filldraw[black] (1.6*\y+18*\x-3.5*\y, 3.1*\y-14*\z-0.3*\y)  node[anchor=west] {\tiny{$i$}};
\filldraw[black] (1.8*\y+18*\x-3.5*\y, 2.8*\y-14*\z-0.3*\y)  node[anchor=west] {\tiny{$r$}};

\filldraw[black] (0*\y+18*\x-3.5*\y, 0*\y+-19*\z-0.3*\y)  node[anchor=west] {\tiny{$(7,3): I \ \textcolor{bole}{[n, v, x, j]}$}};
\draw[] (0*\y+18*\x, 0*\y+-19*\z) -- (0*\y+18*\x,3.36408*\y+-19*\z);
\draw[] (0*\y+18*\x, 0*\y+-19*\z) -- (-3.02264*\y+18*\x,2.7216*\y+-19*\z);
\draw[] (-3.02264*\y+18*\x, 2.7216*\y+-19*\z) -- (0*\y+18*\x,3.36408*\y+-19*\z);
\draw[] (0*\y+18*\x, 3.36408*\y+-19*\z) -- (-1.97736*\y+18*\x,2.7216*\y+-19*\z);
\draw[] (-3.02264*\y+18*\x, 2.7216*\y+-19*\z) -- (-1.97736*\y+18*\x,2.7216*\y+-19*\z);
\draw[] (-1.97736*\y+18*\x, 2.7216*\y+-19*\z) -- (-0.17679*\y+18*\x,1.68204*\y+-19*\z);
\draw[] (0*\y+18*\x, 0*\y+-19*\z) -- (-0.17679*\y+18*\x,1.68204*\y+-19*\z);
\draw[] (0*\y+18*\x, 3.36408*\y+-19*\z) -- (-0.17679*\y+18*\x,1.68204*\y+-19*\z);
\draw[] (0*\y+18*\x, 0*\y+-19*\z) -- (-1.22207*\y+18*\x,1.68204*\y+-19*\z);
\draw[] (-3.02264*\y+18*\x, 2.7216*\y+-19*\z) -- (-1.22207*\y+18*\x,1.68204*\y+-19*\z);
\draw[] (-1.22207*\y+18*\x, 1.68204*\y+-19*\z) -- (-0.17679*\y+18*\x,1.68204*\y+-19*\z);

\filldraw[black] (2*\y+18*\x-3.5*\y, 3.7*\y-19*\z-0.3*\y)  node[anchor=west] {\tiny{$a$}};
\filldraw[black] (1.3*\y+18*\x-3.5*\y, 1.5*\y-19*\z-0.3*\y)  node[anchor=west] {\tiny{$b$}};
\filldraw[black] (3.5*\y+18*\x-3.5*\y, 1.7*\y-19*\z-0.3*\y)  node[anchor=west] {\tiny{$c$}};
\filldraw[black] (2.3*\y+18*\x-3.5*\y, 1.7*\y-19*\z-0.3*\y)  node[anchor=west] {\tiny{$z'$}};
\filldraw[black] (2.2*\y+18*\x-3.5*\y, 3.2*\y-19*\z-0.3*\y)  node[anchor=west] {\tiny{$m''$}};
\filldraw[black] (2.3*\y+18*\x-3.5*\y, 2.6*\y-19*\z-0.3*\y)  node[anchor=west] {\tiny{$n$}};
\filldraw[black] (1*\y+18*\x-3.5*\y, 2.8*\y-19*\z-0.3*\y)  node[anchor=west] {\tiny{$v$}};
\filldraw[black] (3*\y+18*\x-3.5*\y, 2.7*\y-19*\z-0.3*\y)  node[anchor=west] {\tiny{$j$}};
\filldraw[black] (3*\y+18*\x-3.5*\y, 1.4*\y-19*\z-0.3*\y)  node[anchor=west] {\tiny{$x$}};
\filldraw[black] (2.3*\y+18*\x-3.5*\y, 2.1*\y-19*\z-0.3*\y)  node[anchor=west] {\tiny{$v$}};
\filldraw[black] (1.3*\y+18*\x-3.5*\y, 2.5*\y-19*\z-0.3*\y)  node[anchor=west] {\tiny{$n$}};

\end{tikzpicture}
\end{center}
\caption{Two-loop vertex corrections contributing to a pole of leading order $5$ in the S-matrix. Graphs with the same capital letter (A, B, C, D, E, F, G, H, I) are connected by flipping one diagonal. Under each graph its atomic decomposition is reported: lists of letters in square brackets indicate propagators that are cut and represent atoms. Coloured lists correspond to atoms that cancel in the sum while black lists correspond to the non-cancelling atoms. The values of these non-cancelling atoms are written in terms of triangle areas omitting a common factor $i \bigl( \frac{\beta}{\sqrt{h}}\bigr)^5 \frac{\sigma_{ab\bar{c}}}{(\theta-i\theta_0)^2}$.}
\label{2_loops_vertex_corrections_with_flip_symbols_below_the_figures}
\end{figure}%
\clearpage 
}
The two blocks of diagrams
$$
\{(1,1),(1,2),(1,3),(2,1),(2,2),(2,3),(3,1),(3,2),(3,3)\}
$$
and 
$$
\{(4,1),(4,2),(4,3),(5,1),(5,2),(5,3),(6,1),(6,2),(6,3)\}
$$
in figure~\ref{2_loops_vertex_corrections_with_flip_symbols_below_the_figures} are connected by flipping internal propagators. Starting from a generic diagram of the first block we can generate a diagram of the second block acting with a finite number of flips.  In figure~\ref{2_loops_vertex_corrections_with_flip_symbols_below_the_figures} we identified with the same capital letters diagrams connected by flips. 
Sometimes a single type I flip is enough to connect the two diagrams in the different blocks;  examples are the tilings $(1,1)$ and $(5,1)$. According to the convention previously explained the little triangle $\Delta_{aih}$ common to both the tilings, can be equivalently labelled by $\Delta_a^{(1,1)}$ or $\Delta_a^{(5,\bullet)}$. In some other situations we need two flips of type II to connect the pair of diagrams. This second case is represented by the pair of graphs $(2,3)$ and $(5,3)$. To pass from the first diagram to the second one we need to pass from an intermediate non-planar diagram $(7,2)$ acting with two flips of type II: $(2,3)\to (7,2) \to (5,3)$.

For each diagram in figure~\ref{2_loops_vertex_corrections_with_flip_symbols_below_the_figures} we write its cut decomposition under it, representing with different colours the cuts that cancel in pairs.  
The values of the non-cancelling black atoms are reported under the diagrams omitting a multiplicative factor $i \Bigl( \frac{\beta}{\sqrt{h}}\Bigr)^5 \frac{\sigma_{ab\bar{c}}}{(\theta-i\theta_0)^2}$ coming by performing the integration with the additional relation~\eqref{Multiplication_of_f_in_arbitrary_nested_vertex}. It holds that 
\begin{equation}
\label{Sum_rows_1_2_4_5_2loops_vertex}
\begin{aligned}
\sum_{i=1}^3 \vv^{(1,i)}&\rightsquigarrow i \Bigl( \frac{\beta}{\sqrt{h}}\Bigr)^5 \frac{\sigma_{ab\bar{c}}}{(\theta-i\theta_0)^2} \Delta^{(1,\bullet)}_c \hspace{3mm}, \hspace{3mm} \sum_{i=1}^3 \vv^{(2,i)} \rightsquigarrow i \Bigl( \frac{\beta}{\sqrt{h}}\Bigr)^5 \frac{\sigma_{ab\bar{c}}}{(\theta-i\theta_0)^2} \Delta^{(2,\bullet)}_c,\\     
\sum_{i=1}^3 \vv^{(4,i)}&\rightsquigarrow i \Bigl( \frac{\beta}{\sqrt{h}}\Bigr)^5 \frac{\sigma_{ab\bar{c}}}{(\theta-i\theta_0)^2} \Delta^{(4,\bullet)}_c \hspace{3mm}, \hspace{3mm} \sum_{i=1}^3 \vv^{(5,i)}\rightsquigarrow i \Bigl( \frac{\beta}{\sqrt{h}}\Bigr)^5 \frac{\sigma_{ab\bar{c}}}{(\theta-i\theta_0)^2} \Delta^{(5,\bullet)}_c .
\end{aligned}
\end{equation}
As above, we use curly arrows instead of equalities in the relations in~\eqref{Sum_rows_1_2_4_5_2loops_vertex}, since in the sums we are omitting the contributions of the atoms that cancel between the two sums.

We also notice that
\begin{equation}\begin{split}
\vv^{(3,3)}+\vv^{(6,3)} \rightsquigarrow i \Bigl( \frac{\beta}{\sqrt{h}}\Bigr)^5 \frac{\sigma_{ab\bar{c}}}{(\theta-i\theta_0)^2} \Bigl(\frac{\Delta_{j m'' n}+\Delta_{j m v}}{\Delta_{aj}} \ \Delta_{cjx} \Bigr)&= i \Bigl( \frac{\beta}{\sqrt{h}}\Bigr)^5 \frac{\sigma_{ab\bar{c}}}{(\theta-i\theta_0)^2} \Delta_{cjx}\\
&=i \Bigl( \frac{\beta}{\sqrt{h}}\Bigr)^5 \frac{\sigma_{ab\bar{c}}}{(\theta-i\theta_0)^2} \Delta_c^{(3,3)}\\
&=i \Bigl( \frac{\beta}{\sqrt{h}}\Bigr)^5 \frac{\sigma_{ab\bar{c}}}{(\theta-i\theta_0)^2} \Delta_c^{(6,3)}.
\end{split}\label{Sum_rows_3_6_2loops_vertex}
\end{equation}
The sum of \eqref{Sum_rows_1_2_4_5_2loops_vertex} and \eqref{Sum_rows_3_6_2loops_vertex} is easily performed noting that $\Delta^{(4,\bullet)}_c=\Delta^{(3,1)}_c$ and $\Delta^{(5,\bullet)}_c=\Delta^{(3,2)}_c$, so that

\begin{equation*}
\begin{split}
\eqref{Sum_rows_1_2_4_5_2loops_vertex}+\eqref{Sum_rows_3_6_2loops_vertex}
&=i \Bigl( \frac{\beta}{\sqrt{h}}\Bigr)^5 \frac{\sigma_{ab\bar{c}}}{(\theta-i\theta_0)^2} \Bigl(\Delta^{(1,\bullet)}_c+\Delta^{(2,\bullet)}_c+\Delta_c^{(3,3)}+\Delta^{(4,\bullet)}_c+\Delta^{(5,\bullet)}_c\Bigr)\\
&=i \Bigl( \frac{\beta}{\sqrt{h}}\Bigr)^5 \frac{\sigma_{ab\bar{c}}}{(\theta-i\theta_0)^2} \Bigl(\Delta^{(1,\bullet)}_c+\Delta^{(2,\bullet)}_c+\Delta_c^{(3,\bullet)} \Bigr)= i \Bigl( \frac{\beta}{\sqrt{h}}\Bigr)^5 \frac{\sigma_{ab\bar{c}}}{(\theta-i\theta_0)^2} \Delta_{ab},
\end{split}
\end{equation*}
where in the second equality we used 
$$
\Delta_c^{(3,1)}+\Delta_c^{(3,2)}+\Delta_c^{(3,3)}=\Delta_c^{(3,\bullet)}.
$$
A similar simplification could have been obtained by using the fact that 
$$
\Delta^{(6,1)}_c+\Delta^{(6,2)}_c+\Delta^{(6,3)}_c=\Delta^{(6,\bullet)}_c.$$
In this second case we obtain
$$
\eqref{Sum_rows_1_2_4_5_2loops_vertex}+\eqref{Sum_rows_3_6_2loops_vertex}=\Delta^{(1,\bullet)}_c+\Delta^{(2,\bullet)}_c+\Delta_c^{(6,3)}+\Delta^{(4,\bullet)}_c+\Delta^{(5,\bullet)}_c=\Delta^{(4,\bullet)}_c+\Delta^{(5,\bullet)}_c+\Delta_c^{(6,\bullet)}=\Delta_{ab}.
$$
The leading order singularity of the vertex contributing to a $5^{\text{th}}$-order pole in the S-matrix is therefore given by
\begin{equation}
\label{two_loop_vertex_correction_second_order_pole}
\sum_{i=1}^6 \sum_{j=1}^3 \vv^{(i,j)} \rightsquigarrow i \Bigl( \frac{\beta}{\sqrt{h}}\Bigr)^5 \frac{\sigma_{ab\bar{c}}}{(\theta-i\theta_0)^2} \Delta_{ab}=i \Bigl( \frac{\beta^2}{2h}\Bigr)^2 \frac{C_{ab\bar{c}}}{(\theta-i\theta_0)^2}.
\end{equation}
Unlike the one-loop vertex correction found in~\eqref{sum_of_the_three_types_of_vertex_corrections}, the residue of \eqref{two_loop_vertex_correction_second_order_pole} is purely imaginary. This implies that the one-particle reducible part of the S-matrix on the $5^{\text{th}}$-order pole acquires a different sign compared to the $3^{\text{rd}}$-order pole S-matrix~\eqref{one_particle_reducible_sum_two_loops_third_order_pole}. In this case we have
\begin{equation}
\frac{1}{8i \Delta_{ab}} i \Bigl( \frac{\beta^2}{2h}\Bigr)^2 \frac{C_{ab\bar{c}}}{(\theta-i\theta_0)^2} \frac{i}{s-s_0} i \Bigl( \frac{\beta^2}{2h}\Bigr)^2 \frac{C_{ab\bar{c}}}{(\theta-i\theta_0)^2} =i \Bigl( \frac{\beta^2}{2h} \Bigr)^5 \frac{1}{(\theta-i\theta_0)^5},
\end{equation}
that exactly matches the Laurent coefficient of order $(\theta-i\theta_0)^{-5}$
expected from the bootstrapped relation~\eqref{S_matrix_contributions_on_the_poles_expanded_and_containing_all_the_leading_singularities} for $N=2$ and $\nu=i$. 
Surprisingly the one-particle reducible diagrams reproduce the result expected from the bootstrap. In total there are $21 \times 21=441$ such diagrams, since there are in general $21$ different tilings of the vertex, as reported in figure~\ref{2_loops_vertex_corrections_with_flip_symbols_below_the_figures}. Analogously to what happened for the $3^{\text{rd}}$-order pole, we can then start applying flips on each of the one-particle reducible diagrams to generate a network of Feynman diagrams contributing to the $5^{\text{th}}$-order pole. The boundary of the $3^{\text{rd}}$-order pole network was composed of 9 one-particle reducible diagrams ($6$ on the boundary of the disk plus the $3$ separate diagrams $D^{(-2)}$, $D^{(-1)}$ and $D^{(0)}$ in figure~\ref{Network_of_onshell_diagrams_contributing_to_the_third_order_poles}); similarly here we expect a boundary composed by $\sim 441$ different diagrams. Therefore, the internal part of the network contains a tremendous number of graphs. It is remarkable that all the one-particle irreducible graphs internal to the network have to cancel one another and do not contribute to the final answer. We expect the simplification will work in a similar way to that seen in figure~\ref{Bulk_contributions_which_cancel_in_the_third_order_pole_network} and will again be reminiscent of Gauss's theorem in the space of Feynman diagrams. However, due to the huge amount of graphs, we did not manage to make such simplification manifest and we leave the problem open for future investigation.

\section{Conclusions}

We have revisited the Coleman-Thun mechanism for the generation of higher-order poles~\cite{Coleman:1978kk} in the ADE series of affine Toda field theories. 
Even though some of the results discussed in this paper were already known from 
\cite{Braden:1990wx},
the underlying simplification between loop diagrams to reproduce universal residues was not explained in that paper and has been the main focus of our study. 
As observed in~\cite{Braden:1990wx}, Feynman diagrams contributing to higher-order poles
are connected by flipping their internal propagators and can be arranged inside networks.
A naive guess could be that the leading singularity obtained by summing a pair of diagrams connected by flipping an internal propagator cancels.
This is what we might expect from an analogy with Feynman diagrams appearing in two-to-two tree-level processes: in that case, the cancellation between pairs of flipped diagrams is responsible for the absence of singularities in inelastic tree-level amplitudes, and is a necessary condition for these amplitudes to vanish.
However, this guess turns out to be wrong at the loop level where the cancellation mechanism between diagrams is more subtle.

Adopting the technique highlighted in~\cite{First_loop_paper_sagex}, we showed that, in the neighbourhood of a Landau singularity, a loop diagram can be 
written as different sums of products of singular tree-level graphs.
The different ways in which the cuts can be taken depend on the on-shell geometry of the loop, which is determined by the masses of the interacting particles, together with the contour chosen to perform the integral with Cauchy's theorem. 
Only after suitably cutting these loop diagrams into smaller constituents, which we called atoms, were we able to identify an underlying simplification structure: 
we organised the atoms within sets (depicted with different colours in figure~\ref{Bulk_contributions_which_cancel_in_the_third_order_pole_network}) characterised by a common piece but differing by tree-level diagrams connected by flipping internal propagators. 
Most of these sets are composed of pairs of atoms; however, in the network in figure~\ref{Bulk_contributions_which_cancel_in_the_third_order_pole_network} we observe also a big set composed of six atoms (coloured blue).
Sums of atoms cancel separately in the different sets due to the tree-level perturbative integrability of affine Toda field theories. 
In the end, only a small number of surviving atoms, that for $3^{\text{rd}}$-order poles are located at the boundary of the network, reproduce the expected value of the bootstrapped S-matrix on the pole.

How integrability reveals itself in perturbation theory through the simplification of big networks into small numbers of fundamental atoms is fascinating. Even though we have been able to identify this underlying simplification, we based this on a case-by-case investigation performed over different processes and different theories. Many features of these models remain unclear. In particular, a universal explanation of what we observe is missing: at the moment we do not have an algorithm to generate networks of diagrams contributing to singularities of leading orders bigger than three. Moreover, we do not have a clever procedure to cut diagrams into atoms for generic networks. It would be interesting if a theorem existed proving that the internal part of generic networks contributing to higher-order Landau singularities cancel; in this manner it would be possible to compute the results for the residues at the poles just by looking at the boundaries of these networks. 

For $3^{\text{rd}}$-order poles,  the simplification between atoms is reminiscent of a sort of flatness property of the network. 
Indeed in figure~\ref{Bulk_contributions_which_cancel_in_the_third_order_pole_network} we identified six blue atoms differing by diagrams contributing to a five-point tree-level process: these diagrams can be arranged on the vertices of a hexagon (see figure~\ref{fig:5_point_dual_Feyn_diagrams})
and cancel within a closed path.
Going further, while processes with $5$ external particles have collections of cancelling diagrams arranged on the vertices of planar polygons, it has to be possible to arrange tree-level diagrams with $6$ external particles on the vertices of polyhedra: even in this case, the singular part of the sum of all Feynman diagrams comprising this polyhedron has to be null. 
This discussion extends to tree-level processes with multiple external legs where collections of maximally singular diagrams have to lie on the vertices of polytopes, and just as the hexagon in figure~\ref{fig:5_point_dual_Feyn_diagrams} is embedded in the network  depicted in figure~\ref{Bulk_contributions_which_cancel_in_the_third_order_pole_network}, it is reasonable to expect that these polytopes will be embedded in collections of cancelling atoms in network decompositions of higher-order Landau singularities. It would be very interesting to understand this underlying simplification for Landau singularities of arbitrary order.

A natural first step
is to study Landau singularities of order five.
We have shown that the residue of the S-matrix at a $5^{\text{th}}$-order pole expected from the bootstrap can be obtained by summing over only one-particle reducible diagrams: these diagrams are obtained by two vertex corrections of the form shown in figure~\ref{2_loops_vertex_corrections_with_flip_symbols_below_the_figures} connected by an on-shell bound state. To complete our study, it will be important to find the full network of graphs contributing to such poles and prove that the internal part of the network, composed of one-particle irreducible graphs, cancels. This would also require greater precision as to what constitutes the boundary of the network, which at the moment remains unresolved for higher-order poles. Finally, we reiterate that a better connection with the underlying root system and Coxeter geometry should be fundamental to explaining this cancellation mechanism 
in a universal way 
and ultimately understanding perturbative integrability at the quantum level.

\vskip 20pt
\noindent
{\bf Acknowledgments}

\noindent
We thank Gabriele Dian, Ben Hoare and Charles Young for related discussions. This work has received funding from the  European Union's  Horizon  2020  research  and  innovation programme under the Marie  Sk\l odowska-Curie  grant  agreement  No.\ 764850 \textit{``SAGEX''}, and from the STFC under consolidated grant ST/T000708/1 “Particles, Fields and Spacetime”. It was also supported in part by the National Science Foundation under Grant No.\ NSF PHY-1748958 and by 
STARS@UNIPD, under project ``Exact-Holography''.

\vskip 20pt


\end{document}